\input harvmac.tex

\input epsf.tex


\def\figin{\epsfcheck\figin}\def\figins{\epsfcheck\figins}
\def\epsfcheck{\ifx\epsfbox\UnDeFiNeD
\message{(NO epsf.tex, FIGURES WILL BE IGNORED)}
\gdef\figin##1{\vskip2in}\gdef\figins##1{\hskip.5in}
\else\message{(FIGURES WILL BE INCLUDED)}%
\gdef\figin##1{##1}\gdef\figins##1{##1}\fi}
\def\DefWarn#1{}
\def\figinsert{\goodbreak\midinsert}
\def\ifig#1#2#3{\DefWarn#1\xdef#1{fig.~\the\figno}
\writedef{#1\leftbracket fig.\noexpand~\the\figno}%
\figinsert\figin{\centerline{#3}}\medskip\centerline{\vbox{\baselineskip12pt
\advance\hsize by -1truein\noindent\footnotefont{\bf
Fig.~\the\figno:} #2}}
\bigskip\endinsert\global\advance\figno by1}


\def\stokesparameter{\gamma}
\def\wshifted{w_s}
\def\shiftedu{u_s}
\def\shiftedv{v_s}


\lref\AldayHR{
  L.~F.~Alday and J.~M.~Maldacena,
  JHEP {\bf 0706}, 064 (2007)
  [arXiv:0705.0303 [hep-th]].
}

\lref\AldayYW{
  L.~F.~Alday and R.~Roiban,
  Phys.\ Rept.\  {\bf 468}, 153 (2008)
  [arXiv:0807.1889 [hep-th]].
}

\lref\DrummondAUA{
  J.~M.~Drummond, G.~P.~Korchemsky and E.~Sokatchev,
  Nucl.\ Phys.\  B {\bf 795}, 385 (2008)
  [arXiv:0707.0243 [hep-th]].
}

\lref\BrandhuberYX{
  A.~Brandhuber, P.~Heslop and G.~Travaglini,
  Nucl.\ Phys.\  B {\bf 794}, 231 (2008)
  [arXiv:0707.1153 [hep-th]].
}

\lref\AldayMF{
  L.~F.~Alday and J.~M.~Maldacena,
  JHEP {\bf 0711}, 019 (2007)
  [arXiv:0708.0672 [hep-th]].
}

\lref\BonoraAY{
  L.~Bonora, C.~P.~Constantinidis, L.~A.~Ferreira and E.~E.~Leite,
  J.\ Phys.\ A  {\bf 36}, 7193 (2003)
  [arXiv:hep-th/0208175].
}

\lref\McCoyCD{
  B.~M.~McCoy, C.~A.~Tracy and T.~T.~Wu,
  J.\ Math.\ Phys.\  {\bf 18}, 1058 (1977).
}

\lref\ZamolodchikovUW{
  A.~B.~Zamolodchikov,
  Nucl.\ Phys.\  B {\bf 432}, 427 (1994)
  [arXiv:hep-th/9409108].
}

\lref\brandhuber{
  A.~Brandhuber, P.~Heslop and G.~Travaglini,
  Nucl.\ Phys.\  B {\bf 794}, 231 (2008)
  [arXiv:0707.1153 [hep-th]].
}

\lref\DeVegaXC{
  H.~J.~De Vega and N.~G.~Sanchez,
  Phys.\ Rev.\  D {\bf 47}, 3394 (1993).
}

\lref\GaiottoCD{
  D.~Gaiotto, G.~W.~Moore and A.~Neitzke,
  arXiv:0807.4723 [hep-th].
}

\lref\GMNtwo{
  D.~Gaiotto, G.~W.~Moore and A.~Neitzke, to appear
}

\lref\HitchinVP{
  N.~J.~Hitchin,
  Proc.\ Lond.\ Math.\ Soc.\  {\bf 55}, 59 (1987).
}

\lref\DrummondAU{
  J.~M.~Drummond, J.~Henn, G.~P.~Korchemsky and E.~Sokatchev,
  arXiv:0712.1223 [hep-th].
}

\lref\DeVegaXC{
  H.~J.~De Vega and N.~G.~Sanchez,
  Phys.\ Rev.\  D {\bf 47}, 3394 (1993).
}

\lref\JevickiAA{
  A.~Jevicki, K.~Jin, C.~Kalousios and A.~Volovich,
  JHEP {\bf 0803}, 032 (2008)
  [arXiv:0712.1193 [hep-th]].
}

\lref\MaldacenaIM{
  J.~M.~Maldacena,
  Phys.\ Rev.\ Lett.\  {\bf 80}, 4859 (1998)
  [arXiv:hep-th/9803002].
}

\lref\ReyIK{
  S.~J.~Rey and J.~T.~Yee,
  Eur.\ Phys.\ J.\  C {\bf 22}, 379 (2001)
  [arXiv:hep-th/9803001].
}

\lref\AharonyUG{
  O.~Aharony, O.~Bergman, D.~L.~Jafferis and J.~Maldacena,
  JHEP {\bf 0810}, 091 (2008)
  [arXiv:0806.1218 [hep-th]].
}

\lref\BerensteinIJ{
  D.~E.~Berenstein, R.~Corrado, W.~Fischler and J.~M.~Maldacena,
  Phys.\ Rev.\  D {\bf 59}, 105023 (1999)
  [arXiv:hep-th/9809188].
}

\lref\KorchemskayaJE{
  I.~A.~Korchemskaya and G.~P.~Korchemsky,
  Phys.\ Lett.\  B {\bf 287}, 169 (1992).
}

\lref\BassettoXD{
  A.~Bassetto, I.~A.~Korchemskaya, G.~P.~Korchemsky and G.~Nardelli,
  Nucl.\ Phys.\  B {\bf 408}, 62 (1993)
  [arXiv:hep-ph/9303314].
}

\lref\BernAP{
  Z.~Bern, L.~J.~Dixon, D.~A.~Kosower, R.~Roiban, M.~Spradlin, C.~Vergu and A.~Volovich,
  Phys.\ Rev.\  D {\bf 78}, 045007 (2008)
  [arXiv:0803.1465 [hep-th]].
}
\lref\DrummondAQ{
  J.~M.~Drummond, J.~Henn, G.~P.~Korchemsky and E.~Sokatchev,
  arXiv:0803.1466 [hep-th].
}
\lref\KazakovQF{
  V.~A.~Kazakov, A.~Marshakov, J.~A.~Minahan and K.~Zarembo,
  JHEP {\bf 0405}, 024 (2004)
  [arXiv:hep-th/0402207].
}
\lref\PohlmeyerNB{
  K.~Pohlmeyer,
  Commun.\ Math.\ Phys.\  {\bf 46}, 207 (1976).
}

\lref\CecottiRM{
  S.~Cecotti and C.~Vafa,
  Commun.\ Math.\ Phys.\  {\bf 158}, 569 (1993)
  [arXiv:hep-th/9211097].
}

\lref\BershadskyQY{
  M.~Bershadsky, C.~Vafa and V.~Sadov,
  Nucl.\ Phys.\  B {\bf 463}, 420 (1996)
  [arXiv:hep-th/9511222].
}

\lref\BeisertEZ{
  N.~Beisert, B.~Eden and M.~Staudacher,
  J.\ Stat.\ Mech.\  {\bf 0701}, P021 (2007)
  [arXiv:hep-th/0610251].
}

\lref\MiramontesWT{
  J.~L.~Miramontes,
  JHEP {\bf 0810}, 087 (2008)
  [arXiv:0808.3365 [hep-th]].
}

\lref\LuKB{
  H.~Lu, M.~J.~Perry, C.~N.~Pope and E.~Sezgin,
  arXiv:0812.2218 [hep-th].
}

\lref\FreyhultPZ{
  L.~Freyhult, A.~Rej and M.~Staudacher,
  J.\ Stat.\ Mech.\  {\bf 0807}, P07015 (2008)
  [arXiv:0712.2743 [hep-th]].
}

\lref\AnastasiouKN{
  C.~Anastasiou, A.~Brandhuber, P.~Heslop, V.~V.~Khoze, B.~Spence and G.~Travaglini,
  arXiv:0902.2245 [hep-th].
}

\lref\BernIZ{
  Z.~Bern, L.~J.~Dixon and V.~A.~Smirnov,
  Phys.\ Rev.\  D {\bf 72}, 085001 (2005)
  [arXiv:hep-th/0505205].
}

\lref\KomargodskiWA{
  Z.~Komargodski,
  JHEP {\bf 0805}, 019 (2008)
  [arXiv:0801.3274 [hep-th]].
}

\lref\AldayCG{
  L.~F.~Alday,
  Fortsch.\ Phys.\  {\bf 56}, 816 (2008)
  [arXiv:0804.0951 [hep-th]].
}

\lref\AldayHE{
  L.~F.~Alday and J.~Maldacena,
  JHEP {\bf 0711}, 068 (2007)
  [arXiv:0710.1060 [hep-th]].
}

\lref\BarbashovQZ{
  B.~M.~Barbashov, V.~V.~Nesterenko and A.~M.~Chervyakov,
  Commun.\ Math.\ Phys.\  {\bf 84}, 471 (1982).
}

\lref\KruczenskiGT{
  M.~Kruczenski,
  Phys.\ Rev.\ Lett.\  {\bf 93}, 161602 (2004)
  [arXiv:hep-th/0311203].
}

\lref\FG{
  V.V. Fock, A.B. Goncharov,
  arXiv:math/0311149}

\lref\GrigorievJQ{
  M.~Grigoriev and A.~A.~Tseytlin,
  Int.\ J.\ Mod.\ Phys.\  A {\bf 23}, 2107 (2008)
  [arXiv:0806.2623 [hep-th]].
}

\lref\JevickiUZ{
  A.~Jevicki and K.~Jin,
  arXiv:0903.3389 [hep-th].
}

\lref\GiddingsYD{
  S.~B.~Giddings, F.~Hacquebord and H.~L.~Verlinde,
  Nucl.\ Phys.\  B {\bf 537}, 260 (1999)
  [arXiv:hep-th/9804121].
}

\lref\BonelliYT{
  G.~Bonelli, L.~Bonora and F.~Nesti,
  Phys.\ Lett.\  B {\bf 435}, 303 (1998)
  [arXiv:hep-th/9805071].
}

\lref\OoguriME{
  H.~Ooguri and C.~Vafa,
  Phys.\ Rev.\ Lett.\  {\bf 77}, 3296 (1996)
  [arXiv:hep-th/9608079].
}

\lref\SeibergNS{
  N.~Seiberg and S.~H.~Shenker,
  Phys.\ Lett.\  B {\bf 388}, 521 (1996)
  [arXiv:hep-th/9608086].
}

\lref\CecottiME{
  S.~Cecotti and C.~Vafa,
  Nucl.\ Phys.\  B {\bf 367}, 359 (1991).
}

\lref\KruczenskiWG{
  M.~Kruczenski,
  JHEP {\bf 0508}, 014 (2005)
  [arXiv:hep-th/0410226].
}
\lref\SHORT{
  L.~F.~Alday and J.~Maldacena,
  arXiv:0903.4707 [hep-th].
}

\lref\DoreyVP{
  N.~Dorey and M.~Losi,
  arXiv:0812.1704 [hep-th].
}

\lref\DornKQ{
  H.~Dorn, G.~Jorjadze and S.~Wuttke,
  arXiv:0903.0977 [hep-th].
}

\Title{\vbox{\baselineskip12pt \hbox{} \hbox{
} }} {\vbox{\centerline{ Null polygonal  Wilson loops  and
  }
\centerline{
minimal surfaces in Anti-de-Sitter space }
}}
\bigskip
\centerline{Luis F. Alday and Juan Maldacena }
\bigskip
\centerline{ \it  School of Natural Sciences, Institute for
Advanced Study} \centerline{\it Princeton, NJ 08540, USA}

\vskip .3in \noindent
 We consider minimal surfaces in three dimensional
 anti-de-Sitter space that end at the AdS boundary
 on  a polygon given by a sequence of null segments.
 The problem can be reduced to a certain generalized Sinh-Gordon equation  and
 to SU(2) Hitchin equations. We describe in detail the mathematical problem that
 needs to be solved. This problem is mathematically the same as the one studied by
 Gaiotto, Moore and Neitzke in the context of the moduli space of certain supersymmetric
 theories. Using their results we can find the explicit answer  for the area of a
 surface that ends on an eight-sided
 polygon. Via the gauge/gravity duality this can also be interpreted as a certain
 eight-gluon scattering amplitude at strong coupling. In addition, we give fairly explicit
 solutions for regular polygons.


 \Date{ }


\newsec{Introduction }

Recently there has been some interest in
Wilson loops that consist of a sequence of light-like segments.
 These are interesting
for several reasons. First, they are a simple subclass of Wilson loops
 which depend on a finite number
of parameters, the positions of the cusps. Second, they are Lorentzian objects with no obvious
Euclidean counterpart. Finally, it was shown that they are connected to scattering amplitudes
in gauge theories \refs{\AldayHR,\DrummondAUA,\BrandhuberYX,\BernAP,\DrummondAQ},
for a review see \AldayYW .

In this paper we study these Wilson loops at strong coupling by using the gauge/string duality.
One then considers
 the classical equations for a string worldsheet that ends on the null polygon at the boundary of
$AdS$ space.
We consider a special class of
 null polygons   which can be embedded in  a two dimensional subspace, which we can take as an
$R^{1,1}$ subspace of the boundary of $AdS$. For these loops, the string worldsheet lives in an
$AdS_3$ subspace of the full $AdS_d$ space, $d\geq 3$. Since we are merely studying the classical
equations, the solutions can be embedded in any string theory geometry which contains an $AdS_3$
factor. We are interested in the   area of these surfaces.
 This is
a completely geometric problem.

 One motivation for studying these classical solutions is that they will enable us to see how
 to apply integrability to find the quantities of interest. In fact, in the analysis  of
 the spectrum of operators, it was useful to see how integrability determines the solution
 of the classical problem,
  see for example  \KazakovQF .
  Of course, one is  eventually  interested
 in solving this problem for the full quantum theory. Here, we focus simply on the
 classical problem, which, hopefully, will be useful for the solution of the quantum problem in
 the future.

 The area of  the worldsheet  depends  on the positions of the cusps at
 the boundary. Conformal symmetry implies that the area depends only on   cross ratios of these
 positions, up
 to a simple term which arises due to the regulator.
   For null polygons living in $R^{1,1}$ we have six conformal
  generators that move the positions of the cusps. To have a closed null polygon we need an even
  number of cusps, since each cusp joins a left moving with a right moving null line.
  In a situation with  $2n$ cusps, we have
  $2(n-3)$ cross ratios. The cusp positions
   are determined by $n$ $x^+_i$ and $n$ $x_j^-$ coordinates. We have
  $n-3$ cross ratios made out of $x^+_i$ and $n-3$ from the $x^-_i$.
  The first time we have a non-trivial dependence on the cross ratios is for $n=4$ which
  corresponds to an eight-sided null polygon.

 In order to analyze the problem one can use a  Pohlmeyer type reduction
 \refs{\PohlmeyerNB,\DeVegaXC,\JevickiAA} \foot{See also \refs{\GrigorievJQ,\MiramontesWT,\JevickiUZ,\DornKQ}.}.
 This maps the problem
 of strings moving in $AdS_3$ to a problem involving a single field $\alpha$ which obeys
 a generalized Sinh-Gordon equation. We can view this as a sophisticated gauge choice, which is
 similar to a light-cone gauge, where we are left just with the physical degrees of
 freedom. In addition to the field $\alpha$ one has a holomorphic polynomial $p(z)$.
 The worldsheet is the whole complex plane \foot{One can consider also spatial Wilson loops, such at the circular Wilson loop, in this case the world-sheet has the topology of a disk.} and the degree of the polynomial determines the
 number of cusps of the Wilson loop.  The spacetime embedding of the surface
 is obtained by solving an auxiliary linear problem involving the field $\alpha$. This auxiliary
 linear problem displays Stokes phenomenon as $z \to \infty$. In other words,  depending of the
 angular sector in $z$ the solution takes different asymptotic forms.
 The various angular  sectors are associated to the various
 cusps for the Wilson loop. We  explain this in detail and we provide
 formulas for the regularized area in terms of the solution to the generalized sinh-gordon equation.

 The same mathematical problem
  appears in the study of $SU(2)$ Hitchin equations \HitchinVP .
  In fact, it was already pointed out in
 \HitchinVP ,  that the Hitchin equations are related to harmonic maps into $SU(2)$.
  Since $AdS_3 = SL(2)$
 we are not surprised by such a relation.
 Interestingly,  these Hitchin equations also appear in the study of the supersymmetric
 vacua of certain gauge theories \refs{\GaiottoCD,\GMNtwo} .
 For example, we could consider the theory that results from
 wrapping a D4 brane on a two dimensional Riemann surface. The moduli space of vacua of the
 resulting $2+1$ dimensional theory is the same as the Hitchin moduli space. This connection
 is specially useful because Gaiotto-Moore and Neitske have studied this problem
 \refs{\GaiottoCD,\GMNtwo},
  exploiting its
 integrability and obtaining exact solutions in some cases.
 The explicit solution found
 in \GaiottoCD\ can be used to find the area as a function of the cross ratios for the
 simplest non-trivial case, which is the case of an eight-sided null polygon.

 This paper is organized as follows.
 In section two we explain the  reduction to the generalized Sinh-Gordon problem.
 In section three we explore the large $z$ asymptotics of the solution and we
 explain how to relate the spacetime cross ratios to the parameters in the Sinh-Gordon problem.
 In section four we obtain fairly explicit solutions for regular polygons.
 In section five we explain how to regularize the area using a physical spacetime regulator.
 In section six we display the result for the octagonal  Wilson loop  by using the results in
\GaiottoCD\ and in section seven we discuss our results and some
possible future directions. Finally, we defer many technicalities
to various appendices. For a short version of this paper see \SHORT .

\newsec{Sinh-Gordon model from strings on $AdS_3$}

\subsec{Reducing  strings on $AdS_3$ to the generalized Sinh-Gordon equation}

Classical strings in $AdS$ spaces can be described by a reduced model,
which takes into account both  the equations of motion and the Virasoro constraints
\refs{\PohlmeyerNB,\BarbashovQZ,\DeVegaXC}.
$AdS_d$ space can be written as the following surface in $R^{2,d-1}$
 \eqn\emb{\vec{Y}.\vec{Y}=-Y_{-1}^2-Y_{0}^2+Y_1^2+...+Y_{d-1}^2=-1}
 In terms of embedding coordinates, the conformal gauge equations of motion  and Virasoro
 constraints are
 \eqn\eom{\partial \bar \partial \vec{Y}-(\partial \vec{Y}.\bar{\partial} \vec{Y})\vec{Y}=0 ~,~~~~~~
  \partial \vec{Y}.\partial \vec{Y}=\bar \partial \vec{Y}.\bar \partial \vec{Y}=0 }
 We are interested in spacelike surfaces embedded in $AdS$.
We parametrize the world-sheet in terms of complex variables $z$ and $\bar{z}$.
 For the case of $AdS_3$ the above system can be reduced to the generalized sinh-Gordon model. We start by defining
 \eqn\sinhg{\eqalign{e^{2 \alpha(z,\bar{z})} =& {1 \over 2 } \partial \vec{Y}.\bar \partial \vec{Y},\cr
N_a = &  { e^{-2 \alpha} \over 2 } \epsilon_{abcd}Y^b \partial Y^c \bar \partial Y^d ~,
\cr p= & -{1 \over 2}  \vec{N}.\partial^2 \vec{Y}~,~~~~~~~ \bar p=  {1 \over 2}  \vec{N}.{ \bar \partial}^2 \vec{Y}
}}
Notice that $\vec{N}.\vec{Y}=\vec{N}.\partial \vec{Y}=\vec{N}.\bar \partial \vec{Y}=0$ and $\vec{N}.\vec{N}=1$. $\vec{N}$ is a purely imaginary vector whose imaginary part is a time like vector orthogonal to the space-like surface we are considering.
  It can be shown directly from  \eom\
  that $p=p(z)$ is a holomorphic function \foot{For the real solutions considered in this paper, $p(z)$ and $\bar{p}(\bar{z})$ are complex conjugates.
   This condition could in principle be relaxed.}. Later we will see that
  this result arises also as a consistency condition. Let us introduce the following basis of four-vectors
\eqn\hybridbasis{\eqalign{q_1= &\vec{Y}~,~~~~~
q_2= { e^{-\alpha }  } \bar{\partial} \vec{Y}~,~~~~~
q_3={ e^{-\alpha } } {\partial \vec{Y}}~,~~~~~
q_4=\vec{N}~, \cr
q_1^2= & -1~,~~~~~~~~~~~~q_2.q_3=2, ~~~~ q_4^2=1
}}
with the remaining $q_i.q_j=0$.
Notice that there is an internal  $SO(2,2)$ group acting on
the $i$ indices.
Besides, each element of the basis is a space-time vector and there is a
$SO(2,2)$ symmetry associated to space-time as well.
 Using the equivalence between $SO(2,2)$ and $SL(2)\times SL(2)$
 we denote the two $SO(2,2)$ indices,
 internal and space-time,
  by $\alpha,\dot{\alpha}$ and $a,\dot{a}$,   respectively.
  Next, consider the following matrix
\eqn\Wmat{\eqalign{W= {1 \over 2 }
\left(\matrix{   q_1+q_4 &q_2\cr q_3& q_1-q_4 }\right)}}
Being precise, the elements of this matrix have indices $W_{\alpha \dot{\alpha},a\dot{a}}$. The first two indices denotes rows and columns in the above matrix, while the other two are space-time indices. The equations of motion can be written as
\eqn\orthbasis{\eqalign{\partial W_{\alpha \dot{\alpha},a\dot{a}}+(B_z^L)_{\alpha}^{~\beta}W_{\beta\dot{\alpha},a\dot{a}}+(B_z^R)_{\dot \alpha}^{~\dot \beta}W_{\alpha \dot{\beta},a\dot{a}}=0\cr
\bar \partial W_{\alpha \dot{\alpha},a\dot{a}}+(B_{\bar{z}}^L)_{\alpha}^{~\beta}W_{\beta\dot{\alpha},a\dot{a}}+(B_{\bar z}^R)_{\dot \alpha}^{~\dot \beta}W_{\alpha \dot{\beta},a\dot{a}}=0
}}
with the $SL(2)$ connections $B^{L,R}$ given by
\eqn\FGhybrid{\eqalign{B_z^L=\left(\matrix{{1 \over 2} \partial \alpha &- {e^{  \alpha}  }\cr-{e^{-  \alpha} p(z) }  & -{1 \over 2} \partial \alpha } \right),~~~~~ B_{\bar{z}}^L=\left(\matrix{-{1 \over 2} \bar{\partial} \alpha & -{e^{-  \alpha } {\bar p }(\bar{z})  }\cr- e^{ \alpha}    & {1 \over 2} \bar{\partial} \alpha }\right) \cr
B_z^R=\left(\matrix{-{1 \over 2} \partial \alpha &{e^{-  \alpha} p(z)  }   \cr -{e^{  \alpha}  } & {1 \over 2} \partial \alpha }\right),~~~~~
B_{\bar{z}}^R=\left(\matrix{{1 \over 2} \bar{\partial} \alpha & -{e^{  \alpha} }  \cr{e^{-  \alpha} {\bar p }(\bar{z}) }  & -{1 \over 2} \bar{\partial} \alpha }\right) }}
The consistency conditions of the equations \orthbasis\ imply that these connections are flat
\eqn\flatness{\eqalign{\partial B_{\bar z}^{L}-\bar{\partial} B_z^{L}+[B_z^{L},B_{\bar{z}}^{L}]=0 ~,~~~\partial B_{\bar z}^{R}-\bar{\partial} B_z^{R}+[B_z^{R},B_{\bar{z}}^{R}]=0}}
These
imply that $p$ is  a holomorphic function, as claimed above, and that
 $\alpha$ satisfies the generalized sinh-Gordon equation
\eqn\gensinh{\partial \bar \partial \alpha(z,\bar{z})-e^{ 2 \alpha(z,\bar{z}) }+|p(z)|^2 e^{-2 \alpha(z,\bar{z})}=0}
The area of the world-sheet is
simply given by the conformal gauge action expressed in terms of the reduced fields\foot{
If $z = x + i y$,
then $\int dz^2 = \int dx dy $ and  $\partial = { 1 \over 2 } ( \partial_x - i \partial_y) $, etc.}
\eqn\area{ A =4 \int d^2z  e^{2 \alpha }
}
For the solutions in this paper this area is infinite. The proper regularization of
the area is the subject of section five.
 Note that \gensinh\ and \area\ are invariant under conformal transformations provided we transform
 $\alpha$ and $p$ accordingly. This is the original conformal invariance of the theory which
 the homogeneous Virasoro constraints in \eom\ have not broken.

\subsec{Recovering the surface from a solution to the generalized Sinh-Gordon problem}

Given a solution of the generalized sinh-Gordon model,
one would like to reconstruct the classical string worldsheet in
 $AdS_3$. For that purpose, given the connections $B^{L,R}$,
  consider the auxiliary linear problem
\eqn\LS{\eqalign{ \partial \psi^L_{\alpha}+(B_z^L)_{\alpha}^{~\beta}\psi^L_{\beta}=0,~~~~~
\bar{\partial} \psi^L_{\alpha}+(B_{\bar z}^L)_{\alpha}^{~\beta}\psi^L_{\beta}=0 \cr
\partial \psi^R_{\dot \alpha }+(B_z^R)_{\dot \alpha}^{~\dot \beta}\psi^R_{\dot \beta } =0,~~~~~\bar{\partial} \psi^R_{\dot \alpha }+(B_z^R)_{\dot \alpha}^{~\dot \beta}\psi^R_{\dot \beta } =0
}}
Each of these problems has two linearly independent
solutions, which we denote $\psi^L_{\alpha, a}$ , $a=1,2$ and
$\psi^R_{\dot \alpha, \dot a}$, $\dot a =1,2$. Since the connections are in $SL(2)$ we can
define an $SL(2)$ invariant product which will be a constant. Namely we can evaluate it
at any point. We can use this product to normalize the pair of solutions as
\eqn\normalization{\psi^L_a \wedge \psi^L_b \equiv  \epsilon^{\beta \alpha } \psi^L_{\alpha,a} \psi^L_{\beta,b}=\epsilon_{a b},~~~~~\epsilon^{\dot \beta \dot \alpha } \psi^R_{\dot \alpha, \dot a} \psi^R_{\dot \beta,\dot b}=\epsilon_{\dot a \dot b}}
This can be seen by noticing that the left hand side is anti-symmetric in $(a,b)$ and is annihilated by both derivatives. Notice that there is an $SL(2)$ group that acts on the index $a$ transforming the pair
 of normalized solutions into another pair of normalized solutions. There is another $SL(2)$ that acts
 on the index $\dot a$. These are the two $SL(2)$ symmetries of the $AdS_3$ target space.
 The tensor $W_{\alpha \dot \alpha, a \dot a }$ can be written in terms of these solutions as
   \eqn\Wsol{ {W}_{\alpha \dot \alpha, a \dot a}=\psi^L_{\alpha,a} \psi^R_{\dot \alpha,\dot{a}}}
 This can be seen as follows. Notice that if the components of $W$ can be written in this way at one point, then \LS\ imply that the same is true everywhere. In order to show that at one point $W$
  can be written
 in terms of a product of spinors as in \Wsol, we note that each entry in \Wmat\ is a null vector in
 $R^{2,2}$, so they can be written as a product of spinors\foot{Notice that the product of two
 vectors $Y.X$ with the metric in \emb\ is given by $Y.X=
  - { 1 \over 2} Y_{a\dot a} X_{b \dot b} \epsilon^{ab} \epsilon^{\dot a \dot b} $.}. In addition, the inner products among the $q_i$
 obey the equations in \hybridbasis , which ensures the decomposition in \Wsol . This decomposion
   is stating
 that all four null vectors $q_i$ can be written in terms of just two spinors of each kind .
  The explicit form of the solution $Y_{a \dot a}(z,\bar z)$ is simply given by the element $q_1$ in
\Wmat   \Wsol .   More explicitly, we have
\eqn\inversemap{\eqalign{  Y_{a \dot a } =
\left(\matrix{Y_{-1}+Y_{2}& Y_1-Y_0\cr Y_1 + Y_0 & Y_{-1}-Y_{2} }\right)_{a,\dot{a}}=
\psi^L_{\alpha,a} M_1^{\alpha \dot \beta} \psi^R_{\dot \beta ,\dot{a}} ~,~~~~~~~~~M_1^{\alpha \dot
\beta } =  \pmatrix{ 1 & 0 \cr 0 & 1 }
}}
Similarly we can write the expressions for other $q_i$ in \Wmat\ by substituting $M_1$ by $M_i$
where $M_i$ are simple matrices designed to extract the corresponding $q_i$ from \Wmat .
We see that the problem factorizes into the left problem, involving $\psi^L$ and the connection
$B^L$ and the right one. The final formula for the $SL(2)$ group element which parametrizes
$AdS_3$ in \inversemap\ is somewhat analogous to a similar decomposition for the $SL(2)$ WZW model
where the $SL(2)$ group element  $G$
 is expressed as $G= G(z)\tilde G(\bar z)$. In our case, the WZ term is zero and we
have an ordinary sigma model.
We will see below that it might be convenient for some purposes to perform gauge transformations
either on the left problem or the right problem. Such gauge transformations will be restricted
at infinity, so that we cannot remove the flat connection completely. Certain quantities, such
as the inner products \normalization\ are gauge invariant. On the other hand the
expression for the $AdS_3$ coordinates in \inversemap , does  depend on the gauge. In another
gauge the expressions would be given by the same formula in terms of a
 matrix $M_1$. This matrix however would have a different form
which can be computed by acting with the gauge transformations on $M_1$ in \inversemap .
 Fortunately, we will find that we can get most of the important physical
 information from quantities that are gauge invariant.

In appendix F we discuss how worldsheet and spacetime parity are
  realized
 in this description.

\subsec{Introduction of a spectral parameter and Hitchin equations}

It turns out that the left connection $B^L$ can be promoted to a family of flat connections by introducing a spectral parameter. Consider
\eqn\Bgeneric{\eqalign{B_z(\zeta)=\left(\matrix{{1 \over 2} \partial \alpha &- { 1 \over \zeta }{e^{ \alpha} }
\cr-{ 1 \over \zeta } {e^{- \alpha} p(z)  }  & -{1 \over 2} \partial \alpha }\right),~~~~~
 B_{\bar{z}}(\zeta)=\left(\matrix{-{1 \over 2}  \bar{\partial} \alpha &
 -{\zeta}{e^{-  \alpha} \bar{p}(\bar{z})  }\cr-{  \zeta}{e^{  \alpha}  }  &
 {1 \over 2} \bar{\partial} \alpha }\right)}}
The flatness condition $\partial B_{\bar z}-\bar{\partial} B_z+[B_z,B_{\bar{z}}]=0$ is then satisfied for any value of $\zeta$. \foot{One could consider introducing four different  parameters multiplying the off diagonal components. The flatness condition,  allows only two independent
parameters. By performing a gauge transformation of the form $U={\rm diag} (e^{\beta/2},e^{-\beta/2})$ the connections can always be brought to the form \Bgeneric\ .} Notice that both, left and right connections can be obtained from this $B_z(\zeta)$, more precisely
\eqn\leftrightrel{B_z^L=B_z(1),~~~~B_z^R=U B_z(i) U^{-1},~~~U=\left(\matrix{0&e^{i   \pi/4}\cr e^{i 3 \pi/4}&0}\right)}
Notice that we can decompose the connection as follows
\eqn\condec{B_z(\zeta)=A_z+{ 1 \over \zeta }
\Phi_z,~~~~~~B_{\bar z}(\zeta)=A_{\bar z}+{ \zeta} \Phi_{\bar z}}
This agrees with the form used in \refs{\GaiottoCD,\GMNtwo}.
 The zero curvature conditions can be rephrased in terms of the following equations
\eqn\hitchineq{D_{\bar z} \Phi_z=D_{z} \Phi_{\bar z}=0,~~~~~F_{z \bar{z}}+[\Phi_z,\Phi_{\bar z}]=0}
These are the Hitchin equations \HitchinVP, which arise by dimensional reduction of the
four dimensional self duality condition (instanton equations) to two dimensions.
$A$ has the interpretation of a gauge connection in two dimensions and $\Phi$ is a Higgs field.
$D_\mu \Phi=\partial_\mu \Phi_z+[A_\mu,\Phi_z]$ is the covariant derivative and
$F_{\mu \nu}=\partial_\mu A_\nu-\partial_\nu A_\mu+[A_\mu,A_\nu]$ is the field strength.
 These equations have also been studied in the context of matrix string theory, in studying
 scattering processes \refs{\GiddingsYD,\BonelliYT,\BonoraAY} .

The Hitchin equations can be considered for a generic gauge group.
The solutions at hand are a particular case of the Hitchin equations for the $SU(2)$ gauge group.
More precisely,
the generic solutions for the $SU(2)$ case contain also an off-diagonal
contribution to the gauge field $A$.
The configurations we consider here are  the subspace
 which is invariant under a certain $Z_2$
symmetry of the equations.
More precisely, if $\sigma^3$ is the Pauli matrix,
the $Z_2$ symmetry acts as
  $A \to  \sigma^3 A \sigma^3 $ and  $\Phi \to  - \sigma^3 \Phi \sigma^3 $.
  We are projecting onto the $Z_2$ invariant subspace. Thus the moduli space of our
  problem is a subspace of the full hyperkahler Hitchin space. This half dimensional
  subspace is sometimes
  called the ``real section''. The full hyperKahler space can be represented as a torus fibration
  over this ``real section''.

\def\halpha{{\hat \alpha}}

\newsec{The solution for large $|z|$ and the spacetime cross ratios}

The generalized sinh-gordon equation
 is conformal invariant. This is a reflection
of the original conformal invariance of the sigma model, which is
not fixed by setting $T_{zz} = T_{\bar z \bar z } =0$. We could
simplify the equation by defining a new variable $w$ such that
\eqn\defw{ d w = \sqrt{ p(z) } d z } and a similar relation for
$\bar w$. One might think that this would remove $p$ completely
and that we can forget about it. The information in $p$ does not
go away since the variable $w$ has a square root branch cut at
each of the zeros of $p$. We will see that for our problem $p$ is
a polynomial of degree $n-2$, where $2n$ is the number of sides of
the polygon\foot{One can of course consider solutions with
$p(z)$ being a generic, non polynomial, holomorphic functions.
In appendix D we study an example with $p(z) = e^z$.
 }. For such polynomials we end up with the hyperelliptic
Riemann
 surface $y^2 = p(z)$, and then $dw = y dz$ is a one form on the Riemann surface.
 Nevertheless this is a useful change of variables locally.
 In the $w$ variables the generalized sinh-Gordon equation
  \gensinh\
  simplifies to
 \eqn\wsinh{
 \partial_w \bar \partial_{\bar w} \hat \alpha  - e^{2 \hat \alpha } + e^{ - 2 \hat \alpha } =0 ~,~~~~~~~~~~
 \hat \alpha \equiv \alpha - { 1 \over 4 } \log p \bar p
 }
 The expression for the area
  then becomes $A = 4 \int d^2 w e^{2 \hat \alpha } $.
 Thinking about the problem in terms of the $w$ variables is useful for some purposes.
   It is particularly useful to think about the $w$ variables when we
 study the form of the surface at large $|z|$.
 In the $w$ variables the linear problem has the form
 \eqn\linpro{\eqalign{
  & (  \partial_w + \hat B^L_{w} ) \hat \psi
   = ( \partial_{ \bar w} + \hat B^L_{\bar w} ) \hat \psi =0~,
  \cr
   & \hat B^L_{w} = \pmatrix{ - \cosh \halpha & - {i \over 2} \partial_w \halpha - i \sinh \halpha
   \cr  { i\over 2 } \partial_w \halpha -i \sinh \halpha  &  \cosh \halpha }
   ~,~~~~~~~~~~~~
   \cr & \hat B^L_{\bar w} = \pmatrix{ - \cosh \halpha &
    {i \over 2} \bar{\partial}_{\bar w}  \halpha + i \sinh \halpha
    \cr  -{i \over 2} \bar{\partial}_{\bar w}  \halpha +i \sinh \halpha    & \cosh \halpha } ~,
  \cr
  & \hat \psi^L = e^{i {\pi \over 4 } \sigma^3} e^{ i { \pi \over 4 } \sigma^2 }  e^{ {1\over 8 } \log { p \over \bar p} \, \sigma^3 }
   \psi^L
   } }
   where the last equation states that we have made a gauge transformation ($\sigma^3$ is one
   of the Pauli matrices). This transformation was designed to simplify the expression for large
   $z$, and it was chosen with some hindsight.  If we wanted to introduce a spectral parameter
   $\zeta$,  we   would need to divide both hyperbolic functions in $B^L_w$ by $\zeta$ and
   multiply both hyperbolic functions in $B^L_{\bar w}$ by $\zeta $.

 In order to  understand the solution at large $z$ it is convenient to look at
   the particular solution  for $n=2$ (four sides).
 In this case $p=1$ and the $z$ plane and the $w$ plane are equal. The solution is simply
 $\hat \alpha =0 = \alpha$. In order to compute the spacetime quantities we  need
 to solve the linear problem. In this case $\hat B$ is constant and diagonal, thanks to the
 gauge transformation in \linpro .
 In this gauge the two solutions are
 \eqn\twosol{
  \eta_{+}^L = \pmatrix{ e^{    w + \bar w  } \cr 0 }
 ~,~~~~~~~ \eta^L_{-} = \pmatrix{ 0 \cr e^{-   (w + \bar w) }  }
 }
 we reserve the letter $\psi$ to denote generic solutions, while $\eta_{\pm}$ denote these two
 specific solutions.
Note that for $ Re( w) \to + \infty$ the first solution grows while the second decreases.
On the other other hand, the opposite is true for   $Re(w) \to - \infty $.
  The region $Re(w)>0$  defines an anti-Stokes sector
and the region $Re(w)<0$ defines another anti-Stokes sector. In each anti-Stokes sector, one
of the two solutions dominates.  The lines $Re(w)=0$ which separate
them are anti-Stokes lines. Along those lines the dominant solution changes.

We can also consider the right problem. We have seen that the right problem can be obtained
from the left one by introducing the spectral parameter $\zeta$ and then setting $\zeta =i$.
In this way we can find the following two solutions\foot{ For general $\zeta$ the two solutions
are $\eta_+ (\zeta) = \pmatrix{ e^{   (  w /\zeta + \bar w \zeta)}  \cr 0 } $,
 $\eta_- (\zeta) = \pmatrix{0 \cr  e^{ -  (  w/\zeta + \bar w \zeta)}  } $.}
\eqn\chisol{
 \eta^R_{+} = \pmatrix{ e^{  { ( w - \bar w ) \over i } } \cr 0} ~,~~~~~~~
 \eta^R_{-} =  \pmatrix{ 0 \cr e^{ -{ ( w - \bar w ) \over i } } } ~,~~~~~~~ }
Here one solution dominates for $Im(w) \to + \infty$ and the second for $Im(w) \to -\infty$.
Now the anti-Stokes lines are at $Im(w)=0$.
A general solution of the right problem is a linear combination of these two solutions.
These solutions are written in a gauge that differs from the original gauge by $\hat \psi^R =
e^{ i { \pi \over 4 } \sigma^3 } e^{ i { \pi \over 4 } \sigma^2 }  e^{ {1\over 8 } \log { p \over \bar p} \, \sigma^3 } U \psi^R$,
where $U$ is given in \leftrightrel .
In the case of general $n$ (and a general polynomial $p$)
we expect that the solution near each cusp reduces to the solution
for the case we had above. This can be achieved if we demand that $\hat \alpha \to 0$ when
$z$ (or $w$) go to infinity. In addition we demand that $\alpha$ (but not $\hat \alpha$) is finite
everywhere. This is expected to lead to a unique solution for the generalized
sinh-Gordon problem and we discuss some explicit solutions in section four.
For large
$z$ we can set $\hat \alpha =0$ and we recover the above solutions within each anti-Stoke sector.
We will discuss below what happens when we change sectors.
First let us discuss the form of the solution within each sector.

  \ifig\zwcusps{ Stokes sectors in the $z$ plane (a)  and in the $w$ plane (b). The dashed lines
  are Stokes lines for the left problem and anti-Stokes lines for the right problem. The opposite
  is true for dotted lines. In (b) we see one sheet of the $w$ plane. Each quadrant is associated
  to one cusp. Notice that the two upper quadrants (dotted blue line) correspond to a single
  coordinate $x^-$, while the two left quadrants
   (dashed green line) correspond to a single $x^+$ coordinate.
 } {\epsfxsize2.7in\epsfbox{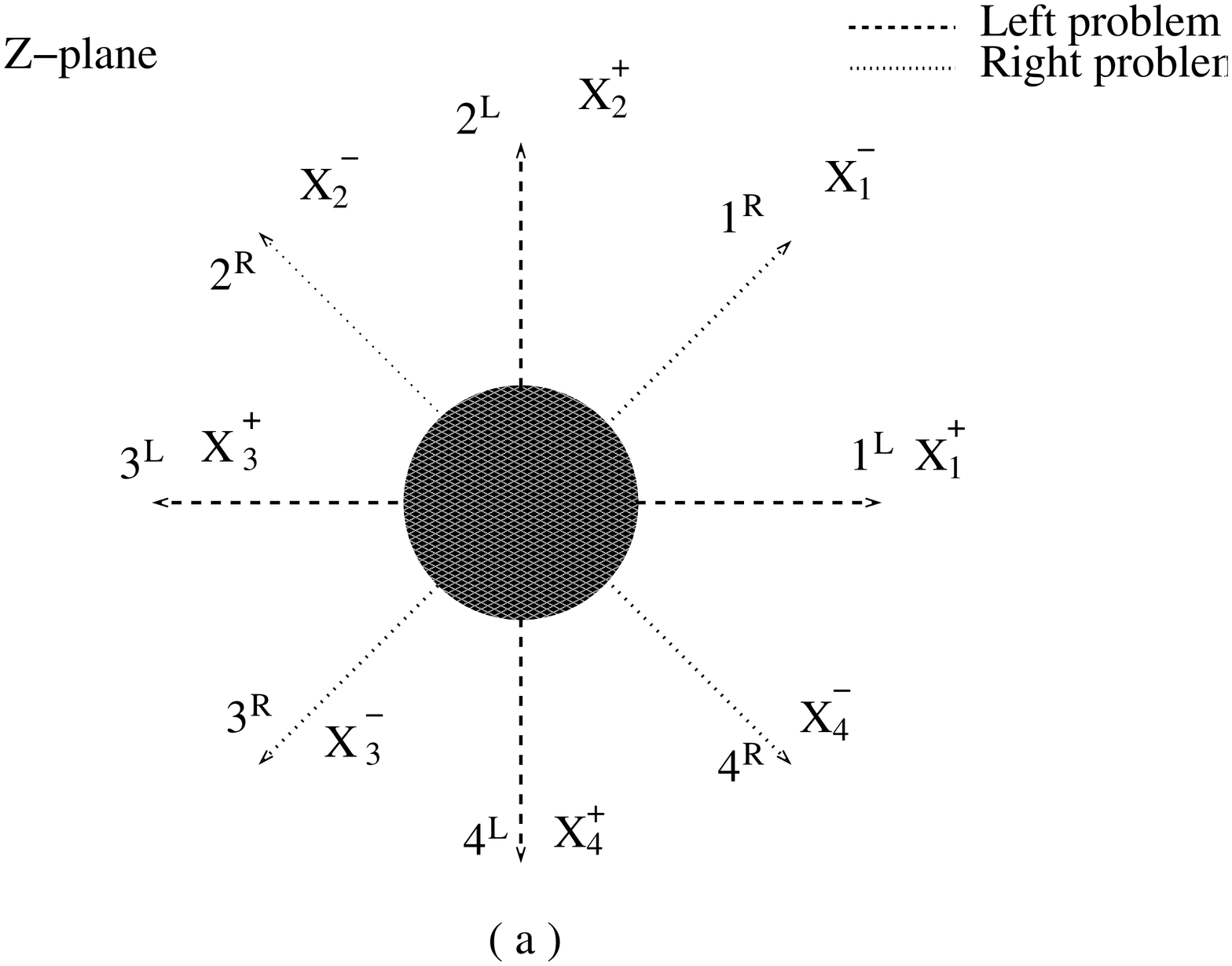} \epsfxsize2.7in\epsfbox{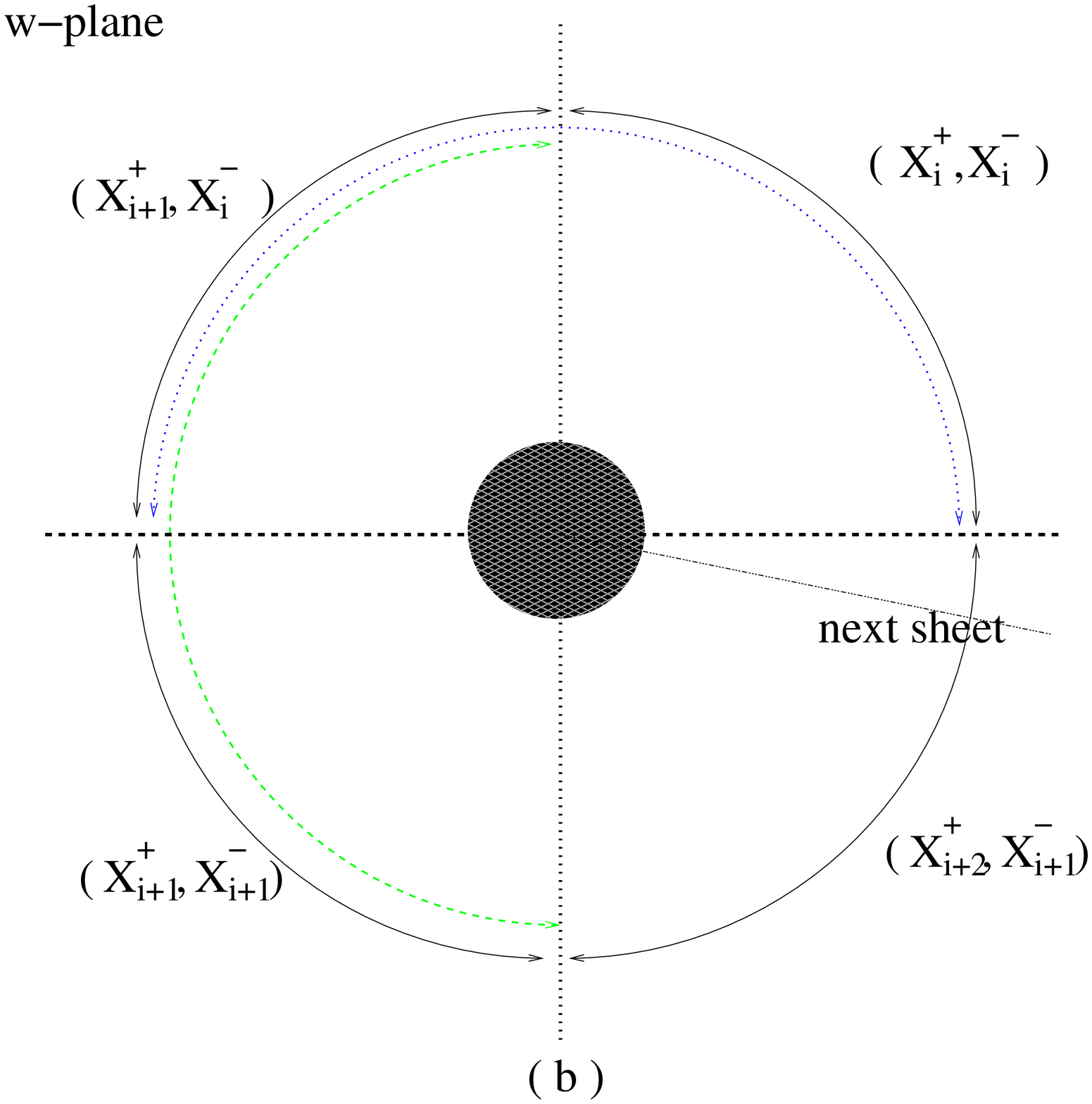}}

\ifig\stcusps{Positions of the cusps on the boundary of $AdS_3$. Each quadrant in $w$ is mapped
to a cusp. As we go from one quadrant to the next we change only $x^+$ or only $x^-$, so we move
along light like lines on the boundary. We only show a portion of the polygon.
} {\epsfxsize2.7in\epsfbox{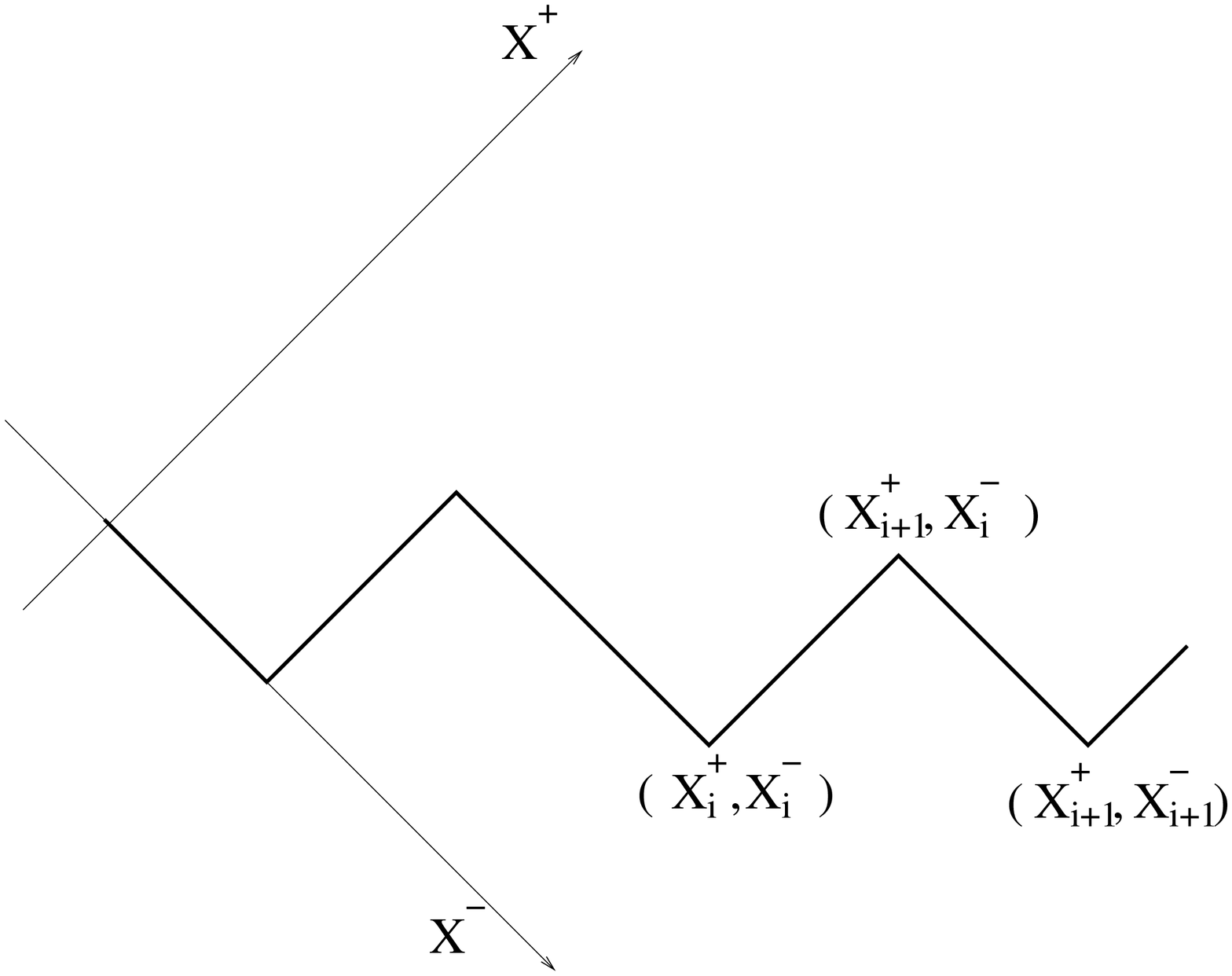}}

When we combine the left and right problems, then we need to divide the $w$ plane into
 quadrants. The first is at  $Re(w) >0$ , $Im( w )>0$ and the rest are simply rotations of
this one. In each of these quadrants the dominant solution is one particular combination of
the solutions described above. We will now argue that each quadrant corresponds to the solution
near a cusp of the Wilson loop.

Within each anti-Stoke sector there
is a big solution and a small solution for the left problem and similarly for the right problem.
The small solution is defined up to an overall rescaling. On the other hand the big solution is not
quite well defined because we can always add a multiple of the small solution.
Furthermore, as we will see below, the big solution shifts by an amount involving the small solution
when we cross a Stokes line. These lines are in the middle of the anti-Stokes sector.
Of course, the big part
of the big solution is what determines the behavior of the string worldsheet for large $z$.
The general form of the worldsheet is given by picking two solutions of the left problem,
$t^L_a = \psi^L_{\alpha \, a}$ $a=1,2$, and two
solutions of the right problem, $\psi^R_{\dot a \dot \beta }$,  and then setting
\eqn\formsof{
Y_{a \dot a } = \psi^L_{ \alpha a } M_1^{\alpha \dot \beta } \psi^R_{\dot \beta \dot a }
}
where $M_1^{\alpha \dot \beta }$ is a matrix that depends on the gauge that we choose for the left and
right problem. For example, in the gauge used in \inversemap\
 $M^{\alpha \dot \beta} = \delta^{\alpha \dot \beta}$.

Since some components of these solutions are going to infinity as we go to large $z$,
we see that
the worldsheet is approaching
the boundary of $AdS_3$. In order to identify the
part of the solution that is growing we can write
a solution $\psi$ as
\eqn\tsolf{
\psi_a = c_a^{\rm big} b + c_a^{\rm small} s
}
where $b$ is the big part of the solution and $s$ the small part and $c^{\rm big}$ and $c^{\rm small}$
 are numerical
coefficients. As we mentioned above we can always
add the small part of the solution to the big one, so the coefficient $c^{\rm small}$ is not so well defined. On
the other hand the coefficient $c^{\rm big}$ is well defined, once we choose a normalization for the small solution
and we set the normalization condition
 $b \wedge s =1$. In this case we can find
 $c^{\rm big}$ as $c^{\rm big} = \psi \wedge s $.
This allows us to write the large part of \formsof\ as
\eqn\bigsol{
Y_{a \dot a } = \psi^L_a \wedge s^L \, \, \psi^R_{\dot a } \wedge s^R  \, \, ( b^L_\alpha M_1^{\alpha \dot \beta} b^R_{\dot \beta} )
}
We see that up to a constant this leads to a spinor product   expression for
 $Y_{a \dot a } \sim \lambda_a \tilde \lambda_{\dot a }$ this implies that,
in this approximation $Y^2 =0$, which is as expected since  the boundary of $AdS$ is
 given by $Y$ with
$Y^2 =0$ and with the identification $Y \sim \gamma Y $ for any
$\gamma$. Note that the target space $SL(2)_L\times SL(2)_R$
transformations act on the $a$, $\dot a $ indices. In other words,
points on the boundary of $AdS_3$ are related to a choice of
spinors $\lambda_a $, $\tilde \lambda_{\dot a }$, defined up to a
rescaling.

We can now choose some coordinates on the boundary, such a Poincare coordinates defined as
 \eqn\poincarrc{
 Y_{-1} + Y_2 = { 1 \over r } ~,~~~~~~~ x^\pm = { Y_1 \pm Y_0 \over Y_{-1} + Y_2 } }
Then we see that
\eqn\boundar{
x_i^+ = { \lambda_2 \over \lambda_{1} } = { \psi^L_2 \wedge s_i^L \over
\psi^L_1 \wedge s_i^L } ~,~~~~~~~~~x_i^- = { \tilde \lambda_{\dot 2} \over \tilde \lambda_{\dot 1} } =
 { \psi^R_{\dot 2} \wedge s_i^R \over
\psi^R_{ \dot 1} \wedge s_i^R }
}
These expressions involve only the solution of the left problem or only the right problem. Both
are gauge invariant under their respective local gauge transformations. We are allowing only
gauge transformations which do not grow at the boundary, otherwise, we would erase the
physical information since we could set the solution to a constant as we have a flat connection.
These expressions do not depend on the normalization of the solutions $s_i^{L,R}$.
We are now in a position to compute
    \eqn\formdiff{
     x^+_i - x^+_j = - { s^L_i \wedge s^L_j \over \psi^L_{1} \wedge s^L_i \psi^L_1 \wedge s^L_j }
     }
     where we used \boundar\ and the fact that $\psi_1$ $\psi_{2}$ is a normalized basis.
    When we pick four points and we form a cross ratio the factors in the denominator in
    \formdiff\ drop out and we get\foot{ This cross ratio can also
    be written in term of a pair of spinors $\lambda $, $\tilde
    \lambda $ which are defined up to rescalings and parametrize
    the boundary of $AdS_3$. Namely we have
    $ { x^+_{12} \,  x^+_{34} \over x^+_{13} \, x^+_{24} } = {
    \langle 12 \rangle \langle 34 \rangle \over \langle 13 \rangle
    \langle 24 \rangle }$, where $\langle 12 \rangle = \epsilon^{ab} \lambda^1_a \lambda^2_b $.
    The minus cross ratios are written in terms of $\tilde \lambda$.}
    \eqn\formcross{
    { x^+_{12} \,  x^+_{34} \over x^+_{13} \, x^+_{24} } = { s^L_1 \wedge s^L_2 \, s^L_3 \wedge s^L_4 \over
    s^L_1 \wedge s^L_3 \, s^L_2 \wedge s^L_4 } = \chi(\zeta =1)
    }
    Where $1,2,3,4$ indicate any four distinct points, not necessarily consecutive.
    Note that the overall factors multiplying each of the $s_i$ drop out.
    The expression in the right hand side are the cross ratios introduced in \refs{\FG,\GMNtwo} .
    The right hand side of \formcross\ can also be considered for any $\zeta$. Here one
    considers simply the small solutions of the problem with the connection in \Bgeneric .
    The
    $x^-$ cross ratios are given by $\chi(\zeta =i)$ which is simply the same expression
    in terms of the small solutions of the right problem. The authors of
    \refs{\GaiottoCD,\GMNtwo}  showed that
    it is useful to
    analyze the cross ratios as a function of $\zeta$ in order to solve the problem. For our
    problem the cross ratios at $\zeta =1$ and $\zeta =i$ play a special role because they
    directly define the physical cross ratios for the spacetime problem.

 Let us now consider a polynomial $p \sim z^{n-2 } + \cdots $. At large $z$ we find that
 $w \sim z^{n/2} + \cdots $. Thus, as we go around once in the $z$ plane we go around the
 $w$ plane $n/2$ times. This implies that the left sector has $n$ anti-Stokes sectors.
 We can label its anti-Stokes sectors
  by $i=1,\cdots , n$. In each sector we have a well defined value of $x^+_i$.
   At the center of each anti-Stokes sector we have a Stokes line where the small
  solution is  smallest and the large solution is largest.
  Once we find the small solution in each region, we can compute all the $n-3$ cross ratios, which
 depend on the $n$ $x_i^+$ points. We have an $SL(2)$ symmetry acting on the  $x^+$ variable and
 that is the reason why we have $n-3$ independent cross ratios.
 The right problem has its own $n$ regions where the solution ends at $x_i^-$.
  The value of $x^+$ changes as
 we cross anti-Stokes lines. An anti-Stokes line of the left problem is a Stokes line for the right
 problem. This means that when $x^+$ changes,  $x^-$ does not change and vice-versa. This means that the different
 cusps are light like separated.
  If we label the $x^-_i$ variables by the index $i=1,\cdots, n$. Then we have that cusps are labeled
  by the pair $(x_i^+,x^-_{i-1})$, $(x^+_i, x^-_i)$, $(x_{i+1}^+,x^-_i)$ and so on.
 Each time we go
  from one cusp to the next only $x^+$ or only $x^-$ changes,
    see \zwcusps\ and \stcusps\

 We end up with $ 2n$ quadrants in the $w$ coordinates corresponding to the $2n$ cusps of the problem.
 The total number of left plus right cross ratios is $ 2(n-3)$, which is also the
 number of non-trivial parameters in the polynomial $p$.  These are counted as follows.
   The coefficient
 of the maximal degree term can be set to one by conformal rescalings of the $z$ coordinates.
 We can remove one other complex coefficient by performing a translation in $z$. Thus we are
 left with $n-3$ complex coefficients, or $2(n-3)$ real ones.
 One might think that it should be possible to
 remove one extra real parameter because we can perform rotations in the $z$ plane. Indeed,
 the generalized
 Sinh-Gordon problem is invariant under such rotations. On the other hand, the values of the
 spacetime cross ratios do  depend on the relative position of the zeros of $p$ and the lines
 defining the Stokes sectors. In fact, we will see explicitly that the final answer for physical
 quantities does depend on the relative orientation of the zeros of $p$ and the Stokes lines.
 The orientation of the Stokes lines was fixed when we set the coefficient of the leading term in
 $p$ to one.

\subsec{ Approximate solutions for large $z$, Stokes lines and Stokes matrices  }

In the case with $p=1$ ($n=2$)  \twosol\ are the exact solutions of the problem.
For general $n$ these are only approximate solutions. These approximate solutions
are a good approximation to the problem only within some angular sectors.
 The problem displays the
Stokes phenomenon. This is just the fact that we cannot analytically continue the approximate
solutions to the full asymptotic region of the $w$ space.
The sectors are separated by Stokes lines where the asymptotic form of the solution jumps.
 Let us describe this more concretely. Consider the left problem in the
  large $w$ region, such as $Im(w)<0$. Within that region \twosol\ is a good
  approximate solution.

Let us consider now what happens when $|w| \gg 1$ and we cross the line where $w$ is real and
positive.   The solution that decreases as $Re(w ) \to + \infty$ is accurately given by
\twosol\ and is the same on both sides of the line.
On the other hand, the large solution has a jump in its small solution component.
More precisely, we choose a basis of solutions which has the asymptotic form in   \twosol \ in
one Stokes sector. We denote these two exact solutions as $\eta^\pm_{\rm before} $.
After we cross the Stokes line, we enter into a new Stokes sector. We can now choose another
basis of solutions which has the asymptotic form in \twosol\ in this new sector. We denote
these two exact solutions as  $\eta^\pm_{\rm after}$.
These two sets of solutions should be related by a simple linear transformation. In fact we have
 \eqn\matris{
 \eta_{a}|_{\rm before } = S_a^{~b} \,
 \eta_{b}|_{ \rm after }  ~,~~~~~~~~~~~~~~~
  S(\stokesparameter) = \pmatrix{ 1 & \stokesparameter \cr 0 & 1 }
 }
 The Stokes matrix acts on the target space  $SL(2)$ index, $a= \pm $,  of the solutions.
 In other words $\eta_\pm |_{\rm before} $
 has a new asymptotic expression in the new sector.
 It is the correct analytic continuation of the solution. This differs from the
 the analytic continuation of the approximate expression by the Stokes matrix. The full exact solution
 is continuous across this line, it is only the approximation that becomes discontinuous.
 The fact that we get this  admixture of the small solution is not important in this region.
 However, as we move to the region where $Re(w)$ is negative, this small solution becomes large and
 then its  effect can become very important. The value of $\stokesparameter$  depends  on the full exact
 solution and we cannot compute it purely at large $w$ (at least without using some tricks).
  In the region where $w$ is real and very negative we have a similar Stokes matrix, but
  of the form $ \sigma^1 S(\gamma) \sigma^1 $ since the large and small solutions are $\eta_-$ and
  $\eta_+$ respectively. Notice that the approximation is good around the anti-Stokes lines.
  In that region the two approximate solutions correctly represent the full solution. From
  the spacetime point of view, the anti-Stokes lines correspond to the segments in the Wilson loop
  while the Stokes line is irrelevant for the behavior of the spacetime solution along that line.
  However, it becomes important for determining the spacetime solution elsewhere.

We can use the Stokes matrices to obtain the correct asymptotic expression of a solution in all regions
once we know it in one region.
These different asymptotic expressions determine the different cusp positions in each sector.
Thus, the Stokes matrices are directly related to the positions of the cusps. We make this more
explicit in appendix B. The information in the Stokes matrices is equivalent to the information
contained in the cross ratios. More precisely, the Stokes matrices can change if we change the
basis. The cross ratios contain the invariant information.

     \subsec{Different presentations for the Wilson loop problem}

 The vacuum expectation value of Wilson loops at strong coupling can be computed
 by computing the area of minimal surfaces in $AdS$ \MaldacenaIM \ReyIK
 \eqn\wilexpec{
\langle W \rangle \sim e^{ - { R^2 \over 2 \pi \alpha' } ( { \rm Area } ) }
 }
where the area is measured in units where the radius of $AdS$ is one. In the application of
these formulas to ${\cal N}=4$ super Yang-Mills ${ R^2 \over 2 \pi \alpha' } = { \sqrt{\lambda}
\over 2 \pi }$. For other theories, like for instance \AharonyUG , we can use the
corresponding relation between the radius
of $AdS$ in string units and the parameters of the field theory. Throughout this paper we
set the radius of $AdS$ to one and we simply talk about the geometric area of the surface.
For different applications one can insert the corresponding overall factors as in \wilexpec .
The overall factor appearing in \wilexpec\ is related to the strong coupling value of the
cusp anomalous dimension.

     Let us discuss in   more detail different possible physical configurations that give rise
     to the same Wilson loop.
       The surface corresponding to the Wilson loop is simplest
     if all cusps are backward or forward directed. In the scattering interpretation
     this corresponds to the case where the momentum transfer between adjacent gluons is spacelike.
     The worldsheet in this case is expected to be a spacelike surface embedded in the Lorentzian
     spacetime.

     \ifig\Roneone{Different pictures for the Wilson loops considered in this paper.
     For a possible scattering configuration one has to include at least one incoming
      right mover and one outgoing left mover, see (a). Notice that the momentum transfer
       is time-like in two of the cusps.
       In (b) we consider a Wilson loop with two lines going
        to infinity. The corresponding  Penrose diagram is shown in (c).
       In (d) we have mapped this Penrose diagram to the cylinder, by identifying its left
       and right vertices.
 } {\epsfxsize4.5in\epsfbox{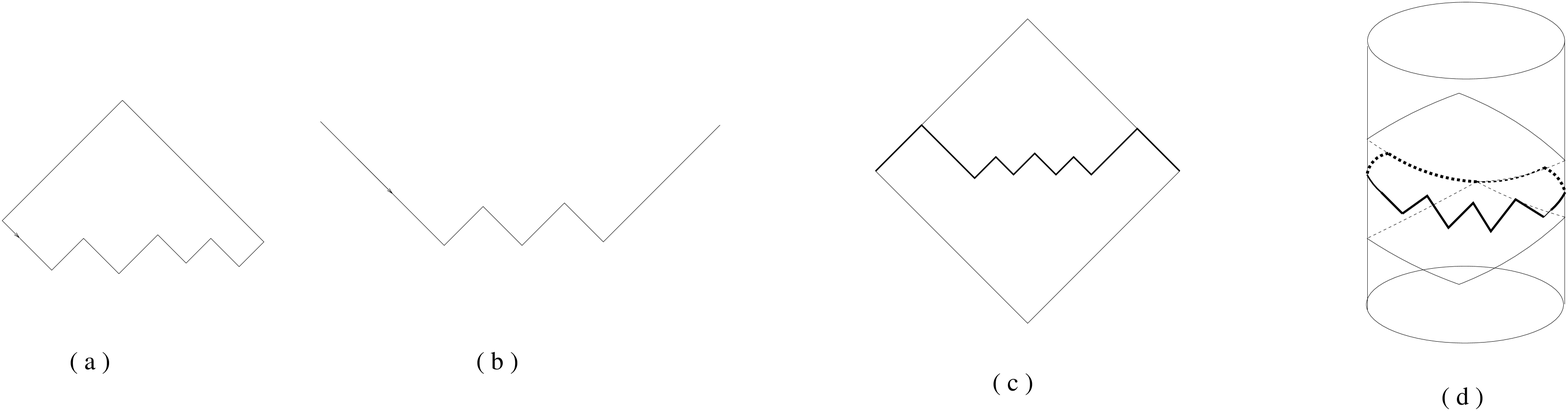}}

   \ifig\cylinder{ We view the two dimensional space as a
   cylinder. We can consider a polygonal Wilson loop going around
   the cylinder.
 } {\epsfxsize1.0in\epsfbox{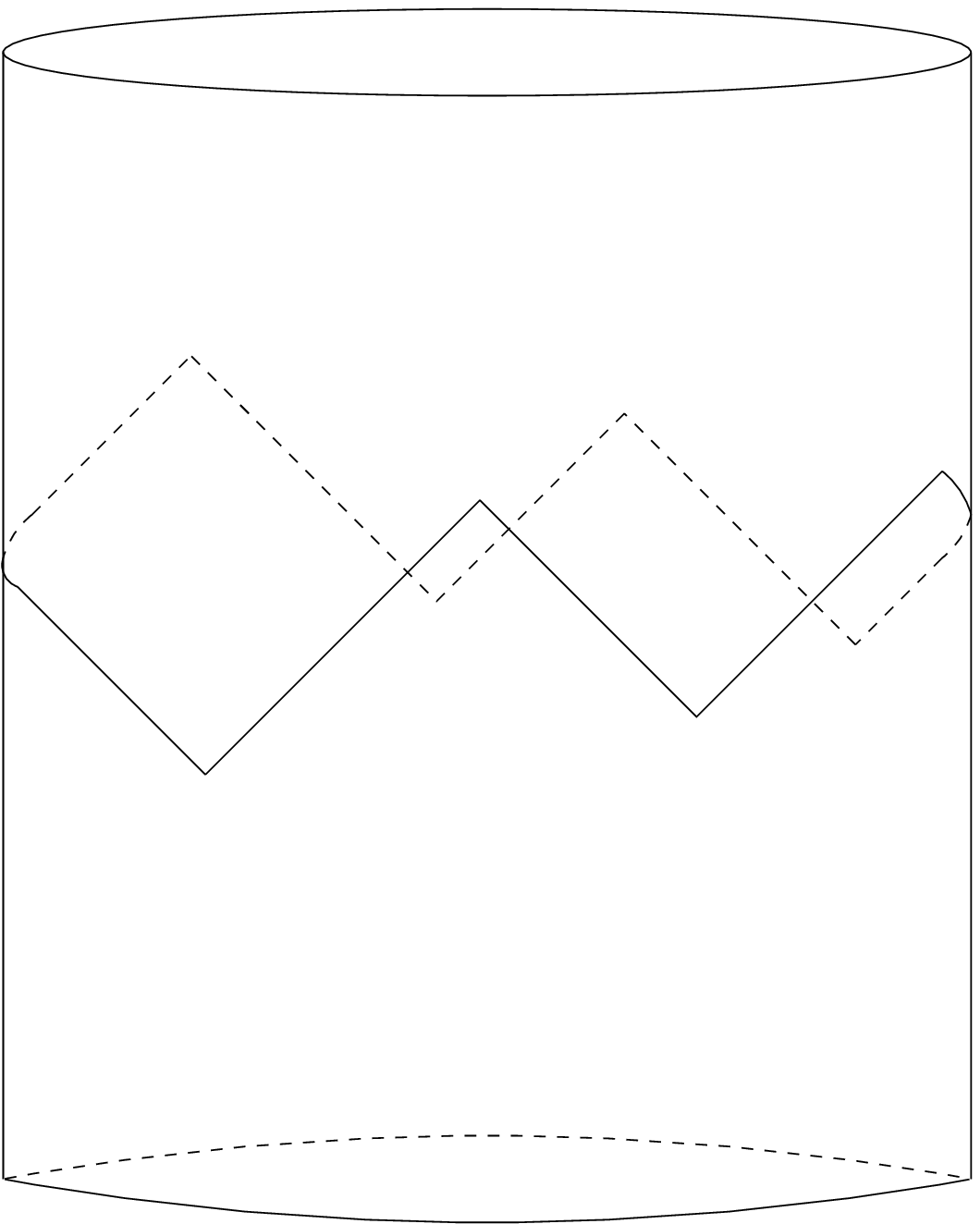}}

   If we consider null polygons in $R^{1,1}$ we find that if we want the polygon to live in
   a compact region of $R^{1,1}$, then we cannot make all cusps forward or backward directed.
   Some of them have to point sideways, see \Roneone (a). In the scattering interpretation
   is is clear why we have this. The bottom part of the jagged polygon in \Roneone (a) corresponds
   to incoming leftmovers and outgoing right movers. This is not a possible scattering configuration
   unless we include at least one incoming right mover and one outgoing left mover.

   If we want to insist in having no sideways cusps in the null polygon, then we can send some points
   to infinity. We are then left with the configuration in figure \Roneone (b), with lines going
   to infinity. If we were to write the Penrose diagram for Minkowski space, then we would find
   that we can have null lines at the boundaries, as in figure \Roneone (c). In this case all the
   cusps are forward or backwards, but three of them are at infinity. Namely, $x_n^+ =x_n^- = \infty$
   so that three of the cusps have at least one infinite coordinate.
      It is natural to consider this configuration on the cylinder, which is the global
      boundary of $AdS_3$.
   In this case we can have a null polygon going around the
   cylinder\foot{We  restrict ourselves to configurations that wrap the cylinder only once.},
   see \cylinder (a).
   If we put a cusp at the point
   corresponding to spatial infinity of the Minkowski patch, then we end up with the configuration
   in \Roneone (c)(d).
     Note that  $SL(2,R)_L \times SL(2,R)_R$ transformations
     allow us to fix the position of three $x_i^+$ and three $x_i^-$. So we can always send some
     points to infinity. However, once we send them to infinity, we might loose the information
     of whether they are closing the contour on the upper side of the Minkowski boundary versus the
     lower side.

      \ifig\embcircle{Special kinematical configuration in $R^{1,2}$ where
      the projection
      of the null polygon to the $x_1,x_2$ plane circumscribes the unit circle.
 } {\epsfxsize2.0in\epsfbox{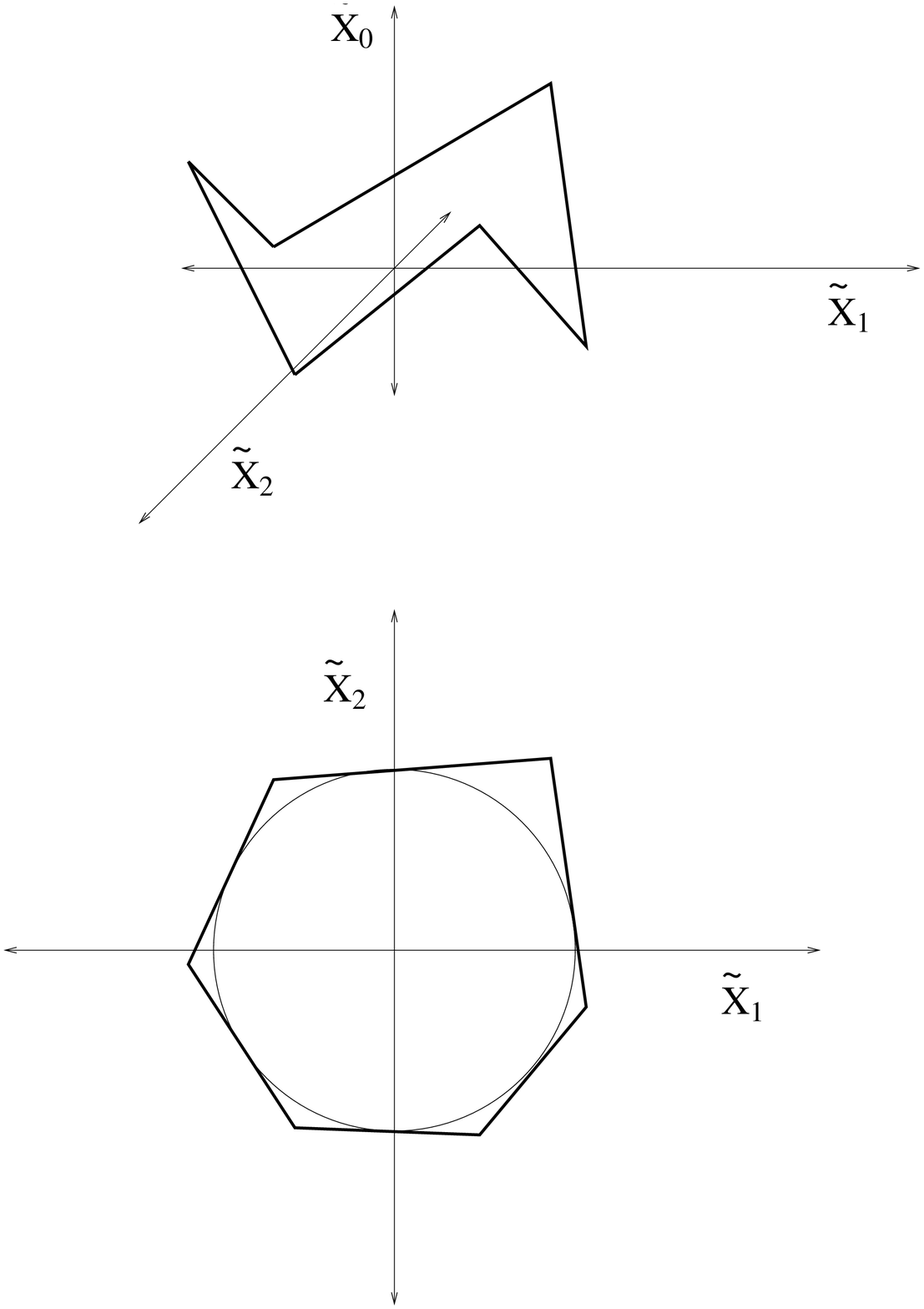}}

     There is an embedding of the null polygon into a bounded region which also has a clear scattering
     interpretation. For this purpose we
      embed the $AdS_3$ space we have been considering
     in a different way inside $AdS_5$ (or actually $AdS_4$).
     We can consider a null polygon which lives in a subspace of
     $R^{1,2}$, the subspace given
     by
     \eqn\hyperb{
     - \tilde x_0^2 + \tilde x_1^2 + \tilde x_2^2 = 1
     }
     This is a subspace of $R^{1,2}$ which is conformal to
     a cylinder. Once we add  the radial coordinate,  the $R^{1,2}$ space leads to
     an $AdS_4$ space (which could be a subspace of $AdS_5$). We will now argue that there
     is an $AdS_3$ subspace of $AdS_4$ which ends on \hyperb\  in $R^{1,2}$.
     We start with $AdS_4$ written in terms of embedding coordinates
     $ - Y_{-1}^2 -  Y_0^2 + Y_1^2 +Y_2^2 + Y_3^2 =-1$. We now take
      $ \tilde r $ to be the radial coordinate
     in $AdS_4$, $ 1/\tilde r= Y_{-1} + Y_3$, and $\tilde x_\mu = \tilde r Y_{\mu}$, $\mu = 0,1,2$. Then
      $  Y_3   = - ( \tilde r^2 + \tilde x_1^2 + \tilde x_2^2 - \tilde x_0^2 -1)/(2 \tilde r)$.
      We see that setting $Y_3 =0$
     gives an $AdS_3$ subspace of $AdS_4$ whose boundary is \hyperb .
    The projection of a null polygon that can be embedded in \hyperb\ is
    given by an ordinary polygon on the   $x_1,x_2$ plane that obeys a special condition.
     All the sides of the polygon should be tangent to the unit circle, see \embcircle (b).

      Thus the null polygonal  configurations that go
       around the cylinder at the $AdS_3$ boundary
      corresponds to a special null polygon  which
     lives in a bounded region of $R^{1,2}$. If we are given the corresponding minimal
     surface in $AdS_3$ we can find the $AdS_4$ embedding of the surface as
     \eqn\adsfouremb{
     { 1 \over \tilde r} = Y_{-1} ~,~~~~~~~ \tilde{x}_0 = { Y_0 \over Y_1} ~,~~~
     \tilde{x}_{1,2} = { Y_{1,2} \over Y_{-1}}
     }
     We use tilde to denote the $AdS_4$ Poincare coordinates,
      in order to differentiate them from the $AdS_3$ Poincare coordinates.
      In this coordinates it is easy to consider null polygons
       which live in a bounded domain of $R^{1,2}$ where
     all the cusps are future or past directed.
     Using these changes of coordinates and \inversemap\
     we can easily construct solutions in any of these pictures.

     Though we can map these configurations to each other, once we introduce the regulator, these
     give slightly different answers because the natural  regulator is slightly different in
     the different cases. For example, for  configuration that live in $R^{1,1}$ we can
     regulate the radial poincare $AdS_3$ coordinate $r$ which is $1/(Y_{-1} + Y_2)$.
     On the other hand the natural regulator for the embedding into  $R^{1,2}$
     in \embcircle\ is $\tilde r = 1/Y_{-1}$.
       The various answers should be related by the Weyl anomaly for null Wilson loops, which  could
       probably be deduced from \DrummondAU .

\subsec{Connection to the results of Gaiotto-Moore-Neitzke }

Introducing a spectral parameter, $\zeta $, and thinking about the analytic structure of
the gauge invariant information contained in the flat connection is a standard tool for
analyzing the solutions of integrable models. For example, if one considers strings
moving in $AdS$, the worldsheet is a cylinder and then one can consider the eigenvalues of
the holonomy of the flat connection around the cylinder \KazakovQF .
The analytic structure of
these eigenvalues characterizes the solution.
In our problem, the gauge invariant information is contained in the cross ratios.
So studying the cross ratios as a function of $\zeta $ is  a way to solve the problem.
Indeed, this is precisely what Gaiotto, Moore and Neitzke   do in \refs{\GaiottoCD,\GMNtwo}.
They considered essentially the same
  mathematical problem. They were motivated by   study the hyperkahler moduli
space of certain three dimensional field theories with ${\cal N}=4 $ supersymmetry (8 supercharges).
  Those theories can  arise from
wrapping D4 branes on Riemann Surfaces \GMNtwo .
 If we consider two D4 branes we get an $SU(2)$ gauge field plus a Higgs field (due to the
 twisting   \BershadskyQY ).
The classical Higgs branch moduli space of vacua of these theories is the moduli space
of the Hitchin equations on the corresponding Riemann surface. We do not have the full Hitchin
moduli space because our problem amounts to a projection of the Hitchin problem, which is sometimes
called a real section. In our case the moduli space is simply Kahler. The problem we consider
corresponds to an orbifold of the above brane configuration which leaves only a $U(1)$ subgroup
of $SU(2)$ and the charged Higgs fields as mentioned in the previous section .

Gaiotto, Moore and Neitzke
 have   exploited  the analytic structure of
the cross ratios as a function of $\zeta$ and have written a  Riemann-Hilbert problem
whose solution determines the metric in moduli space.
The moduli space is parametrized by the coefficients of the polynomial,
$z_i$, where $z_i$ are taken to be the zeros of the polynomial, $p= \prod_{i=1}^{n-2} (z-z_i)$,
with $\sum z_i =0$.
In appendix C we show that
from the expression for the metric one can compute the area  as
\eqn\metricarea{
A \sim \sum_i (z_i \partial_{z_i} + \bar z_i \partial_{\bar z_i}) K ~,~~~~~~~~~
\partial_{z_i} \partial_{
\bar z_j} K = g_{z_i \bar z_j }
}
In other words, we first find the Kahler potential that leads to
the known metric $g_{z_i \bar z_j}$
 and then we
take the derivatives in \metricarea . Note that the problem is
invariant under an overall rotation $z_i \to e^{i \varphi} z_i$.
In the above formula we are assuming that we write $K$ in an
invariant fashion. This fixes the freedom of Kahler
transformations. Equivalently, we can think of the area as the
moment map for this rotation symmetry. Having written $K$ in this
rotational invariant fashion we see that this moment map is $ D
\sim \sum z_i
\partial_{ z_i}  K = \sum_i \bar z_i
\partial_{\bar z_i} K $, where we used the rotational invariance
of $K$.

In \GaiottoCD , this procedure is carried out explicitly for a quadratic polynomial, which is
the first non-trivial case. It is found that the metric corresponds to that of a
 four dimensional ${\cal N}=2$ theory with a single hypermultiplet compactified on
  a circle, which had already been computed in \refs{\OoguriME,\SeibergNS}. Using this
result  we will obtain the area for the eight sided null polyogn  in section six .

Let us finally mention  that the connection $B(\zeta)$ introduced in \Bgeneric\
 and the usual
flat connection which appears in the discussions of general
principal chiral  models  \KazakovQF\
differ by a gauge transformation. The
gauge transformation is  determined by the solution of the linear
problem \LS . We explain  this explicitly in appendix A.

\newsec{Regular polygons}

\subsec{Regular polygons and Painleve transcendentals}

In this section we focus on a particular class of solutions that
possesses a $Z_n$  symmetry,  which can be studied in some
detail. The simplest polynomial of degree $n-2$ is the homogeneous
polynomial
 \eqn\regsol{p(z)= z^{n-2},~~~~~~ d w = dz \sqrt{p}
~ ~~  \longrightarrow~~~ w= { 2 \over n} z^{n\over 2 } }

The equations for $\alpha$ or $\hat \alpha$ are rotational
invariant. Since we expect a unique solution with the given
boundary conditions we find that the solution will also be
rotational invariant, so we have $\hat \alpha(|w|)$.  We find
  \eqn\painl{\hat \alpha '' +{\hat \alpha '  \over |w|}= 8
\sinh(2\hat \alpha  ),
 }
 This is a
particular case of the Painleve III equation  and its
solutions have been extensively studied in \McCoyCD . We must
supplement this equation by the appropriate boundary conditions.
As already mentioned, we are interested in solutions where $\hat \alpha$  decays
at infinity. Besides, $\alpha$ is regular everywhere, which
implies a logarithmic singularity for $\hat \alpha$ near the
zeroes of $p$. More precisely, for the case at hand
 \eqn\painlsing{\hat \alpha (|w|)=-{n-2 \over n}\log |w|+regular,~~~~|w|
\rightarrow 0
}
 In \McCoyCD\ it is shown that exists a one
parameter family of solutions, parametrized by $n$ in \painlsing ,
free of singularities for $|w|>0$ and which decays exponentially at
infinity. We can also find the behavior of the solution at
infinity
 \eqn\painsol{\eqalign{\hat \alpha (|w|) \sim { 2 \over \pi }  \,   \cos { \pi \over
 n}  \, \,  K_0(4 |w|) \sim { \cos { \pi \over n} \over \sqrt{ 2 \pi |w| } } e^{ -  4 |w| }
  ~,~~~~~{\rm for }~~~ |w| \to \infty
}}
 where $K_0(t) $ is a Bessel function which decays exponentially for large $t$.
  Once the solution is found, the regularized area should be
 computed. In order to do that, we divide the area into two pieces
 \eqn\regarea{{  A}=4\int d^2 w e^{2\hat \alpha} =4\int d^2 w+4\int
 d^2 w (e^{2\hat \alpha}-1)
  }
 The first piece needs to be regularized, and the exact answer
depends on the details of the regularization. On the other hand,
the second piece is finite, and it can be computed without
introducing any regularization. For the case at hand
 \eqn\finitearea{{ A}_{Sinh} \equiv
   4\int d^2 w (e^{2\hat \alpha}-1) ={4 \pi n  }\int_0^\infty  \rho d \rho
  (e^{2\hat \alpha(\rho)}-1) ~,~~~~~~~~~\rho = |w|
}
 where we introduced a factor of $n/2$ since the $w$ plane is
 covered $n/2$ times.  The integrand was studied in
 \ZamolodchikovUW , where it was shown that
 \eqn\Wdef{ e^{2 \hat  \alpha }-1= { 1 \over 4} (W''+{W' \over |w|})
 }
   where $W$ admits
an expansion in terms of multiple integrals. Notice that $W$ is
defined up to a solution of the homogeneous equation of the r.h.s.
of \Wdef , such freedom can be used in order to require that $W$
decays exponentially at infinity. The small $|w|$ behavior was
analyzed in \ZamolodchikovUW where it was shown that
\eqn\Wsmall{
W(t)=
{4 ( 1-1/n)^2-1 \over 4 } \log {2 \over |w| }+... } When written in
terms of $W$, the integrand of \finitearea\  is a total
derivative, hence the result can be computed from the behavior of
$W$ at $w=0$. We obtain
 \eqn\regareafirst{{  A}_{Sinh}={\pi \over
4 n}(3n^2-8 n+4)
 ~,~~~~~~~~~ A_{Sinh} (n=3) = { 7 \pi \over 12 } ~,~~~~~A_{Sinh}(n=4) = { 5 \pi \over 4 }
 }
where we have recorded a couple of special cases that will be important later.

\subsec{Finding the spacetime coordinates}

In order to understand the space-time features of the above
solution we need to solve the linear problem associated to the
flat connection.
We will not be able to find the whole surface explicitly, but we will be able to
find it along some lines and we will show that it has a $Z_n$ symmetry.

 It is more  convenient to work directly in the
$z-$plane, and consider, for instance, the radial holonomy,
between the origin of the $z-$plane and an arbitrary point at a
large distance. The radial and angular connections are simply
 \eqn\radrho{
  B_\rho=e^{i \phi} B_z+e^{-i \phi}B_{\bar{z}},~~~~B_\phi=i z B_z-i \bar{z} B_{\bar z}
 }
For both, left and right connections, where we have writen $z=\rho e^{i \phi}$.
  The solutions under consideration have an additional symmetry.
    Under a shift $\phi \rightarrow \phi +2\pi/n$, we obtain ``shifted"
    connections $\hat B_\rho$ related to the initial connections $B_\rho$ by
 \eqn\Znsym{  B^L_\rho(\phi + 2 \pi/n) = H B^L_\rho(\phi) H^{-1},~~
~B^R_\rho(\phi+ 2 \pi/n)=H^{-1} B^R_\rho H,~~~~H \equiv
\left(\matrix{ e^{i {\pi \over n}} & 0 \cr 0 & e^{-i {\pi \over
n } }}\right)
 }
 This can be seen by looking at \radrho\ and \FGhybrid .
 Suppose that we set the boundary conditions such that at the origin
 $\psi^L_{\alpha a } = \delta_{\alpha a } $ (the identity matrix). We can then compute
$\psi^L$ from the origin to a radial distance $\rho$. Then
\Znsym\ translates into $\psi^L_{\alpha a }(\phi + 2 \pi/n) = ( H
\psi^L(\phi)  H^{-1} )_{\alpha a } $ and a similar expression for $\psi^R $.
  Using \inversemap
for the spacetime coordinates we find
 \eqn\Znsymst{ \eqalign{ Y (\phi + 2 \pi/n) = & H^{-1} Y H,~~~~
Y_{a,\dot a}= \left(\matrix{Y_{-1}+i Y_0 &
 Y_1-i Y_2\cr Y_1+i Y_2 & Y_{-1}-i Y_0 }\right),
 \cr Y_{a,\dot a}( \phi + 2 \pi/n) = &\left(\matrix{Y_{-1}+i Y_0 &
 e^{-{2 \pi i \over n}}(Y_1-i Y_2)\cr e^{2 \pi i \over n}(Y_1+i Y_2) & Y_{-1}-i Y_0
 }\right)}}
 Hence, a rotation by a $2\pi/n$ angle in the world-sheet,
 corresponds to a rotation in the $Y_1,Y_2$ plane, with the
 $Y_0$ and $Y_1$ coordinates left untouched. Hence the solution
  possesses a $Z_n$ symmetry, which is of course to be expected.
   Notice that the form for  $Y_{a,\dot a}$ in \Znsymst\
   is related to the one in \inversemap\ by a simple target space $SL(2)^2$
   transformation
 \eqn\Znsycoo{\eqalign{
   \left(\matrix{Y_{-1}+i Y_0 &
 Y_1-i Y_2\cr Y_1+i Y_2 & Y_{-1}-i Y_0 }\right)= & V \left(\matrix{Y_{-1}+ Y_2 &
 Y_1- Y_0\cr Y_1+ Y_0 & Y_{-1}- Y_2 }\right) V^{-1} = \pmatrix{
 \cosh \rho e^{i \tau} & \sinh \rho e^{- i \varphi}
 \cr
 \sinh \rho e^{ i \varphi} & \cosh \rho e^{- i \tau }  }
 \cr
 V= &{1 \over \sqrt{2}}\left(\matrix{1 &
 i \cr i & 1 }\right)
  }}
 where we have also given the form of the matrix $Y_{a \dot a }$
  in $AdS_3$
 ``global'' coordinates where $ds^2_{AdS_3} = d \rho^2 - \cosh^2
 \rho d\tau^2 + \sinh^2 \rho d\varphi^2 $. In this coordinates we
 clearly see that the $Z_n$ symmetry maps  $\varphi \to
 \varphi + 2 \pi/n$. Note that $\phi$ is the worldsheet angular
 coordinate while $\varphi$ is the target space angular
 coordinate.

  \ifig\regularoct{Example of null
  polygon whose projection to the $(\tilde{x}_1,\tilde{x}_2)$
     plane is a regular polygon, in this case an octagon, with the unit circle inscribed on it. The point $A$ corresponds to the middle point between two cusps, located at $\tilde{x}_1=1,~\tilde{x}_2=0$. We first compute the holonomy between the origin $O$ and $A$ and then the holonomy between $A$ and $B$, which corresponds to the cusp located at $\tilde{x}_1=1,~\tilde{x}_2=\tan {\pi \over 2n}$
 } {\epsfxsize1.8in\epsfbox{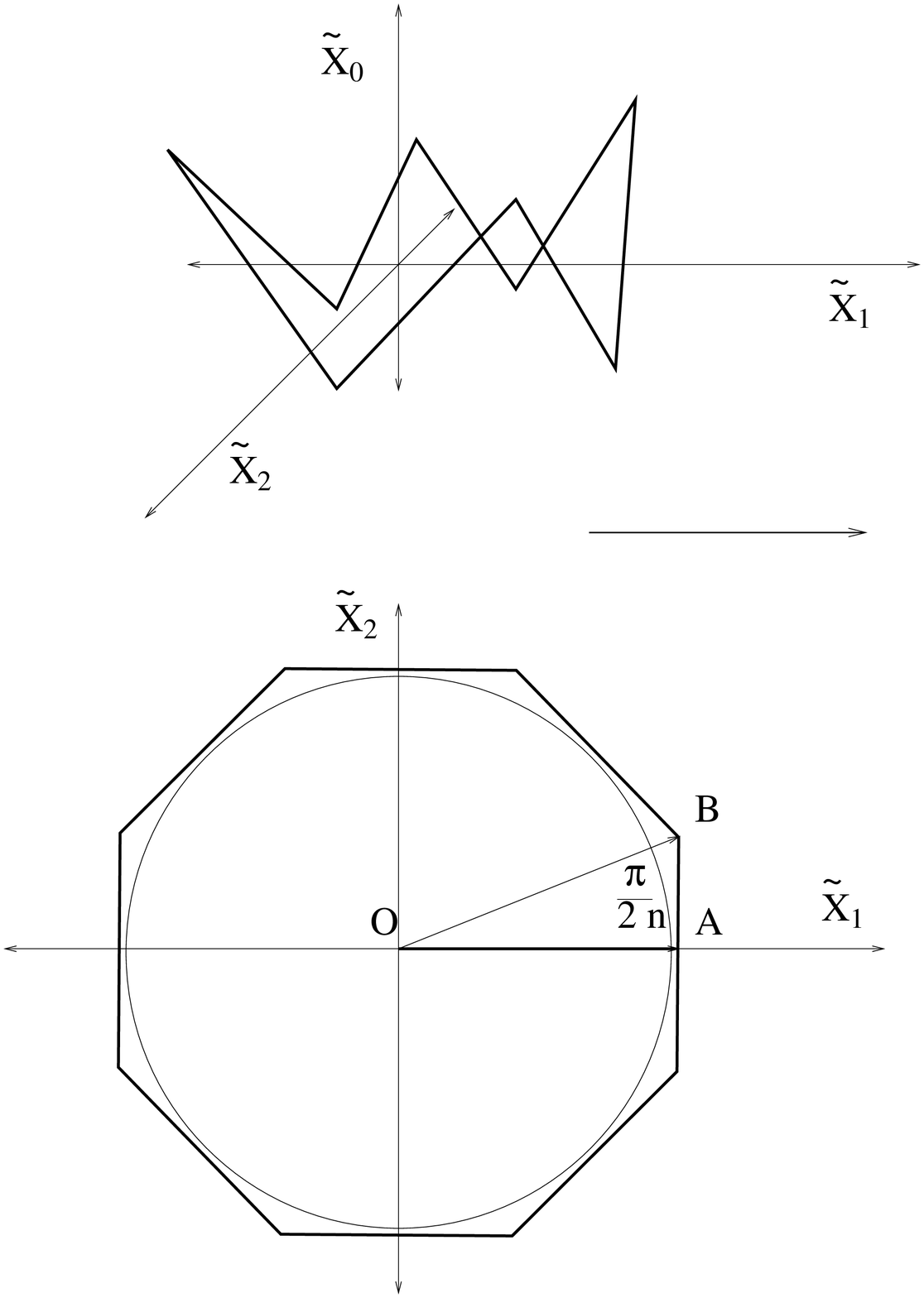}}

 When we embed these  solutions in $AdS_4$ as in \adsfouremb , we have
 a null polygon in $R^{1,2}$ which projects to an ordinary regular polygon
 with $2n$ sides in the $\tilde x_1, \tilde x_2$ plane, see \regularoct .

With our choice of spinors at the origin of the worldsheet, we see
that the origin maps to the ``center" of the polygon, at
$Y_0=Y_1=Y_2=0$ and $Y_{-1}=1$, in this case $Y_{a \dot a}$ is
simply the identity. Given the symmetry of the problem, we can
focus on a given angular region, for instance $\phi \in (0,{\pi
\over 2n})$. It turns out that the radial holonomy can be easily
computed along $\phi=0$, we find
 \eqn\radialhol{\eqalign{ \psi^L _{\alpha a}(\phi =0) = & e^{ c(\rho) \sigma^1 }
 ~,~~~~~~~~~~ \psi^R_{\dot \alpha \dot a }(\phi=0) = e^{ s(\rho) \sigma^1 }
\cr
   c(\rho) = & \int_0^\rho 2 dr r^{n/2-1}\cosh({\hat
 \alpha(r)}) ~,~~~~~~~~   s(\rho) = \int_0^\rho 2 dr r^{n/2-1}\sinh({\hat
\alpha(r)})
 }}
 where $\hat \alpha $ is evaluated at $|w(r)|$ via \regsol , and $r= |z|$.
 It is clear from the form of these matrices that the solution is mapping a line with
 $Y_0=0= Y_2$ and $Y_1$ is increasing from zero to infinity. This implies that $\tilde x_1$ is
 going from zero to one, and projecting on to the segment $OA$ marked in \regularoct .
 As $r \to \infty$ the solution maps to the  boundary point  $\tilde x_0=\tilde x_2=0$ and
 $\tilde x_1=1$ in the  Poincare
 coordinates of the $AdS_4$ embedding \adsfouremb . This is the point $A$ in \regularoct , and
 it corresponds to the
  middle point between two cusps.
  In order to understand  the boundary of the world-sheet away from this point,
 we must compute the angular holonomy. Since we are interested in computing it at a
 large distance from the origin, we can safely assume $\hat \alpha=0$, which greatly
 simplifies the computation, since we can use the approximate form of the
 solutions \twosol .  In computing the full holonomy, the following identities
  for the asymptotic
   values of the function in \radialhol\ are useful\foot{These identities can be shown by computing the radial holonomy from
 the origin along the paths at $\phi=0$ and $\phi=\pi/n$ and then the angular holonomy
 from each of these two to $\phi=\pi/2n$. For large values of the radial coordinate the
 leading part of both holonomies should coincide.}
 \eqn\identities{\eqalign{
 e^{s(\rho = \infty) }= &
{1 \over \sqrt{\tan{\pi/(2n)}}},~~~~
\cr
e^{c(\rho) }  \sim &
{ e^{4{\rho^{n/2} \over n}} \over \sqrt{\sin{\pi/n}}} = { e^{ 2 |w| } \over \sqrt{ \sin \pi/n } } ~,~~~~~
\rho \gg 1
}}
 For $\phi$ strictly bigger than zero and very large $\rho$ we obtain a very simple result
 \eqn\Ycusp{Y=
  { 1 \over \sqrt{8} \sin\pi/n} e^{ w + \bar w + { w - \bar w  \over i }  } \pmatrix{ e^{ i { \pi \over 2 n} } & e^{ -i { \pi \over 2 n} }
  \cr e^{ i { \pi \over 2 n} } & e^{ -i { \pi \over 2 n} } } ~,~~~~~~ w = {2 \over n} r^{n/2} e^{ i n \phi/2}
  }
   In  global coordinates \Znsycoo\ this corresponds to a cusp at the boundary of $AdS_3$
   at $\tau = \varphi  = { \pi \over 2 n}$.
   There is  another cusp at $\tau = \varphi = - { \pi \over 2 n }$ and so on.
   In the $AdS_4$ embedding \adsfouremb\
  this corresponds to $\tilde x_1=1$ and $\tilde x_0 = \tilde x_2 =
   \tan{ \pi \over 2 n }$ which is where we expect the cusp
  for the regular polygon with $2n$ sides, see point B in \regularoct\ .
   Notice that from the world sheet point of view, the cusp is not located at a point, but on a whole angular region.
    This already happened for the four-sided null polygon,  $n=2$,  \refs{\KruczenskiGT,\AldayHR},
     and is of course expected.
     We can now compute the regularized area. We can introduce the regulator that is natural from
     the point of view of the $AdS_4$ embedding \adsfouremb .
     This regulator involves the $AdS_4$ radial coordinate $\tilde r = 1/Y_{-1}$.
     We can regularize the integral using dimensional regularization, with $\epsilon <0$,
  \eqn\diva{ \eqalign{
   & {A}_{cutoff}= 4\int \tilde{r}^{-\epsilon}|p|d^2z= 4 \int \tilde{r}^{-\epsilon} d^2 w =
   \cr
 &=  { 4  \over ( 4 \sqrt{2} \sin { \pi \over 2 n } )^{\epsilon }}  2 n   \int_{\rm 1st~ Quadrant}
 d^2 w e ^{  \epsilon ( w + \bar w + { w - \bar w  \over i } )}  =
 { 2 n  \over ( 4 \sqrt{2} \sin { \pi \over 2 n } )^{\epsilon } \epsilon^2 }
  }}
  The sine term can be rewritten in terms of $s_{I,I+1}= d^2_{I,I+2}$, which is the square of the sum of two
  consecutive sides of the null polygon.
  For the regular polygon is  $s_{I,I+1}=16 \sin^2{\pi \over 2n}$.
     This has the generic expected form for  the divergent term
      \refs{\KorchemskayaJE,\BassettoXD,\BernIZ}.
 Thus the only non trivial contribution to the regularized area comes from \regareafirst\ .

An interesting limit of the solutions analyzed here is $n \rightarrow \infty$ .
In this case, the boundary of the regular polygon approaches a
circular Wilson loop. The area   contains a divergent piece, proportional to the ``perimeter",
$n$,  and a finite  contribution
 \eqn\regareacircle{ {A}_{Sinh}= {3 \over 4} \pi n-2\pi+{\cal O}(1/n) ~,~~\longrightarrow
 A_{renormalized}=- 2 \pi
  }
This is the renormalized area for the circular Wilson loop \BerensteinIJ\ .
 Notice that the terms involving the
divergences \diva\  are also linear in $n$, and should also be dropped.

\newsec{ Regularizing the Area }

In this section we explain how to regularize the area. We consider
a physical regularization which corresponds to placing a cutoff on
the radial $AdS_3$ direction. Writing the $AdS_3$ metric as
$ds^2 = (dx^+ dx^- + dr^2 )/r^2$ we put a cutoff that demands that
$r \geq \mu $ \foot{
There is a small problem with this regularization. If we put some points at infinity, which
is necessary is we want a real surface, then this regulator fails to regularize the area near the
points at infinity. A regulator that does not have this problem is the one natural in the
embedding \adsfouremb , which is ${ 1 \over Y_{-1} } = \tilde r > \mu$. Since the formulas are
simpler with the $r \geq \mu $ regulator, we ignore this problem. The problems with
this regulator  involve  only the three cusps that are at infinity, and the dependence of the answer
on these is fixed by the Ward identities. However, our cavalier attitude will lead to formulas
containing logarithms of negative numbers, which introduce terms involving $i\pi $'s. We ignore
all such terms and the final answers for the remainder function will be correct. }.
 Since the worldsheet approaches the
boundary when $z$ (or $w$) go to infinity, we see that this
cutoff    renders the area finite
since it does not allow arbitrarily large values of $z$ (or $w$).
In other words the regularized area is given by
\eqn\regar{
A = 4 \int_{r(z,\bar z) \geq \mu }  d^2z e^{2 \alpha}
}
In order   to cutoff the integral at large $|z|$ we need to know the
asymptotic behavior of the radial $AdS_3$ coordinate $r(z,\bar z)$.
This appears
to require a full explicit solution to the problem. However, we also know that the
asymptotic form of the solution also determines the position of the cusps, which
in turn determine the kinematic invariants such as the distance between the cusps.
Indeed we will find that
 most of the dependence on the form of the explicit solution $r(z,\bar z)$
  can be reexpressed in terms of the kinematic invariants.
Since the problem is conformal invariant one would have naively  expected that the area would
depend only on the conformal cross ratios. However, the introduction of the regulator
spoils the conformal symmetry. After subtracting the divergent piece, the finite left over answer is
not conformal invariant.   It can
depend on   distances between cusps which are Lorentz invariants but  do not
form cross ratios. In the above expression, \regar ,
 such dependence can only arise through the
form of the explicit solution $r(z,\bar z)$ which appears in the regularization procedure.
This finite piece which is not conformal invariant is constrained to obey a certain
anomalous conformal Ward identity \DrummondAU .

In order to extract the dependence on the regulator it is convenient to rewrite
\regar\ as
\eqn\regarsep{ \eqalign{
A = & 4 \int d^2z  ( e^{2\alpha} - \sqrt{ p \bar p} ) + 4
\int_{r(z,\bar z) \geq \mu } d^2z \sqrt{p \bar p} =
A_{Sinh} + 4 \int_{\Sigma} d^2 w
\cr
A_{Sinh} = &4 \int d^2z ( e^{2 \alpha} - \sqrt{ p \bar p} ) = 4 \int d^2 w (e^{2  \hat \alpha} -1)
} }
We have taken the regulator away in $A_{Sinh}$ since it is finite. This   term
seems complicated to compute. It involves only the solution to the Sinh-Gordon
problem. It depends only on   the coefficients of the  polynomial $p(z)$.
The same coefficients determine the
spacetime cross ratios. Thus $A_{Sinh}$ depends only on the spacetime cross ratios.
 When all the zeros of $p$ are widely separated
  we expect that $A_{Sinh} \to (n-2) { 7 \pi \over 12}$ which
      is the sum of the contributions at each zero.  We expect this because $\hat \alpha$ is a
      massive field which is forced to go to infinity at the branch points but is expected to
      go  to zero rapidly
      as we go away from the branch points.  We then expect to get small corrections
      which are of the form $e^{ - 4 |w_i - w_j|}$ where $|w_i -w_j|$ are the distances between branch
      points.

 The second term in \regarsep\ involves an integral over $w$. We are integrating
 over a complicated region.  Since we have a Riemann surface
we have some structure of cuts which depends on the form of the polynomial. In addition, the
boundary of the integration region at large $|w|$ depends on the explicit solution $r(z,\bar z)$.
It is convenient to disentangle these two complications in the following way.
First we note that the piece that depends on the solution
 $r(z,\bar z)$ involves the Riemann surface at large $|w|$. In this region
the Riemann surface is simpler. In the case of $n$ odd is simply a $n/2$ cover of the
$w$ plane, which we already encountered for the regular polygon.
In the case with $n$ even we also cover the plane $n/2$ times, but, in addition, we undergo
a shift $w \to w+ \wshifted$ once we go around $n/2$ times. Thus there is a single branch cut that
survives at infinity.

\subsec{Regularizing the area when $n$ is odd}

Let us first discuss in more detail the   odd $n$  case.
In this case we can cleanly split the integral $ \int d^2 w$ into two pieces.
One involves  the region that is sensitive to the
branch cuts in $w$ and the other is the integral at very large values of $w$ which
depends on the solution for $r(z,\bar z)$. We write
\eqn\separ{
4 \int_{\Sigma} d^2 w = A_{periods} + A_{cutoff}
}
where $A_{periods} $ is given the part depending on the branch cuts and comes from the
region with finite values of $w$. While $A_{cutoff}$ is the part that involves only the
large $w$ region of the Riemann surface and depends explicitly on the cutoff.
 More precisely, $A_{periods}$  is the
difference between the integral with the full structure of
 branch cuts and the integral for the regular
polygon, where all the branch cuts have been collapsed to the origin.
To compute the difference we
cut off both integrals at the same value of $|w|$ (not the same value of $z$).
This difference is finite and it can be expressed in a simple way by choosing a
basis of compact electric and magnetic cycles $\gamma_r^e$, $\gamma^{m, r}$ which obey
$\gamma^{e}_r \wedge \gamma^{m, s} = \delta_s^r$.
 We can denote by
  $w^e_s$ and $w^m_r$ the periods of the one form
   $dw = \sqrt{p} dz$ along these cycles.  Here $r,s =1,\cdots, { n-3 \over 2 } $.
  We then obtain, see appendix B,
   \eqn\aprep{\eqalign{
   A_{periods}= & {4  i   } \sum_{r=1}^{n-3 \over 2} \bar w^e_r  w^{m,r} -   w^e_r \bar w^{m,r} =
   {4  i   }  \sum_{r} \oint_{\gamma^e_{\, r}} \sqrt{ \bar p}
   \oint_{\gamma^{m, r}} \sqrt{ p } - \oint_{\gamma^e_{\, r}} \sqrt{p}
   \oint_{\gamma^{m,r}} \sqrt{ \bar p }
   \cr
    = &  \sum_r
   ({  w^e_{\, r} - \bar w^e_{\, r} \over i  })({  w^{m,r} + \bar w^{m,r}  }) -
   ({  w^e_{ \, r} + \bar w^e_{\, r}   } )
    ({  w^{m,r} - \bar w^{m,r} \over i   })
      }}
      The structure of $A_{periods}$
 is very similar to the structure we   obtain in cases where the same
Riemann surface appears in the description of the vector moduli space of ${\cal N}
=2$ supersymmetric
theories in four dimensions, where $A_{periods}$ would be the Kahler potential that
is expressed in terms of the prepotential of the ${\cal N}=2$ theory.

    Given the polynomial $p$, we can directly compute $A_{periods}$. Expressing this in terms of
    the spacetime cross ratios could be hard, since the relation between the coefficients of
    $p$ and the spacetime cross ratios could be difficult.

      \ifig\wareaodd{We have to compute the area in the $w-$plane
      of the region bounded by the lines $r=const.$, shown in the figure.
 } {\epsfxsize3.0in\epsfbox{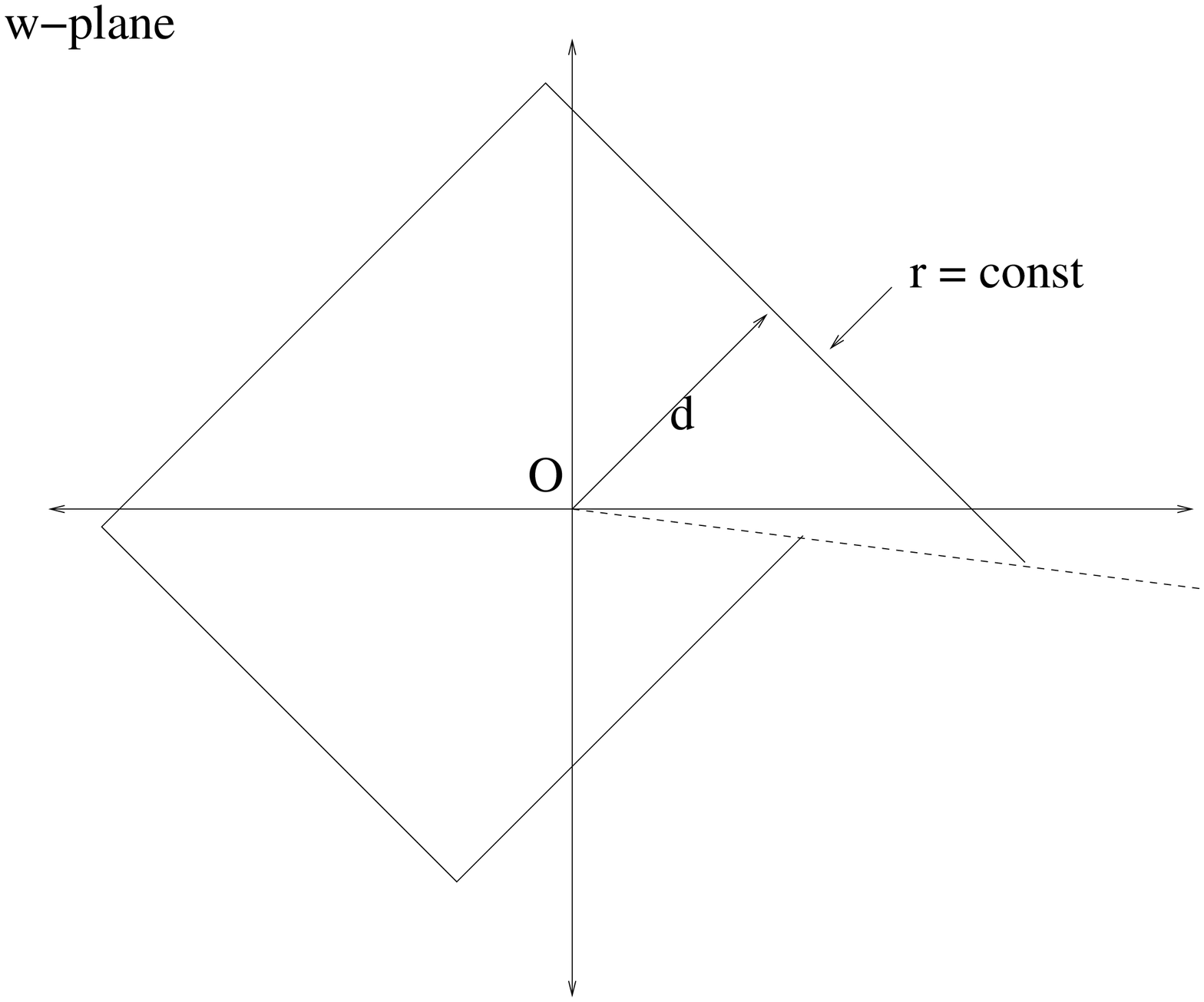}}

    Finally we are left with the integral $A_{cutoff}$.
    In this integral we define the $w$ variable as in the regular polygon.
    We then express the solution $r(z,\bar z)$ at large $w$ in terms of the  basis of
    solutions in  \twosol . A particular  solution of the left problem has an expression
    of the form
    \eqn\formexp{
    \psi = c^+ \eta_+ + c^- \eta_{-} ~,~~~~~~~~~\eta_+ = \pmatrix{ e^{ w + \bar w   } \cr 0}
    ~,~~~~~~\eta_- = \pmatrix{ 0 \cr e^{-( w + \bar w)   }  }
    }
    with different constants $c^\pm$ in different Stokes sectors. They change when we go from
    one stokes sector to the next according to the Stokes matrices. These depend on the full
    solution to the problem and would be difficult to compute.
    We have a similar expression for the right  problem. So in each Stokes sector we have
    $c_a^{L \pm}$ and $c_{\dot a}^{R \pm }$ because we are tracking a pair of solutions.
    These expressions are telling us how to cutoff the $w$ integral. For example, in a quadrant
    where both the real and imaginary part of $w$ are positive and large the expression for the
    spacetime coordinate $r$  has  the approximate form
    \eqn\approxr{
    { 1 \over r } \sim   c_{1}^{L +} c_{\dot 1}^{R +} e^{ { w + \bar w   } +
    { w - \bar w \over i   } }
     }
    The integral is cutoff at some line whose precise position depends on the
     product of these coefficients, see \wareaodd .

      The same is true at the various other cusps where we
     have similar lines. The coefficients are different in different sectors.
      However, the final answer for the area depends only on a specific combination
     of these coefficients which is such that it can be expressed in terms of spacetime
     distances between the different cusps. We leave the details of this computation for
     appendix B but we record here just the final answer
     \eqn\areaf{\eqalign{
A_{cutoff}  = &  A_{div} + A_{BDS-like}
\cr
A_{div} = & \sum_{J=1}^{2 n} { 1 \over 8}
\left( \log { d_{J,J+2}^2 \over \mu^2 } \right)^2 + g   \log { d_{J,J+2}^2 \over \mu^2 }
  }}
  where $d_{J,J+2}$ is the distance between the cusp labelled by the corresponding
   indices\foot{The capital index $J$ runs over all cusps. In terms of the index
   $i=1,\cdots, n$ which label the plus and minus
coordinates of the cusps we have that as $J=1,2,3$ we go over the cusps
at $(x^+_1,x^-_1),~ (x^+_2,x^-_1),~ (x^+_2,x^-_2),~ \cdots $.}.
  In the scattering interpretation, it is related to the square of the sum of two consecutive
  momenta. $A_{div}$ contains all the  $\mu$ dependence and its form is expected from
  general arguments, see {\it e.g.} \refs{\KorchemskayaJE,\BassettoXD,\BernIZ}.
  $g$ is a constant that characterizes the subleading divergences.
  It depends on the regularization scheme. It
  was computed in a specific regularization scheme in
\AldayHR . We will not discuss it further here\foot{We will also not discuss a finite
constant which is also regulator dependent and  is multiplied by the total
number of cusps $2n$.}. Finally the second term in
\areaf\ is
\eqn\bdslike{\eqalign{
A_{BDS-like} = &   \ell^+_i M_{ij} \ell^-_j={1 \over 2 }
\sum_{i=1}^n \sum_{j=i}^{n+i-2}(-1)^{i+j}\ell_i^+ \ell_j^-+
{1 \over 4 } \sum_{i=1}^n \ell_i^+(\ell^-_{i-1}-\ell_i^-)
  \cr
  \ell_i^+ \equiv  & \log(x^+_{i+1} - x^+_i ) ~,~~~~~~~~~~\ell_i^- \equiv  \log(x^-_{i+1} - x^-_i )
  ~,~~~~~~\ell^\pm_{n+i} \equiv \ell^\pm_i
  }}
This depends on the distances between the cusps and not purely the cross ratios.
It obeys the anomalous special conformal Ward identities derived in  \DrummondAU .
(For a derivation at strong
coupling see \KomargodskiWA\ and \AldayCG ).
In our case the special conformal generators are very simple in terms of the $x^\pm$
coordinates, $K^\pm = (x^\pm)^2
\partial_\pm $.
The anomalous Ward identities from \DrummondAU\ state that
\eqn\anomward{
K^+ A_{BDS-like} = \sum_{i} (x_i^+)^2 \partial_{x^+_i} A_{BDS-like} =
   {1 \over 4}  \sum_{i=1}^n  (x^+_{i+1} - x_i^+) \log \left({ x^-_{i+1} - x^-_i \over x^-_{i} - x_{i-1}^-}\right)
}
and a similar expression for $K^-$.
This is an important constraint, but when there are nontrivial cross ratios
it does not determine $A_{BDS-like}$ uniquely because we can always add
a function of the cross ratios. Of course, \bdslike\ obeys \anomward .

We know that there exists a solution to \anomward\ that arises when we do weak coupling computations.
Namely, it is the expression that arises at one loop, which is a particular solution of
the Ward identity equations. Just for comparison with \bdslike\   we
simply record here its form for our kinematic configuration
\eqn\resabd{
A_{BDS} = - { 1 \over 4 } \sum_{i=1}^n \sum_{ j=1,   j \not = i,i-1 }^n  \log { x^+_j - x_i^+ \over x^+_{j+1} - x^+_i }
 \log { x^-_j - x^-_{i-1} \over x^-_j - x^-_i }
 }
 One popular way to write the full answer is
 \eqn\usualexp{
 A = A_{div} + A_{BDS} +  R(\chi)
 }
 where $R$ is a finite remainder function which is a function of the cross ratios and
 contains the non-trivial information.
  We can easily write the answer in this way
 by simply saying that
 \eqn\finites{
 R = (A_{BDS-like} - A_{BDS}) + A_{periods} + A_{Sinh }
 }
 The difference $(A_{BDS-like} - A_{BDS})$ depends only on cross ratios and it is
 written explicitly in appendix E . This is the final expression for the remainder
 function.
   The complicated part of the problem is to compute $A_{Sinh}$ and also to
express the coefficients of the polynomial as a function of the spacetime cross ratios.
Of course, in the end we simply want to compute the area as a function of the cross ratios.
If one devised a method that does not need the coefficients of the polynomial, that would be
better!. The final expression \finites\ depends only on the cross ratios and one can send some
points to infinity if one wanted.

Finally, note that the formula \metricarea\ is expected to give us $A_{Sinh} + A_{periods}$ in
a  regularization which introduces a cutoff in the $w$ plane and throws away the divergent terms.
   In other words,
    the left hand side of \metricarea\ should be $A_{Sinh} + A_{periods}$.

\subsec{Regularizing the area for $n$ even }

\ifig\wplaneevenref{Structure of the $w$ surface for $n$ even. In this case we have $n=8$ (a
16-gon) in a
specific configuration with a nonzero $w_s$ but with other periods vanishing. The surface has four
sheets in this case and it is missing a sliver. Here we have placed half of the missing sliver in
the first sheet and half in the last sheet.
 } {\epsfxsize3.0in\epsfbox{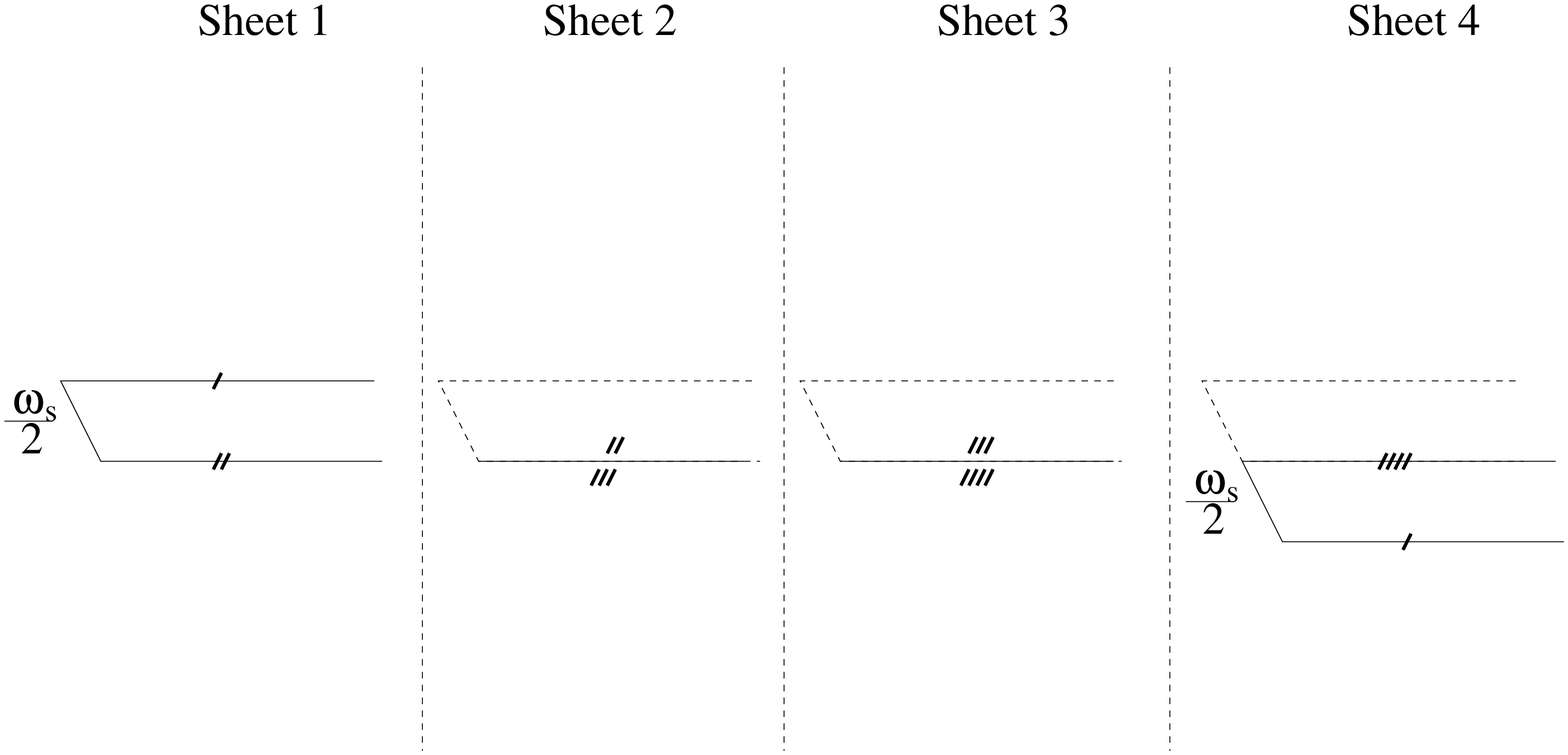}}

The case with $n$ even is a bit more complicated.
In this case we find that
 $\sqrt{p} \sim z^{n/2 -1} + \cdots + { \tilde m \over   z } + \cdots $, where we defined $\tilde m $
  to
 be the combination of the coefficients of the polynomial $p$ which multiplies the $1/z$ term in
 this expansion.
 The term that goes like $1/z$ leads to a
logarithmic term in $w$ as we integrate the one form $dw = \sqrt{p} dz$. Thus, when $n$ is even
$w \to w + \wshifted$ as we go around $n/2$ times, where $\wshifted = i 2 \pi \tilde  m $.
The fact that the information of one of the coefficients of the polynomial survives at large $w$
has one nice consequence. It allows us
to compute a particular combination of spacetime cross ratios in a simple way, see appendix B,
 \eqn\crosssimp{ \eqalign{
  e^{ {  \wshifted + \bar \wshifted  } } = & - { ~ x^+_{23} ~ x^+_{45} ~x^+_{67} \cdots ~ x^+_{n1 } \over
  x^+_{12} ~ x^+_{34} ~ x^+_{56} \cdots x^+_{n-1,n} }
  \cr
   e^{ {  \wshifted - \bar \wshifted  \over i  } } = & - { ~ x^-_{23} ~ x^-_{45} ~x^-_{67} \cdots ~ x^-_{n1 } \over
  x^-_{12} ~ x^-_{34} ~ x^-_{56} \cdots x^-_{n-1,n} }
}}
The complication arises when we try to split the answer into a form similar to
\separ . We cannot   replace the $w$ space  by a simple multicover as for the case of
the regular polygon at infinity. In fact
the $w$ plane is missing a sliver, see \wplaneevenref .

 When we try to separate the two
pieces we   introduce a fictitious dependence on each term under shifts of the origin
in the $w$ plane. This is reflecting the fact that the full answer involves the knowledge of
the precise length of the sliver. In other words, if we select an origin for the $w$ plane,
then the solution at infinity in the
region of the first cusp has the form $ c^+ \eta_+ + c^- \eta_- $. The value of $c^+$
depends on the full solution of the problem. Alternatively, this can also be expressed in
terms of the value of the Stokes matrix at the first cusp, see appendix B.
Let us denote by $\stokesparameter_1$ the
off diagonal coefficient for the Stokes matrix on the first cusp. In order to define this quantity
we need to choose an origin in the $w$ plane. Let us choose it at one of the
zeros of $p$. Thus $w$ is defined by integrating $dw = \sqrt{p} dz$ starting from this particular zero.

Then the final answer contains the following pieces
\eqn\nevenfinal{
4 \int_{r(z, \bar z) \geq \mu }
 d^2 w = A_{div} +  A_{BDS-like-even} + A_{periods} + A_{extra}
}
$A_{div}$ is the same as in \areaf .
$A_{BDS-like-even}$  has a structure similar to \bdslike , except that
now the first cusp is treated in a special way
\eqn\abdslikeeven{\eqalign{ A_{BDS-like-even}  = &
 \sum_{i,j} \ell^+_i \hat M_{ij}^{(1)} \ell^-_j -  { 1 \over 2 }
  ( { \wshifted - \bar \wshifted \over i }) \ell^+_1  +
   { 1 \over 2 }    ( { \wshifted  + \bar \wshifted  })\hat \ell^-_1-(\sum_{i=1}^n (-1)^i \ell_i^+ )^2-(\sum_{i=1}^n (-1)^i \ell_i^- )^2
   \cr
 \sum_{i,j}  \ell^+_i \hat M_{ij}^{(1)} \ell^-_j = &
   { 1 \over 2 }  \sum_{i=1}^n \sum_{j =i}^{n}   (-1)^{i+j} \ell^+_i \ell^-_{j}
    - { 1 \over 2 }  \sum_{i=1}^n \sum_{j =1}^{i-1}   (-1)^{i+j} \ell^+_i \ell^-_{j}
  -
 { 1 \over 4 } \sum_{i=1}^n \ell^+_i ( \ell^-_i + \ell^-_{i-1} )
 }}
 where
 \eqn\hateldef{
  \hat \ell_i^\pm  \equiv   \log ( x^\pm_{i+1} - x^\pm_{i-1} ) - \ell_i^\pm - \ell^\pm_{i-1} =
  \log { ( x^\pm_{i+1} - x^\pm_{i-1} ) \over (x^\pm_{i+1} - x^\pm_i ) (x^\pm_{i } - x^\pm_{i-1} )}
   }
   Note that only $\hat \ell^\pm_1$ appear in \abdslikeeven . In addition, the matrix $M^{(1)}$ treats
   the first cusp in a special way.
We also have that
\eqn\aextra{ \eqalign{
  A_{extra} = &    - { 1 \over 2} { ( \wshifted + \bar \wshifted)   }\log \stokesparameter_1^R
  +  { 1 \over 2 } { ( \wshifted - \bar \wshifted) \over i }
  \log \stokesparameter_1^L
  }}
   $A_{extra}$
   takes into account the total shift in $w$ that has accumulated in the
  exact solution relative to the approximate solution. It represents
   the extra area that the sliver in \wplaneevenref\  has   gained due
  to this shift in $w$.
 The Stokes parameter $\stokesparameter_1^L$
  can also be viewed as a ``magnetic'' cross ratio $\chi^m$.
  This point of view arises
  naturally if we consider the even case to be a limit of the odd case when we take one
  of the points to infinity. This is not a proper cross ratio and its definition depends
  on a choice of cusp as well as a choice of origin for $w$. These also appeared in
   \GaiottoCD .

 \ifig\cycleseven{ (a) Pairs of cycles for the case that $n$ is even (here $n=6$ or a 12-gon).
 One of the cycles is non compact.
(b) Cycles for the case that $n$ is odd.
 We can choose them all compact. Here we represented the $n=5$ case  (a $10$-gon).
 }{\epsfxsize4.5in\epsfbox{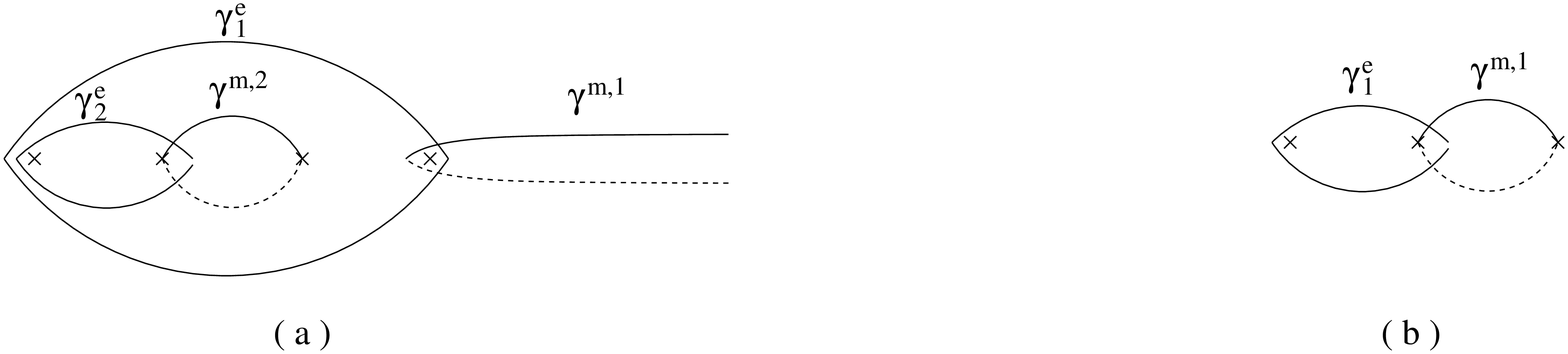}}

   $A_{periods}$ is similar to
  \aprep\ except that now one of the magnetic cycles is non-compact.
  We choose it to be a cycle that goes to infinity starting from
 the zero of the polynomial $p$ that we chose above
  as the origin for $w$, see \cycleseven. This choice of zero was important for providing the correct
  normalization for the Stokes parameter in \aextra .
  We take the dual electric cycle to be an integral
 around all the zeros of $p$. Of course this integral just gives $ \wshifted = w^e_1$. Notice that
 the integral along the magnetic cycle is infinite, $w^{m, 1 } = \infty$.
 We then complete this basis of cycles to
a complete cannonical basis of cycles $\gamma^e_r $, $ \gamma^{m,r} $, with
$r = 2, 3,\cdots, { n - 2 \over 2}$. Then we have
\eqn\aprepeven{ \eqalign{
A_{periods} = &  { 4 i   } \sum_{r=2}^{n-2 \over 2} \bar w^e_r   w^{m,r} -   w^e_r \bar w^{m,r} =
    4 i \sum_{r} \oint_{\gamma^e_r} \sqrt{\bar p}
   \oint_{\gamma^{m, r}} \sqrt{ p } - \oint_{\gamma^e_r} \sqrt{p}
   \oint_{\gamma^{m,r}} \sqrt{  \bar p }
   \cr
    = &  \sum_r
     ({  w^e_r - \bar w^e_r \over i   })({  w^{m,r} + \bar w^{m,r}   }) - ({  w^e_r + \bar w^e_r   } )
    ({  w^{m,r} - \bar w^{m,r} \over i  })
      }}
  which is similar to the expression for the odd case \aprep , except that the first cycle,
  the cycle  that goes
  around all the poles is missing together with its dual non-compact cycle. Of course \aprepeven\
  is finite.

 The extra term \aextra\ depends on the choice of the first cusp.
   This choice of a first cusp is also
  present in our expression for \abdslikeeven . If we treated the second cusp in a special way,
  then we would have a change in both of these terms and these changes would cancel out.
  Indeed, one
 can also check that even though the individual terms are not cyclically invariant the sum of
 $A_{extra} + A_{BDS-like}$ are cyclically invariant once we change the choice for the special cusp in
 both at the same time.
  Similarly, there is a dependence of $\stokesparameter_1$ on the choice of a special zero of $p$ where
  we defined the zero of $w$. This choice also figures in the definition of \aprepeven\ where
  we chose the same zero of $p$ to define the divergent  magnetic period. These two choices need
  to be the same. In this way the point we choose does not matter.

\newsec{The Octagon}

In this section we carry out the computation of the various terms in the
area for the octagonal Wilson loop.
We take the polynomial to be
\eqn\polynom{
p(z) = z^2 - m , ~~~~~~\bar p(\bar z) = {\bar z}^2 - \bar m
}
 Then the shift in the $w$ plane at infinity is given by
\eqn\wzer{
\wshifted = - i \pi m  = \oint_{|z| \gg \sqrt{|m|} }  dz \sqrt{z^2 - m}
}
 \ifig\octcrossratios{The octagon configuration has two cross-ratios, $\chi^+$ and $\chi^-$ .
  In the figure, we have sent the points $x_4^{\pm}$ to infinity. Besides, we have fixed $x_1^\pm=0$ and $x_3^\pm=1$. With this choice $\chi^\pm={1-x_2^\pm \over x_2^\pm}$.
   Notice that, as $\chi^\pm$ goes from zero to infinity, $x_2^\pm $ goes between one and zero.
 }{\epsfxsize3in\epsfbox{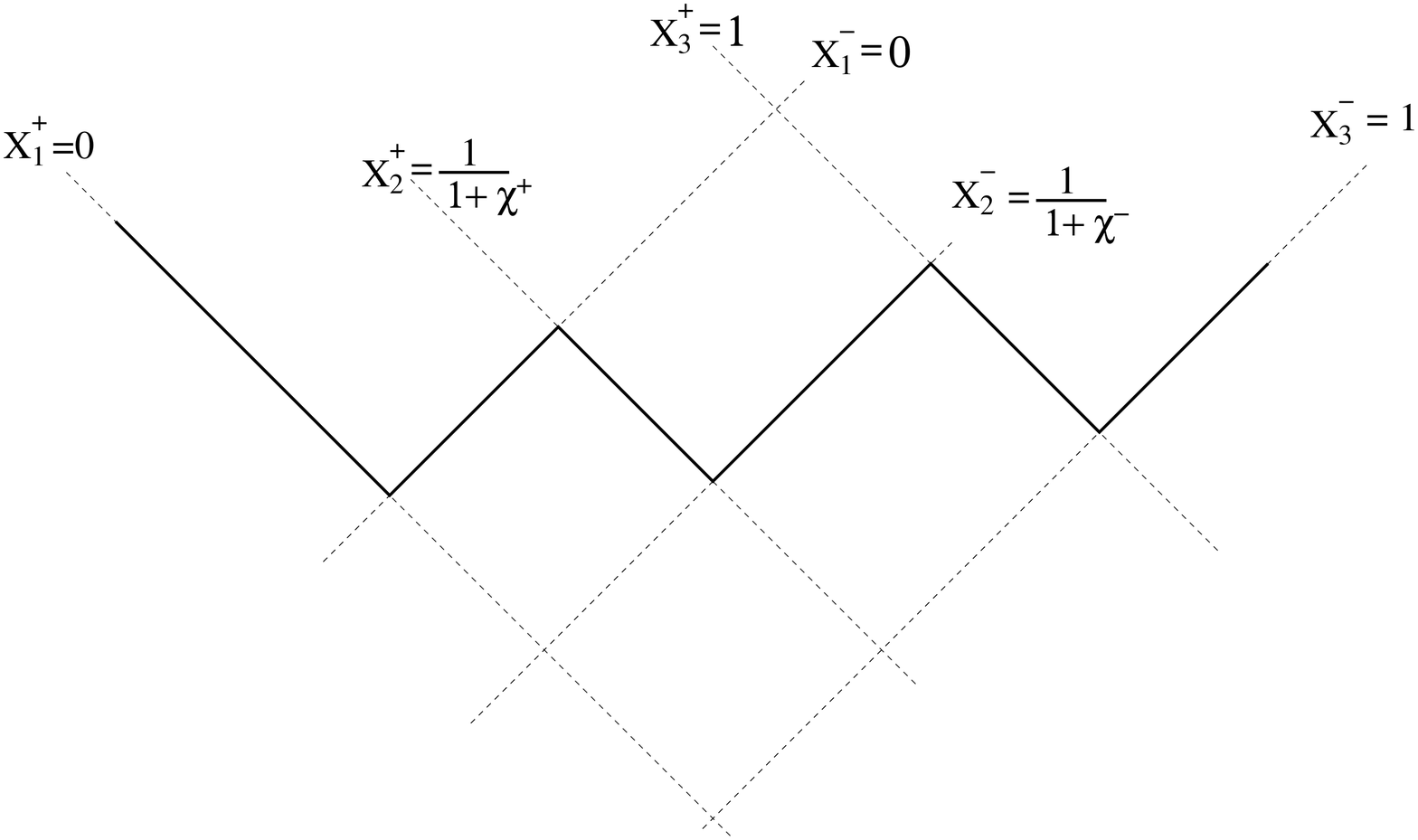}}
%
%
 In this case we have four $x_i^+$ values and
four $x_i^-$ values and two  cross ratios, one depending on $x^+_i$ and one on $x^-_i$.
 We can write them  using \crosssimp\
as
\eqn\singlc{ \eqalign{
\chi^+ \equiv & e^{ \wshifted + \bar \wshifted } =  e^{ \pi ( { m - \bar m \over i } ) } =
{ ( x_4^+ - x_1^+)  ( x_3^+ - x_2^+)
 \over ( x_4^+ - x_3^+ ) (x_2^+ - x_1^+)  }
\cr
\chi^- \equiv & e^{ \wshifted - \bar \wshifted \over i } =  e^{ - \pi ( { m + \bar m  } ) } =
 { ( x_4^- - x_1^-)  ( x_3^- - x_2^-)
 \over ( x_4^- - x_3^- ) (x_2^- - x_1^-)  }
}}
In this case the relation between the parameter of the polynomial, $m$, and the
spacetime cross ratios \singlc\ can be determined exactly and in a simple fashion.

We can now use the results in \GaiottoCD\ in order to compute the various terms in the
area that we discussed above. Our problem is mathematically the same as the problem
considered in \GaiottoCD . The problem considered in \GaiottoCD\ arises when one considers the moduli
space of a theory containing a single charged hypermultiplet that is becoming light at one
point in moduli space. The theory is a four dimensional ${\cal N}=2$ supersymmetric theory
compactified on a circle to three dimensions, in the regime that the radius of the circle is
finite. Then one considers the vector multiplet moduli space, which is a four dimensional
hyperkahler space, as a torus fibration over a two dimensional Kahler space parametrized
by $m, \bar m$.
We will use the following information from \GaiottoCD .
They give us the
explicit form for the Kahler metric $g_{m \bar m}$, using previous computations
in \refs{\OoguriME,\SeibergNS} and the explicit form
for what they call ``magnetic cross ratio'', which is basically the same as the Stokes coefficient
$\stokesparameter_1$ introduced above.
Notice that \GaiottoCD\ does not give us the explicit solution of
the Sinh-Gordon equation. In fact, they can    compute the answer by understanding the
analytic structure of the cross ratios as a function of $\zeta$. This is what one expects
to be able to do in the general case, as explained in \GaiottoCD .
Thus we read off the following two expressions from \GaiottoCD
\eqn\metricf{
 \partial_{m} \partial_{\bar m} K =    g_{m \bar m} = A
 \sum_{n=-\infty}^{\infty} {1 \over \sqrt{|m|^2   +(n+1/2)^2 } }  + {\rm const}
}
and
\eqn\gammaone{ \eqalign{
\log \stokesparameter_1 (\zeta)  =  { e^{ - i \phi} \zeta \over \pi } \int_{-\infty}^\infty dt { e^{ t }
\over e^{ 2 t } + \zeta^2  e^{- i 2 \phi } }  \log\left( 1 + e^{ -2 |m| \pi \cosh t }  \right)
\cr
 \log \stokesparameter_1^L = \log \stokesparameter_1(\zeta =1) ~,~~~~~~~~~~ \log \stokesparameter^R_1 = \log \stokesparameter_1(\zeta=i)
 ~,~~~~~~m = |m| e^{i \phi}
 }}
where $A$  is a  constant  we will fix soon. We have written things in terms of the absolute value
of $m$ and its phase \foot{
In the formulas of \GaiottoCD\ we should
 set $R=1, ~ a=-i m$, which is just a matter of conventions, and $ q =1$ can be set
 by comparing the approximate expression for $\stokesparameter_1$ which we would get in
 what is called the ``semi flat'' approximation in \GaiottoCD .
 Here $\stokesparameter_1(\zeta) = \chi^m(\zeta)/\chi^m_{sf}(\zeta) $.}.

\subsec{Evaluation of $A_{Sinh}$ for the octagon }

Let us start with the first term, $A_{Sinh} = 4 \int d^2 z ( e^{ 2 \alpha} - \sqrt{ p \bar p } )$.
 As explained above we can obtain this from the metric \metricf\ by first finding the
 Kahler potential by integrating \metricf\ twice and then using
  \metricarea .
 This gives us
\eqn\kahlerder{\eqalign{
 4 \int d^2 z e^{ 2 \alpha } \sim & (m \partial_m +\bar m \partial_{\bar m})K
  \cr
 4 \int d^2 z e^{ 2 \alpha }  = & -16 A |m|  \sum_{n =1}^\infty {(-1)^n \over n \pi  } K_1(2\pi |n   m|)  -4 A m \bar m \log m \bar m +   \hat c_2  m \bar m+
     \hat c_3 }}
 Which depends only on three undetermined parameters.
 The constants $\hat c_2 , ~\hat c_3$   in \kahlerder\ are divergent. This divergence
 arises from the integral in the large $z$ region where we can approximate $e^{ 2 \alpha} \sim
 \sqrt{ p \bar p }$. Thus we will also compute the integral
 \eqn\Areasec{
 \eqalign{\partial_{|m|} \partial_m \partial_{\bar m} 4 \int \sqrt{p \bar p}d^2 z=&
  -{\pi \over  |m|}  ~~~\longrightarrow
   \cr  4  \int \sqrt{p \bar p}d^2 z=&
   -{\pi \over 2} m \bar m \log m \bar m +  \tilde c_1  \log m \bar m  + \tilde c_2 m \bar m  + \tilde  c_3 }}
   where $\tilde c_2, ~\tilde c_3$ are again divergent constants, whose divergent parts should
   be the same as in \kahlerder , and $\tilde c_1$ is an integration constant which we expect to
   be zero, but we leave it for the time being.
 The final result is then
 \eqn\answeroct{\eqalign{ 4 \int d^2z(e^{2 \alpha}-\sqrt{p \bar p})= &
 -16 A |m|  \sum_{n =1}^\infty {(-1)^n \over n \pi  } K_1(2\pi |n   m|)+
 \cr + &
  {1 \over 2} m \bar m (\pi-8 A)\log m \bar m + c_1  \log m \bar m + c_2 m \bar m  +c_3 }}
where we have denoted by $c_i$ all the finite integration
constants.
 These are all finite since the left hand side is finite.
  For large $|m|$ we expect $A_{Sinh}$ to approach a constant for the following reason.
  In this case the zeroes of $p$ are far apart from each other,
  and the solution is simply the linear superposition of two ``single-zero" solutions.
  The single zero solution is the case of the   hexagon.
   Hence we expect the above result to approach twice its value for the regular hexagon.
   Since the sum of Bessel functions decays exponentially for large $|m|$, the above discussion
   implies $c_1 = c_2=0$ and $A=\pi/8$, fixing almost all the constants.
    Finally, from the results for the regular hexagon
   \regareafirst
    , we have $c_3={7 \pi \over 6}$.
     The final answer is then
 \eqn\finaloct{ A_{Sinh} = 4 \int d^2z(e^{2 \alpha}-\sqrt{p \bar p})=-   2 |m|
 \sum_{n =1}^\infty {(-1)^n \over n    } K_1(2\pi |n   m|)+{7 \pi \over 6} }

Note that some care is needed in applying \metricarea\ since one has to subtract the divergent
pieces properly and possibly add suitable constants to the answer.

Having fixed all the constants,
the behavior for small $m$ is then fixed by the sum of Bessel functions.
 We obtain
 \eqn\finaloctf{-  2  |m|  \sum_{n=1}^\infty {(-1)^n \over n  }
 K_1(2\pi |n   m|)=
 {\pi  \over 12 }+  \pi  |m|^2 \log |m|
 + o(|m|^2)
 }
which we will use later as we perform a couple of checks.

 Notice that the exponential decay in \finaloct\ goes as
 $e^{ - 2 \pi |m| } \sim e^{ - 4 { |\wshifted| \over 2  } } $ where ${ |\wshifted| \over 2 }$ is the distance, in
 the $w$ plane between the two branch points. We see that the exponential decay is precisely what
 we expect, since the $\hat \alpha $ field has mass equal to four.

 Notice that the final result for $A_{Sinh}$ is independent of the phase of $m$. This is is expected
 since the sinh gordon problem is independent under a rotation of $m$. In fact, this is a general feature
 for any $n$, the Sinh-Gordon part of the problem is invariant under rotations in the $z$ plane.

\subsec{Evaluation of $A_{extra}$ for the octagon}

 Let us now turn to the evaluation of $A_{extra}$. Using \gammaone \aextra
 we find
  \eqn\termequ{ \eqalign{
  A_{extra}  = &  - { \pi \over 2}
  (  m + \bar m ) \log \stokesparameter_1^L -  { \pi \over 2 }   { m - \bar m \over i } \log \stokesparameter_1^R
  \cr
  = &
 F(|m|,\phi)=
  -|m| \int_{-\infty}^\infty dt
   {\sinh ( t + i 2 \phi )\over \sinh(2 t + 2 i \phi)} \log \left(1+e^{-2|m| \pi \cosh t} \right)
    }}
 This function is defined by the integral expression when $\phi \in ( 0 , \pi/2) $ and by analytic
 continuation for other regions.
\ifig\poles{ Poles in the integrand of \termequ\ whose position
changes when we change $\phi$. As we increase $\phi$ the poles
move downwards and can cross the integration contour. In that case
we move the integration contour to obtain the analytically
continued function.
 }{\epsfxsize1.5in\epsfbox{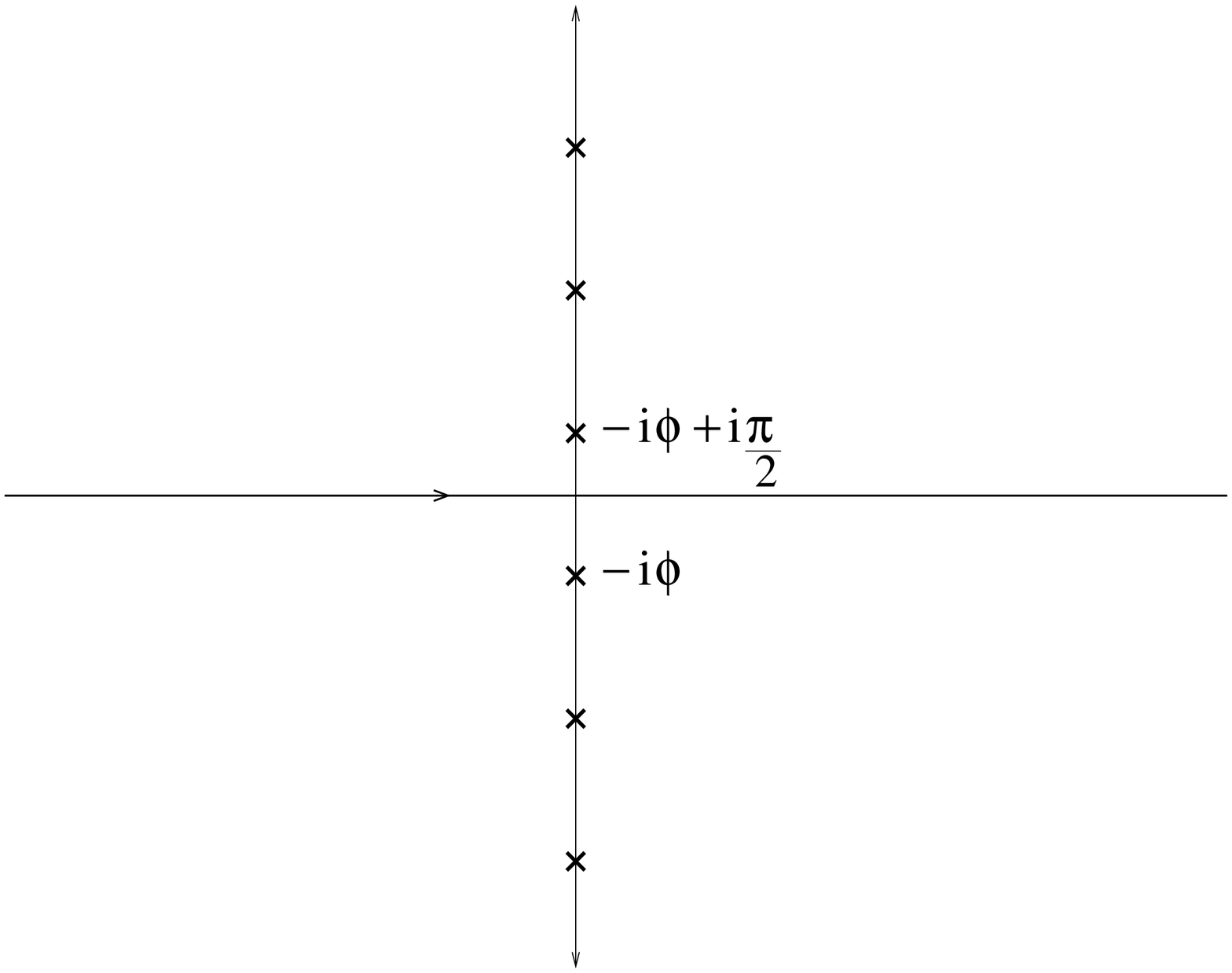}}
\ifig\polescrossing{ As we increase the value of $\phi$, a pole
can cross the real line.
 The deformed contour can then be written as the sum of a contour over the real line
  plus a contour encircling the pole.
  }{\epsfxsize3.0in\epsfbox{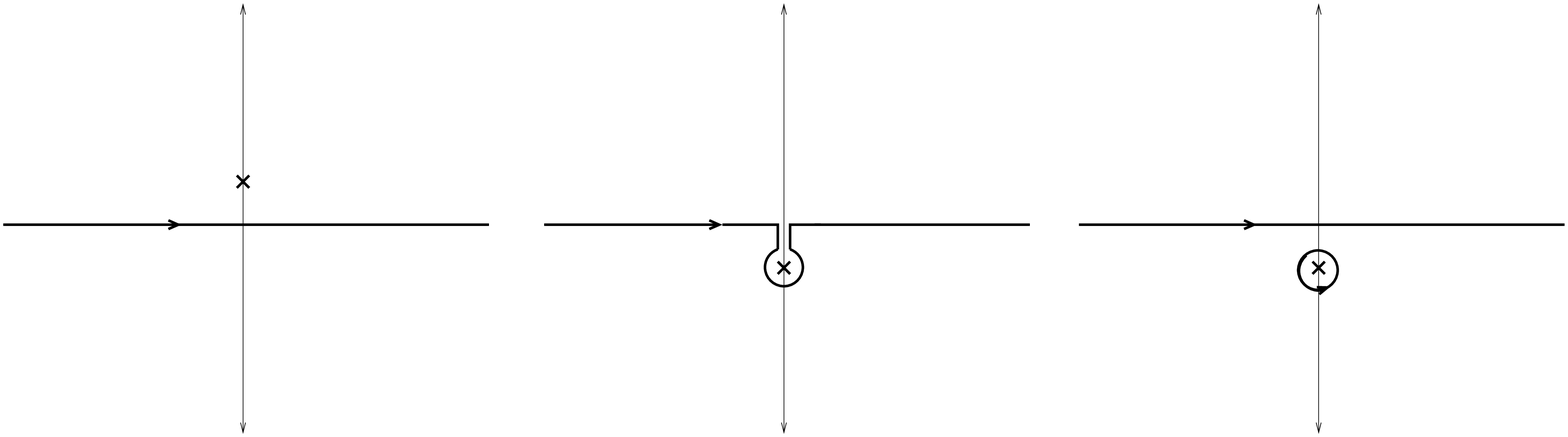}}

 It can be checked that the function so defined is periodic as we
 take $\phi \to \phi + 2 \pi $. This is not obvious since there are poles that cross the integration
 contour as we increase the value of $\phi$, see \poles\ . It turns out that the sum of the residues of
 all poles cancel once we take $\phi \to \phi + 2 \pi$.
%
 %
 Let us be more explicit. Note that the integrand in \termequ\ is
 periodic under $\phi \to \phi + {\pi \over 2}$. This corresponds to going
 from one quadrant to the next. In fact, we can check that
 \eqn\equpo{\eqalign{
  F(|m|, \phi + \pi/2) =& F(|m|, \phi) + P(|m|,\phi)
\cr
 P(|m|, \phi ) = &  \pi |m| \sin \phi \log ( 1 + e^{ - 2 |m| \pi
 \cos \phi } )
 }}
 where $P$ comes from evaluating the residue at the pole at $t_p = -
 i (\phi + \pi/2) + i \pi/2 = -i \phi $. Which is the pole that is crossing the
 integration contour as we increase $\phi$ from the first region
 into the next.
We can then see that $F( \phi + 2 \pi) = F(\phi) + P(\phi) + P(
\phi + { \pi\over 2} ) + P( \phi + \pi ) + P(\phi + {3 \pi \over
2}) = F(\phi )$ since \eqn\sumpo{  P(\phi) + P( \phi + {\pi \over
2}) + P( \phi + \pi ) + P(\phi + { 3 \pi \over 2}) = 0 } where we
suppressed the $|m|$ in the argument for simplicity. 

 It turns out that $F$ has a non-analytic piece in $|m|$ for small $m$ which goes as
  \eqn\fna{F(|m|,\phi)|_{\rm non~analytic}=-{ \pi }|m|^2  \log |m|
  }
 This non-analytic piece is cancelled by the non-analytic term in
  \finaloctf\ so that the full answer is smooth at small $|m|$ as
  expected.
 Except for \fna , the full function \termequ\ is analytic in $|m|$. Naively,
 we expected the Stokes parameters
 to be analytic in $m$. However, since the precise definition of the Stokes parameters depends
 on the origin we choose in the $w$ plane, we can introduce some non-analytic dependence due to
 this choice of origin. In particular, if we choose one of the zeros of the polynomial as the origin,
 then we see that the non-analytic behavior in \fna\ arises from the semiclassical or WKB expression
 for the ``magnetic coordinate'' which is given by the integral between that zero and infinity.
 We can write this  in terms of the magnetic period
 \eqn\intergf{
 w^m_1  \sim 2 \int_{\sqrt{m} } dz \sqrt{ z^2 - m } \sim  {  m \over 2 }  \log m + m ({ \rm constant})
 }
 where the constant is divergent and it
 comes from cutting off the integral at a large
 value of $z$. Here we see a  non-analytic behavior.
  This integral  also appears in setting the origin of the $w$ plane, which is
   involved in defining the Stokes parameter $\stokesparameter$. This is the reason we
    get a non-analytic piece in $\stokesparameter$.
    Thus \intergf\ is the origin of the non-analytic piece in \fna .

 We can also check the expression for $\stokesparameter_1$ by computing it for the regular octagon. In that
 case we need to compute the $m\to 0$ limit of \gammaone . We see from \gammaone\ that for $m=0$
 we get $\log \stokesparameter_1 = \log \sqrt{2} $ which is in agreement with the result for the
 regular  octagon (see appendix B).

  As a non trivial check, we can look at the full answer in the limit $m\to 0$. In this
  case $A_{extra}$ goes to zero due to the factor of $|m|$ in \termequ .
  On the other hand we see from \finaloct\ and \finaloctf\ that $A_{Sinh}$ goes
  to a constant.  Namely, $A_{Sinh} \to  {7 \pi \over 6} + { \pi \over 12}  = {5 \pi \over 4}$,
   which is the answer for the
 regular octagon
  \regareafirst .

 \subsec{ Summary of the final answer for the octagon }

 The final answer for the octagon has the from
 \eqn\finalan{ \eqalign{
 A = &  A_{div} + A_{BDS-like-even} + A_{Sinh} + A_{extra}
 \cr
 A = &  A_{div} + A_{BDS } + R
 \cr
 R = & -{ 1 \over 2} \log(1 + \chi^-) \log ( 1 + { 1 \over \chi^+ } )
+
  { 7 \pi \over 6 } +  \int_{-\infty} ^{\infty}
 dt  { |m| \sinh t  \over \tanh(2 t + 2 i \phi) }
 \log\left( 1 + e^{ - 2 \pi |m| \cosh t } \right)
 \cr
 R = & -{ 1 \over 2} \log(1 + \chi^-) \log ( 1 + { 1 \over \chi^+ } )  + { 7 \pi \over 6 } + { 1 \over 2} \int
 dt    { ( \bar m e^t - m e^{-t} ) \over
  \tanh 2 t }
 \log\left( 1 + e^{ -   \pi  ( \bar m e^{t} + m e^{-t} ) } \right)
\cr
 }}
 $\chi^\pm$ and $m, ~ \bar m$ are defined in terms of the two spacetime cross ratios
 in \singlc .
 The first term in $R$ arises as the difference between $A_{BDS-like-even} - A_{BDS}$ and is computed
 in appendix E. The second term results from combining $A_{Sinh} + A_{extra}$. For this particular
 case $A_{periods}$ is zero since there are no other cyles.
We have written \finaloct\ as an integral expression and combined it with
\termequ . In the last term we shifted the integration variable to emphasize the dependence
of the function on
  $m$ and $\bar m$ and the integration contour is determined by the previous expression.

%
Even thought it is not explicit, the expression \finalan\ is
periodic under $\phi \to \phi + { \pi \over 2 } $. This can be
made more explicit by rewriting the integral $I$ in \finalan\ as
\eqn\integr{ I = I_{\rm periodic} - { 1 \over 4 } \left[
 3 P(\phi) + 2 P (\phi + { \pi \over 2 } ) + P( \phi + \pi ) \right]
 }
 Where $I_{\rm periodic} $ is defined through this formula\foot{ Since $A_{Sinh}$ depends only
 on $|m|$ the discussion regarding the poles  crossing   the contours as we move $\phi$
  is the same as
 in \equpo }.
  We can check using
 \equpo , \sumpo\ that $I_{\rm periodic}(\phi + { \pi \over 2 } ) = I_{\rm periodic}( \phi)$.
 We can then combine the second term in \integr\ with the first term in \finalan\ to write
 the remainder function as
 \eqn\remainper{\eqalign{
 R = & - { 1 \over 2 } \log \left[2\cosh( |m| \pi \sin \phi)\right] \log\left[
  2 \cosh( |m| \pi \cos \phi )\right]  + { 7 \pi \over 6}
 + I_{periodic }( |m|, \phi)
 \cr
R = & - { 1 \over 8 } \log \left[ { ( 1 + \chi^+)^2 \over \chi^+ }
\right] \log\left[
   { ( 1 + \chi^-)^2 \over \chi^- } \right]  + { 7 \pi \over 6}
 + I_{periodic }( |m|, \phi)
 }}
Written in this way we see that the remainder function has a
manifest symmetry under changing $\phi \to \phi + \pi/2$ and $\phi
\to - \phi $. This implies that it is invariant under $ \chi^+ \to 1/\chi^-$, $\chi^- \to \chi^+$
and also under $\chi^+
\leftrightarrow \chi^-$.    This is a consequence of spacetime parity,
the cyclicity  and conformal invariance. These   symmetries  were broken by
our choice of a special cusp during the regularization procedure.
  Here we are checking explicitly that the
final full answer has all the expected symmetries.

\subsec{Soft and collinear limits  }


\ifig\doublesoft{ Double soft limit where two consecutive segments are going to zero, $\epsilon^+ \to 0$ and
$\epsilon^- \to 0$.
  }{\epsfxsize3in\epsfbox{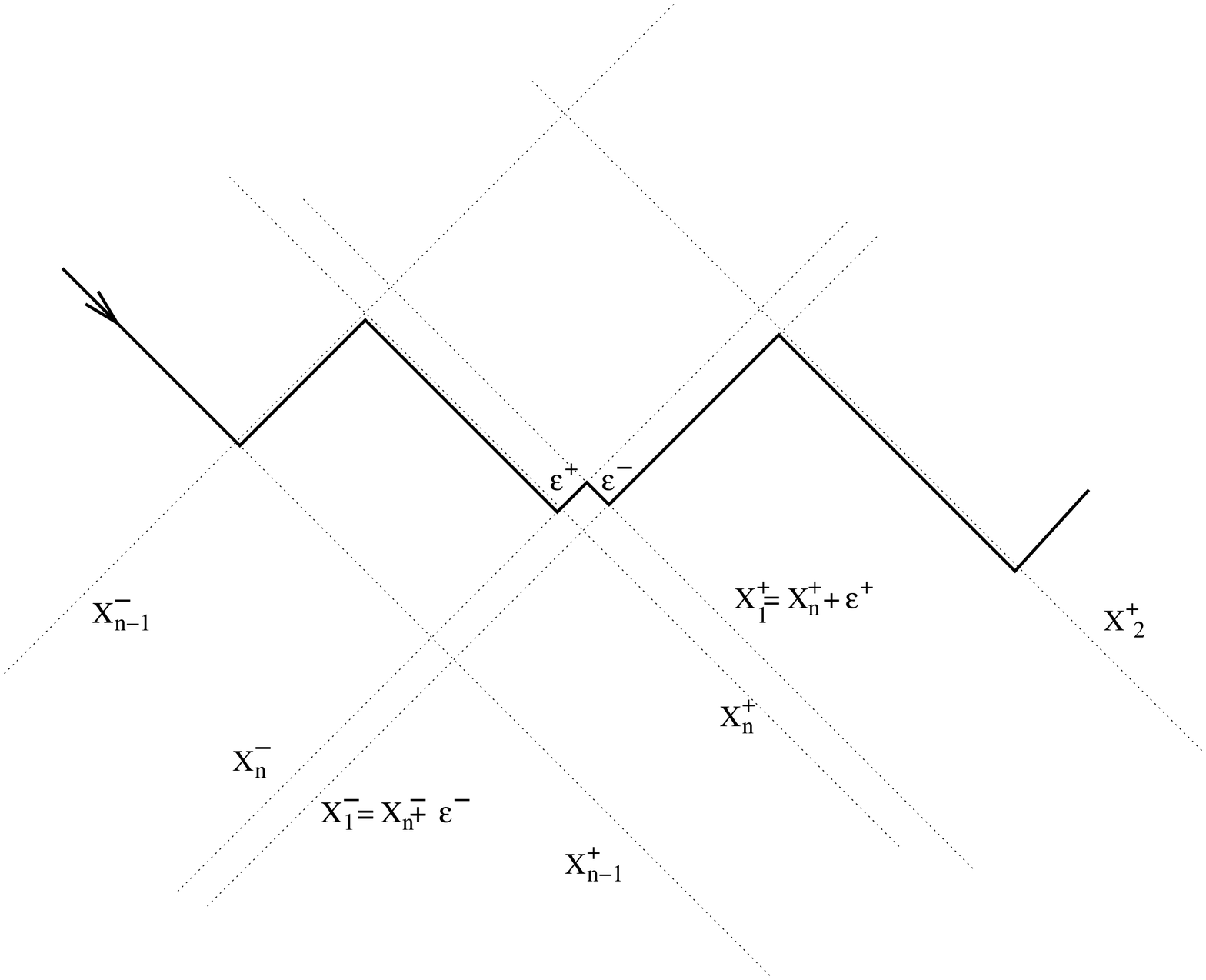}}

 \ifig\softcollinear{ Soft collinear limit where  only one segment is  going to zero, $\epsilon^+ \to 0$.
 In this limit the $x^-_n$ coordinate disappears from the limiting answer, up to the soft-collinear factor.
  }{\epsfxsize3in\epsfbox{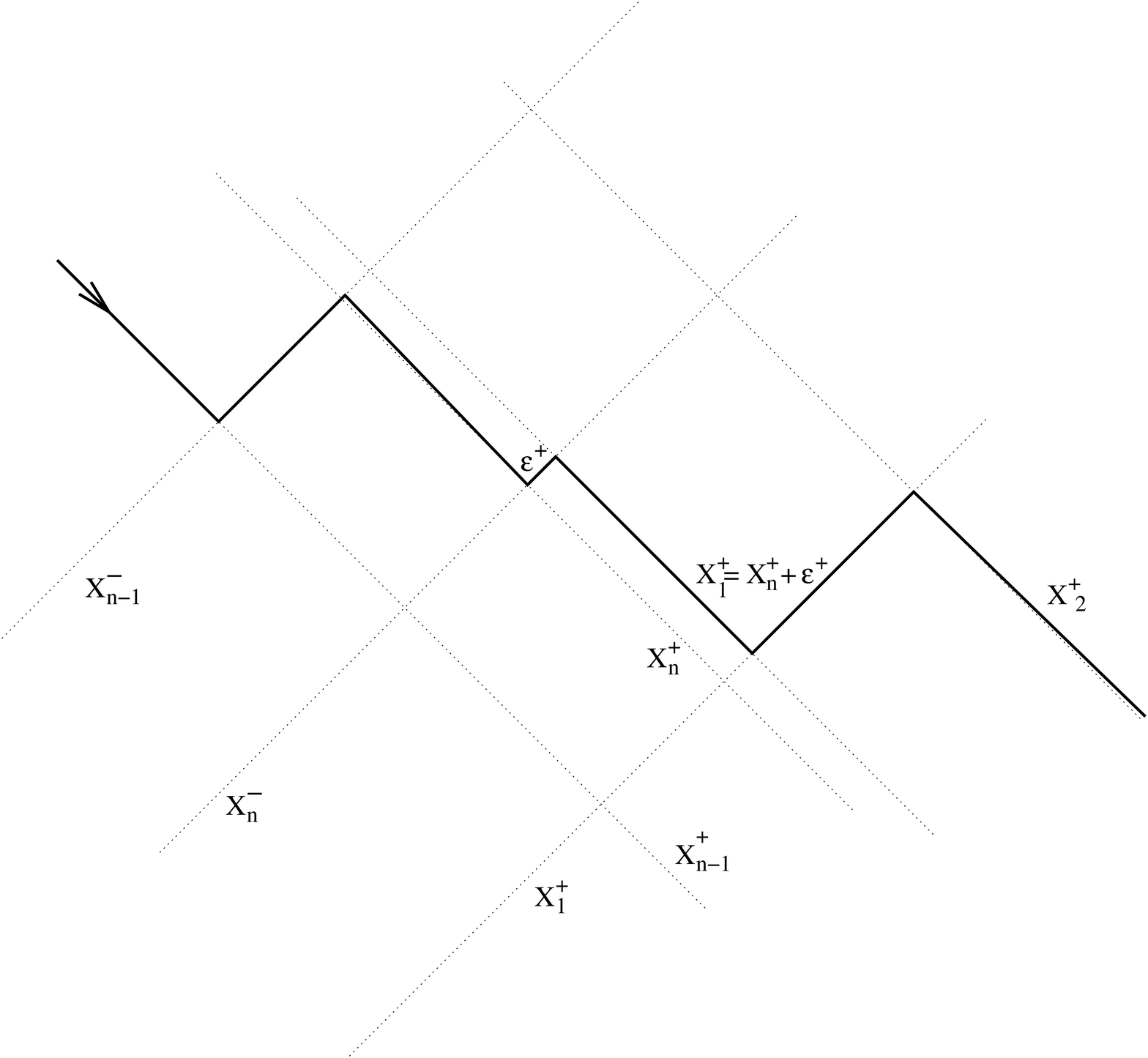}}

 Here we study the final answer as a function of $m$ for large $|m|$.
Let us  first recall the expression \singlc\ \eqn\crossrm{ \chi^+
= e^{ -  \pi i ( m - \bar m) } = e^{  2 \pi |m| \sin \phi }
~,~~~~~~~ \chi^- = e^{ - \pi (m + \bar m) } = e^{ -2 \pi |m| \cos
\phi } } The large $|m|$ region corresponds to a Wilson loop
degenerating in some way. It is convenient to divide the $m$ plane
into four quadrants.
 In the regions within the quadrants both $\chi^+$ and $\chi^-$ are going to some extreme values
 (zero or infinity) and they correspond to a situation where two consecutive segments are becoming very
 small, see
 \doublesoft .
 If we take $|m| \to \infty$ along the lines dividing
 the quadrants, then only one null segment is going to zero, see
 \softcollinear .
 This can be called a ``soft collinear" limit.

 For example, let us consider the first quadrant, where $\phi \in (0, \pi/2)$. In this region
 $\chi^+ \to + \infty$ and $\chi^- \to 0$.
Then the integral in \finalan\ can be approximated by the saddle point at $t=0$
 and we get terms
 that are exponentially suppressed as $e^{ - 2 \pi |m | }$ from $A_{Sinh} + A_{extra}$.
 We also get exponentially small terms from the first term in the remainder function $R$ in
 \finalan . Thus the full remainder function $R$ goes to a constant ($ 7\pi/6$)
  up to exponentially small
 terms.

The general expected behavior for the area in the double soft limit is the following.
  If we denote by $\epsilon^\pm$ the two separations that are becoming
 small, say $x^\pm_{1} - x^\pm_{n} = \epsilon^\pm$
 then in this limit the area factorizes as
 \eqn\amplif{
  A^{2 n}(x^\pm_1, \cdots , x^\pm_n)  \to  { 1 \over 4 } \log ( {\epsilon^+ \over   x^+_{1,2}} ) \log ({ \epsilon^- \over
  x^-_{n-1,1}} ) + A^{ 2n - 2 }(x^\pm_1, \cdots , x^\pm_{n-1} )
  }
   This behavior  is determined by conformal symmetry plus
  the assumption that only the coordinates explicitly appearing in
  \amplif\ could be involved in the soft factor. Of course, we
  find this behavior already in $A_{BDS}$, which is a solution of
  the conformal Ward identities. Thus the remainder function
  should go to zero (or a constant) in this limit and we see that
  it indeed does so.
    There can be a constant that we can add to the right
  hand side of
  \amplif\ which we are not keeping track of, since it is related to the precise way
  we regularize each of the cusp divergences. Once we choose a particular regularization, we
 can compute this constant. Thus, we should not assign a deep significance to
  the ${ 7 \pi/6}  $ in \finalan\ until we define precisely the way we subtract the cusps.
  This should not be too difficult to do at strong coupling, but we leave it for the future.

  Let us now discuss the case when $|m|$ goes to infinity along the neighborhood
  of $ \phi \sim { 3 \pi \over 2 }$.
  In this case $\chi^+ \to 0 $ and $\chi^-$ remains finite \crossrm\ .
  In general, such a limit could be   achieved
  by taking a small value of $x^+_1 -x^+_n$ while keeping $x^-_{1,n}$  finite.
   (In our case $n=4$, but
  we keep the discussion general).
  In this case dual conformal symmetry implies that
  \eqn\coppyap{\eqalign{A^{2n}  &\rightarrow
- { 1 \over 4 } \log \left( { \epsilon^+ \over x^+_{1,n-1} }
\right) \log (1 + {  z^-} )  - { 1 \over 4 } \log { \epsilon^+
\over x^+_{2,1} } \log (1 + { 1 \over z^-} ) + O(\epsilon^+) +
A^{2(n-1)} \cr &  ~~~~~~~~ z^-
  \equiv {  x_1^--x_{n}^- \over x_n^- - x_{n-1}^- } ,~~~~~~~~~~0<z^-
  < \infty
}} In the limit, the dependence of the area on $x_n^-$
disappears. Namely the area  $A^{2n -2}$ does not involve
$x_n^-$, the only explicit dependence on $x_n^-$ is through the
explicit appearance of $z^-$ in \coppyap .

  Again this behavior of the area  is saturated by $A_{BDS}$ so that the remainder function
  should go to zero (or a constant) in this limit. Indeed, the remainder function continues
  to go zero exponentially in this region too.  This discussion also implies that this should be
  a general feature of remainder functions for any number of particles in this
  limit.

  The conclusion is that the remainder function we obtained has the correct double-soft and
  soft-collinear limits. This behavior is correctly captured by
  $A_{BDS}$ and the remainder function goes to zero (or a
  constant) in both of these limits.

\subsec{ Wall crossing and the soft-collinear limit}


\ifig\wallcrossing{ The wall crossing phenomenon is related to a change in the coefficients of the expansion in
terms of $\epsilon^+$ when we change $x^-_n$. In (b) we see the middle range for $x^-_n$. In (a) $x_n^-$ is going
into an extreme limit where $x_n^- \to x_{n-1}^-$. In (c) we see the other extreme limit when $x_n^- \to x_1^- $.
  }{\epsfxsize3in\epsfbox{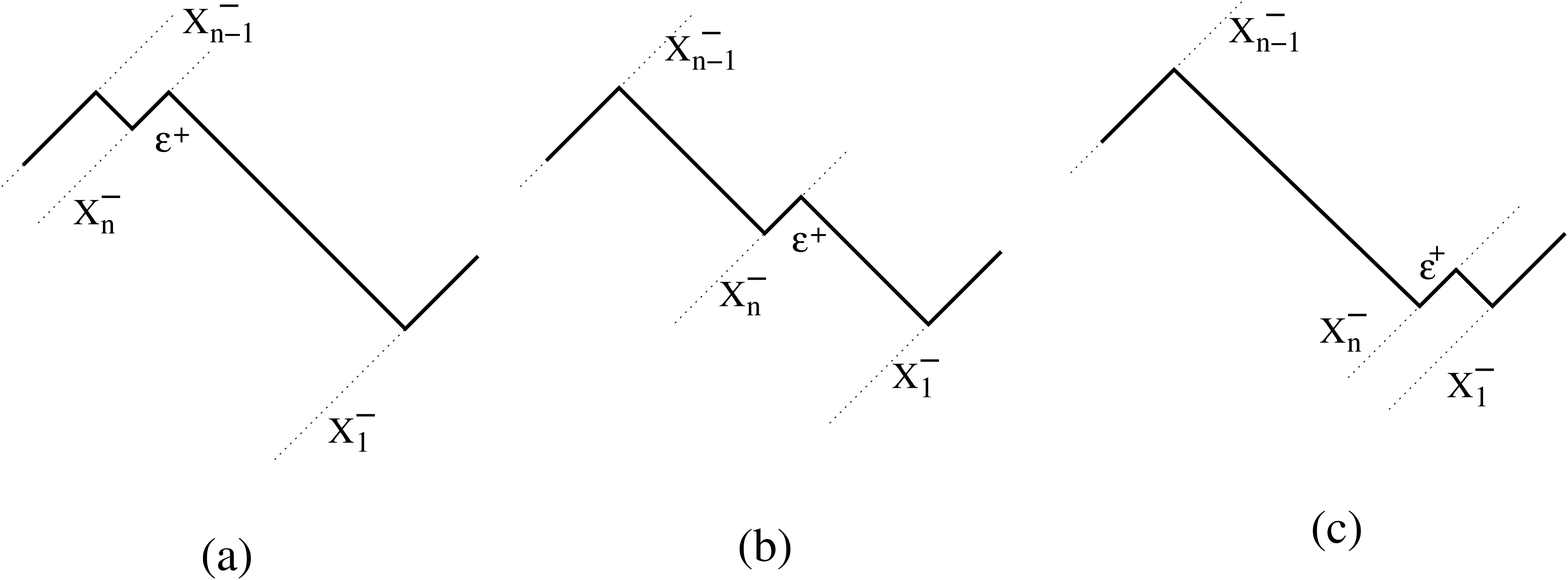}}

 In this subsection we discuss in more detail the soft collinear limit.
  We have
 already seen that this limit works correctly.  However, we want to make contact with the
 Wall crossing phenomenon discussed in \GaiottoCD .

 In this particular case the wall crossing phenomenon amounts to the fact that
 $\gamma $ in  \gammaone\ has different asymptotic limits as $|m | \to \infty $ in different
 quadrants. This is basically the same reason that $F( |m| , \phi)$ receives a pole contribution
 \equpo , this new pole contribution implies that $F$ has a slightly different expansion
 when we cross from one quadrant to the next.
 We have already seen that the
 full answer is actually periodic when we change $\phi \to \phi + \pi/2$.
 What happens is that the first term in \finalan\ also displays a Stokes phenomenon, which
 precisely cancels the one from the integral.
  It thus seems   that there is nothing new to discuss.

However, let us be a bit more explicit. The wall crossing
phenomenon is a statement about very small terms in the limit. In
our problem it is a statement about the higher order in
$\epsilon^+ $ terms in the expansion in \coppyap .
 More explicitly, we are expanding the small $\epsilon^+$ limit of the area  as
\eqn\resampl{ A^{2 n} = - { 1 \over 4 } \log \epsilon^+  \log { (1
+ z^-)^2 \over z^- }   +  {\rm finite} + \sum_{k=1}^\infty
c_k(z^-) ( \epsilon^+ )^k } where the finite piece contains also
the $A^{2n-2}$ area.
 One would normally throw away the higher powers of $\epsilon^+ $ in this expansion.
 However, the wall crossing phenomenon described in \GaiottoCD\ is contained precisely in
 these terms.

  The coefficients
 of this expansion depend on the minus location of the segment that is going to zero, which is
 $x^-_n$ with our choices, see figure \wallcrossing . The coefficients can
 also depends on the rest of the cross ratios. Note that the variable $x_n^-$
 disappears from $A^{2n - 2}$.
 Let us focus on the dependence of the coefficients
 on the minus location of the plus segment that is going to zero. Namely we  vary
 $x^-_n$ keeping everything else fixed. Note that changing $x^-_n$ between
 its two extreme values $x^-_{n-1} < x^-_n < x_1^-$ we have that $\log z^-$ changes between
 infinity  and minus infinity \coppyap .
 In principle this $x_n^-$ dependence could be complicated.
 The wall crossing phenomenon is the statement
 that  as we change $x^-_n$ from one extreme value to the other, so that $\log z^- \sim \log\chi^-$
 changes from
 minus to plus   infinity,
  then $c_k(q^-)$
 changes in a very specific way. Even though we have no information on the precise values of
 $c_k$, we do know how they change from a very small value of $\chi^-$ to a very large value of
 $\chi^-$. By a piece that changes, we mean a piece whose
 asymptotic value at large $\log \chi^-$ is not the analytic
 continuation of the behavior from very negative values of $\log
 \chi^- $.
  In our particular case we find that
  for very large $|\log \chi^-|$ we have the asymptotic
  behavior
 \eqn\totchag{
  \sum_{k=1}^\infty  c^{jumps}_k ~ { (\chi^+)^k } = -
   { 1 \over 4 }  |\log \chi^- | ~ \log ( 1 +  \chi^+   )
  ~,~~~~~~~~~~ - \log \chi^+ \gg |\log \chi^- | \gg 1
   }
  where we have expressed $\epsilon^+$ in terms of the $\chi^+$ cross
  ratio. Note that $\chi^+ \to 0$.
  This formula is only expressing the asymptotic behavior
  of the coefficients for large $|\log \chi^-|$ and it is
  capturing the change in the coefficients. We are not saying
  anything about terms that are possibly  linear in $\log \chi^-
  $, for example.

 What we have written here is the total change in the area. It turns out that the
 whole change in the coefficients
 comes from the BDS term  $A_{BDS}$. The  function $R$ does not display
 a  change of this kind.

 We interpret this as saying that the total change in the coefficients $c_k(q^-)$ as
 we change $\chi^-$ is the same at weak and strong coupling.
 This
  change is coming purely from $A_{BDS}$ and it would be nice to understand if it is being
  fixed by dual conformal symmetry.

  It turns out that   the detailed behavior of each of the coefficients $c_n$ in \resampl ,
  when we increase $\chi^-$, is actually different at weak and strong coupling. In fact the
  first term in \finalan\ is subtracting the weak coupling behavior and the integral is
  putting in the correct strong coupling behavior. This happens in such a way that
  the total change in the coefficients is the same at weak and strong coupling.
 In particular, at weak coupling the change in behavior occurs over
  a small  region of order one in $  \log \chi^-$. While at strong coupling
the change occurs over a larger region of size $ \Delta ( \log
\chi^- )
 \sim {  \sqrt{ \log \chi^+ } }$.
This corresponds to an angular scale $\Delta \phi \sim 1/|m|$ at
weak coupling while we have $\Delta \phi \sim 1/\sqrt{|m|}$ at
strong coupling. Thus the function is changing faster at weak
coupling than at strong coupling. This particular strong coupling
behavior can be understood by taking a scaling limit of the
integral in \finalan\ where we take $|m| \to \infty$ with $ \tilde
\phi = \sqrt{ |m|} \phi$ and $\tilde t = \sqrt{|m|} t$ kept fixed.

 Since the wall crossing phenomenon was
  crucial for \GaiottoCD ,  it would be interesting to know
  if these observations help in  the computation
 of the Wilson loops for all values of the coupling.
  In other words, one would like to generalize
 an  analysis analogous to the one in
 \GaiottoCD ,  for  all values of the coupling.

\newsec{Conclusions and discussion }

In this paper we have studied   classical strings on $AdS_3$ focusing on configurations that
end on the boundary on a polygon with null edges.
  We have explained how to map the problem into the
generalized Sinh-Gordon equation. This problem involves a holomorphic polynomial $p(z)$.
One then is supposed to solve the generalized Sinh-Gordon problem for each $p(z)$.
In order to find the spacetime embedding of the solution we need to solve an auxiliary linear problem.
This linear problem is everywhere smooth except at infinity, where it displays the Stokes phenomenon.
In each Stokes sector the spacetime solution goes to infinity. Thus, it goes to the boundary of
$AdS_3$. The degree of the polynomial determines the number of sectors and the thus the number of cusps.
  The
coefficients of this polynomial parametrize the configuration.  We explained how to
obtain the spacetime cross ratios in terms of the solutions to this linear problem.
 This map seems to
be complicated in general. We have also explained how to regularize the area, splitting the
computation into several parts. In general,  the difficult part of the problem is to compute the
spacetime cross ratios as a function of the parameters of the polynomial, $p$. If these
cross ratios are also computed as a function of the spectral parameter, then one can also find
the area in a simple way.   The problem is mathematically
equivalent to the one considered in \refs{\GaiottoCD,\GMNtwo}, based on a
different physical motivation. Using
the results obtained in \GaiottoCD\ we could find the full result for an octagonal  Wilson loop.
The final result is given in \finalan . In addition we have considered the solutions for
regular polygons and we have also checked that in the limit of a large number of sides, $n\to \infty $,
we recover the result for the ordinary circular Wilson loop.

The structure of the problem and its integrability  was explored in
detail
in \GaiottoCD\ and \GMNtwo .
 In particular,  Gaiotto-Moore-Neitzke reduce the problem to a certain Riemann Hilbert
problem involving the spacetime cross ratios as a function of the spectral parameter.
One could wonder if there are further simplifications
for our problem and whether one can find other solutions.

Given the close connection between the problem considered in \GaiottoCD\ and the problem
considered here it seems likely that there might be a way to relate them directly by
a physical construction.

The problem considered in  \refs{\GaiottoCD,\GMNtwo}
 is more general than the one considered in this paper.
In our case the linear problem had a single irregular singularity at infinity. One could add
other singularities \GMNtwo . It is possible that the more general problem in \GMNtwo\ could enable
one to find solutions also for correlation functions of operators, or even perhaps, problems
involving higher genus surfaces in the classical limit. It would be nice to do this explicitly.

In \GMNtwo\ the problem is solved in an approximate way when the zeros of the polynomial are
widely separated. In this case, the periods give a good approximation for certain cross ratios.
It is likely that in this regime the remainder function will be small. We saw this in the
case of the octagon. It would be nice to see if it is a general feature.

 It would be nice to study further the equations in \GaiottoCD\ so as to reformulate
 the problem in terms of Bethe equations for some particles moving on the worldsheet.
 In this problem there is a very clear candidate particle, which is the massive excitation
 that the field $\alpha$ is describing. In fact, this particle has a clear spacetime meaning
 which can be elucidated by considering the four-sided null polygon. The corresponding surface and
 configuration were studied in detail in \refs{\KruczenskiGT,\AldayMF}.
 One could consider a new vacuum for the
 spin chain that corresponds to the state of a high spin operator. Then it seems natural to
 consider impurities propagating on this new vacuum, as discussed briefly in \FreyhultPZ .
 This is the vacuum that might lead to the simplest description for
  Wilson loops, or gauge theory amplitudes. It  seems likely that this $AdS_3$ problem
 is a full consistent subsector of the ${\cal N}=4$ gauge theory.
 If one understood how to rephrase \GaiottoCD\ in terms of Bethe equations, then one could
 derive the solution for the full problem by inserting the correct quantum scattering phase for
 the corresponding excitations, such as the phase in  \BeisertEZ .

One of the motivations for the analysis in \GaiottoCD\ was to understand further the wall crossing
phenomenon. In our case, this phenomenon arises when we consider a soft-collinear limit where one
null segment becomes very small. As we change the location of the small segment
from one cusp to the next,    there is
a change in the coefficients of the expansion in terms of the parameter that is becoming very small.
  We have found that this change has the same form
  at weak
and strong coupling, up to an overall  factor due to the cusp anomalous dimension (see \wilexpec ).
This suggests that it should be the same at all values of the coupling. It would be nice to
find a direct argument for this fact. Notice that this involves subleading terms in the
soft-collinear  expansion, these are terms that one would normally throw away in computing the
 soft-collinear limit\foot{ What we call soft-collinear here corresponds to taking the size of only
 one null segment to zero, keeping everything else fixed. }.

It is interesting that the problem with spectral parameter $\zeta$ also arises as the physical
problem on
  $AdS_3$ with a WZ term \LuKB .
If the WZ term is very small, then $\zeta$ is close to one.
But if the WZ term approaches the WZW limit then we find that $\zeta$ goes to zero or infinity,
depending on the sign of the WZ term . Since these limits also play an important role in setting
boundary conditions for the Riemann Hilbert problem in
\GaiottoCD , perhaps the WZW version of this problem, which should be more solvable, could
be particularly useful.

One obvious problem is to generalize this structure to the case of a surface in $AdS_5$ (or $AdS_4$),
as opposed to $AdS_3$. The Pohlmeyer-type reduction is know for these cases
\refs{\DeVegaXC,\MiramontesWT}.  In all cases
there is a single holomorphic function, but we get three fields instead of one field $\alpha$
as we had here.

So far we discussed Wilson loops. If one is interested in amplitudes, there is
  extra information
that has to do with the particle polarizations at the boundary of the worldsheet. This information
is not visible in the classical problem we have been considering.

 The particular kinematic configurations that we have considered here are also expected to lead
 to simplifications in Wilson loops or scattering amplitudes at weak coupling. This looks like
 the kinematics of a 1+1 dimensional theory, but the theory is still four dimensional. The
 two loop perturbative result for an $n$ sided Wilson null polygon was computed in \AnastasiouKN ,
 it would be nice to see if the result can be made more explicit with this particular kinematics.

 It would also be interesting to know if there is any relation
 between the solutions in this paper  and the multispike solutions in
 \refs{\KruczenskiWG,\DoreyVP} and various references therein. The
 solution for $n=2$ (four cusps) is an analytic continuation of
 the solution with two spikes. However, this does seem to be the
 case for other solutions ($n >2$).

{ \bf Acknowledgments }

We are very grateful to Davide Gaiotto, for a large number of important suggestions,
for his detailed explanations of \GaiottoCD\ and for giving us a draft of \GMNtwo .
We also thank N. Arkani-Hamed, D. Gross, V. Kazakov and  P. Vieira  for discussions.

This work   was  supported in part by U.S.~Department of Energy
grant \#DE-FG02-90ER40542.



\appendix{A}{ Relation between the flat connection of the reduced model and the one of the
original model}

In this appendix we consider the relation between the flat connection with spectral parameter
$\zeta$ introduced in \Bgeneric\ and the one that is more familiar from the study strings in $AdS$,
such as the one considered in \KazakovQF , whose expansion in powers of the spectral parameter gives the
infinite number of conserved charges.
  In order to understand the relation one can  start from an explicit solution for $Y$ in \inversemap\ and construct the usual current from it $J = - d G G^{-1} $, where we think of
$Y_{a \dot a} = G_{a \dot a} $ as an $SL(2)$ group element.
It turns out that $J$ can be obtained from a gauge transformation of $B$. Since both are
flat connections, this is not surprising. Our objective here is to find the explicit
transformation between the two. For this purpose it is convenient to think of
$\psi_{\alpha a}$ as an $SL(2)$ matrix. Its determinant is one because of \normalization .
Moreover $\psi^{a \alpha} = \epsilon^{\alpha \beta} \epsilon^{a b} \psi_{\beta b}$ is the
inverse of $\psi_{\alpha a}$. Namely, \normalization\ is equivalent to
 $\psi^{a \alpha} \psi_{\alpha b} = \delta^a_b $. The expressions for the group element and
 its derivatives can now be written as
 \eqn\groupelm{
 \eqalign{
 G_{a \dot a }= Y_{a \dot a}  =&  q_1 = \psi^L_{\alpha a} M_1^{\alpha \dot \beta}
 \psi^R_{\dot \beta , \dot a } , = (\psi^L)^t  M_1 \psi^R ~~~~~
 G^{-1} = (\psi^R)^{-1} M_1^{-1} [ (\psi^L)^t]^{-1}
 \cr
 \partial G = & 2 e^{\alpha} q_3 = 2  e^{\alpha}
 (\psi^L)^t M_3 \psi^R ~,~~~~~~ M_3 = \pmatrix{ 0 & 0 \cr 1& 0 } ~,~~~~~M_1 =  \pmatrix{ 1 & 0 \cr 0 &1 }
 \cr
 \bar \partial G = & 2 e^{\alpha } q_2 =2 e^{\alpha } (\psi^L)^t M_2 \psi^R ~,~~~~~~ M_2 = \pmatrix{ 0 & 1\cr 0 & 0 }
}}
%
%
 %
 %
%
Using these expressions we can compute the left currents
\eqn\leftcu{ \eqalign{
(J_z)_a^{~b} = &  - \partial G G^{-1} = -  2 e^{\alpha} (\psi^L)^t M_3 M_1^{-1} [(\psi^L)^t ]^{-1} =
 2  e^{\alpha}
\psi^L_{2 a} \psi_{2}^{L ~b}
\cr
(J_{\bar z})_a^{~b}  = &  - \bar \partial G G^{-1} = - 2 e^{\alpha } (\psi^L)^t M_2 M_1^{-1} [(\psi^L)^t ]^{-1} =  - 2  e^{\alpha}
\psi^L_{1 a} \psi_{1}^{L~b}
}}
where I have used the expressions for $M_1$ and $M_{2,3}$ in a particular gauge. Notice that
$\psi^R$ cancels out.
  \leftcu\ is the ordinary left flat connection of the $AdS_3$ sigma model.  Notice that it depends only on ``left" spinors, $\psi^L$. It is well known that one can consider a one parameter of flat connections, or a Lax pair, by introducing a spectral parameter
\eqn\Laxorig{L_z=c_z J_z,~~~~~L_{\bar{z}}=c_{\bar z}  J_{\bar z},~~~~2 c_z c_{\bar z} - c_z  - c_{\bar z} =0}

Let us now turn our attention to the connection with spectral parameter introduced in \Bgeneric\ .
We then perform a gauge transformation
\eqn\gaugetransf{ \eqalign{
B(\zeta) &  \rightarrow \tilde  B(\zeta)=h^{-1}B(\zeta)h+h^{-1}d h~,~~~~~~~~~~ \psi^\zeta \to \tilde \psi^\zeta = h^{-1} \psi^\zeta ~,
\cr h=& h_\zeta \Omega ~,~~~~~~~~~~ h_\zeta = \left(\matrix{\zeta^{1/2} &
 0 \cr 0 & \zeta^{-1/2} }\right)
 \cr \Omega_\alpha^{~ \, a}=&
  (\psi^L)_{\alpha}^{~a} = \pmatrix{ - \psi^L_{1,2} & \psi^L_{1,1} \cr
  - \psi^L_{2,2} & \psi^L_{2,1} }
    ~,~~~~~~~~~~ d \Omega=-B(\zeta = 1) \Omega
 }}
 $\Omega$ is simply given by the solution of the left problem. Note that $\Omega$ is independent
 of $\zeta$. Of course, for $\zeta =1$ such a gauge
 transformation sets  the connection to zero.  $h_\zeta $
 is chosen so that $h_\zeta^{-1} B_z h_\zeta $  depends on $\zeta$
  only through a $\zeta^2$ factor in the upper
 right hand corner. Thus, the $\zeta$ dependent term in $h^{-1}_\zeta B(\zeta) h_\zeta $
    is  proportional to  the matrix  $M_3$ which
 appeared in the definition of $J_z$ \leftcu . In addition,  note that for $\zeta =1$ the
 gauge transformation $\Omega$ sets the
 connection to zero. Thus, after the gauge transformation,
  the full connection is
 proportional to $\zeta^2 -1$.
 A similar thing happens with $B_{\bar z}$.
 Thus we have that
 \eqn\transf{ \eqalign{
 \tilde B_z(\zeta) =&  - { (\zeta^{-2} -1  ) } e^{\alpha }
   \psi_{ \, a}^{L~ \alpha} \pmatrix{ 0 & 1 \cr 0 & 0 }_{\alpha}^{~ \beta} \psi_{~\beta}^{L~\, b} =
   - { (\zeta^{-2} -1 ) }e^{\alpha}  \psi^L_{\,a \alpha} \pmatrix{ 0 & 0
\cr 1 & 0 }^\alpha_{~\beta}  \psi^{L \, \beta b}
\cr
 \tilde B_z(\zeta) =&  { \zeta^{-2} -1 \over 2  } J_z
\cr
\tilde B_{\bar z} (\zeta) =&  - { (\zeta^{2} -1 ) } e^{\alpha}
\psi_a^{L~ \alpha} \pmatrix{ 0 &0 \cr 1 & 0 }_{\alpha}^{~ \beta} \psi_{\beta}^{L~ b} =
- { (\zeta^{2} -1 ) }e^{\alpha}  \psi^L_{\, a \alpha} \pmatrix{ 0 & 1
\cr 0 & 0 }^\alpha_{~\beta}  \psi^{L \, \beta b}
\cr
\tilde B_{\bar z} (\zeta) =& { \zeta^{2} -1 \over 2 } J_z
}}
Which has exactly the form \Laxorig\ . Hence, the spectral
 parameters in the reduced and original models coincide.
  Notice that in \transf\ the ``poles" are at
  $\zeta \rightarrow 0$ and $\zeta \rightarrow \infty$.
  A more standard choice is $c_z=-{1 \over 1+\lambda}$ and
  $c_{\bar z}=-{1 \over 1-\lambda}$, where $\lambda={ \zeta^2 +1 \over \zeta^2 -1 }$,
  the poles are then at $\lambda=\pm 1$   \KazakovQF .
Thus we see that a gauge transformation by an $h$ given by a solution to the left problem
of the left connection of the reduced problem gives us the usual left connection of the
sigma model.

 The same discussion carries over for the right connection given by
 $ J \sim G^{-1} d G$. The relation between the left and the right connections is simply a
 gauge transformation by the group element $G$.

\appendix{B}{Solution for large $|w|$ and regularization of the area}

 In order to regularize the area it is necessary to understand the solution of the linear problem
 at large $z$ or $w$. In this appendix we examine this problem in more detail and we provide a derivation for
 the formulas quoted in the main text.

 Let us first discuss some general issues regarding the solutions at large $w$. Let us consider
   the left problem first. The problem involves Stokes sectors.
 We denote the Stoke lines by the index $i=1,\cdots, n$. The first line is when
 $w$ is real and positive, the second when $w$ is real and negative, the third when $w$ is real and positive but
 on the second sheet, and so on. We move from one sheet to the next in an anticlockwise fashion.
  We  label Stokes sectors by two
 consecutive indices $[i,i+1]$.
 In each Stokes sector a simple basis for two approximate solutions is
 \eqn\simples{
  \eta_{+ \, [i,i+1]} = \pmatrix{ e^{   w + \bar w  } \cr 0 } \sim  \pmatrix{ e^{
   \int \sqrt{p} dz + \int   \sqrt{ \bar p } d\bar z   } \cr 0 }~,~~~~~~
  \eta_{- \, [i,i+1]} =
 \pmatrix{ 0 \cr e^{ - (w + \bar w) } }
 }
 In different sectors we define similar  functions which are given by the same expressions
 as in \simples , but with
  $w$ is defined by analytic continuation from the previous sector. In order to define the variable
  $w$ we just perform an
  analytic continuation  through all the sheets obtained using the polynomial $p$, without any
  information of the Sinh-Gordon problem yet.
 We will sometimes suppress the sector index.

 We now imagine that we have
  solved the Sinh-Gordon problem. In addition, imagine that we have also solved
   the linear problem   \LS .
 An exact solution can be approximated as
 \eqn\exacs{
  \psi  = c_{[i,i+1]}^+ \eta_{+,[i,i+1]} + c^-_{[i,i+1]} \eta_{-,[i,i+1]}
  }
  The numbers $c^\pm$ depend on the Stokes sector.
  As we cross the $i+1$ Stokes line we have that
  \eqn\jumps{
   \eta_{a \, [i,i+1]}^{\rm exact} = S_{a}^{~b} \eta_{b , [i+1,i+2]}^{\rm exact}
   }
   Where the left hand side is the exact analytic continuation of the solutions that have the approximate
   expressions \simples\ in the stokes sector $[i,i+1]$, while the right hand side
   involves the exact solutions which have the expressions \simples\ in the $[i+1,i+2]$ sector.
    The indices $a,b$ run over $\pm$ in \simples .
   This implies that the numbers $c^a$ in \exacs\ change as
   \eqn\changec{
   c_{[i+1,i+2]}^b = c^a_{[i,i+1]} S_{a}^{~b}
   }
   as we change sectors.
   For a Stokes line along the positive real axis we have $S_p = \pmatrix{ 1 & \stokesparameter \cr 0& 1} $. While
   for one along the negative real axis we have $S_n = \pmatrix{ 1 & 0 \cr \stokesparameter & 1 }$, where
   $\stokesparameter$ is generically different for each line.

   Let us now understand what should happen when we go
   around once in the $z$ plane (or $n/2$ times in the $w$
   plane). Naively, we would expect the solution to go back to the original solution.
   This is not quite right because when we went to the variables where the connection was constant at
   infinity we performed a gauge transformation
    \linpro\
   \eqn\fullg{
   \hat \psi =e^{ i { \pi \over 4 } \sigma^3 }
    e^{i { \pi \over 4 } \sigma^2 } e^{ { 1 \over 8 } \log { p \over \bar p} \sigma^3 }
    \psi
   }
   where $\psi$ is the original variable of the problem in
    \FGhybrid .
   This same gauge transformation diagonalizes the connection for any value of $\zeta$, so that
   this also covers the right problem after we do the transformation
   \leftrightrel .
   Since the gauge transformation \fullg\ is not single valued as we go  once around the
   $z$ plane, we find that $\hat \psi$ is not single valued. Of course, the original
   variable $\psi$ is single valued, since the connection is smooth everywhere.
   We find that under a full rotation in the $z$ plane $\hat \psi$ changes as
   \eqn\changetps{
   \hat \psi \to e^{i { \pi \over 4 } \sigma^3}
    e^{i { \pi \over 4 } \sigma^2 } e^{ { 1 \over 8 } 4 \pi i (n-2)  \sigma^3 }
   e^{- i { \pi \over 4 } \sigma^2 } e^{ - i { \pi \over 4 } \sigma^3 }
   \hat \psi  = e^{  i { \pi \over 2} (n-2) \sigma^2 }
   \hat \psi
   }
    In addition, in the case that $n$ is even we have the shift in the $w$ coordinate as
    we go around $w \to w + \wshifted$. In other words, $w_{n+1} = w_1 + \wshifted$ where $w_1$ is the
    $w$ coordinate on the first sheet in the first region and $w_{n+1}$ is the result of
    going around $n/2$ times in the $w$ plane (or once in the $z$ plane).
     This implies that after we go around $n/2$ times we should
    identify the original spinor $\psi(w_1) $ in the first patch with
    $\psi^{exact}(w_{n+1}) = \psi^{exact}(w_1 + \wshifted)$ in the last
    patch.  In other words, we require
    \eqn\condit{
    \hat \psi^{exact}(w_{n+1}) = \hat \psi^{exact}(w_1+\wshifted) =
    e^{ i { \pi \over 2 } (n-2) \sigma^2} \hat \psi^{exact}(w_1)
    }
    where the left hand side is the exact analytic continuation of the solution through all
    the sheets of the Riemann surface.
   Taking both of these effects
   into account the we find that the product of all Stokes factors should obey
\eqn\stokesident{ \eqalign{ &
   S_p(\stokesparameter_1)S_n (\stokesparameter_2) S_p(\stokesparameter_3) \cdots
   S_p(\stokesparameter_n)   = e^{  i { \pi \over 2} \sigma_2 ( n-2) } = i (-1)^{ n-3 \over 2 } \sigma^2
 ~~~~~~n~~{\rm odd}
 \cr
 & S_p(\stokesparameter_1)S_n(\stokesparameter_2)\cdots
  S_n(\stokesparameter_n) e^{   \sigma^3  (  \wshifted + \bar \wshifted )  }    =
   e^{  i { \pi \over 2} \sigma_2 ( n-2) }
  = - (-1)^{ n/2   } ~~~~~~n~~{\rm even}
 }}
 where $\sigma^i$ are the ordinary Pauli matrices. In the first equation, the one for $n$ odd,
  we should choose the origin of the $w$ plane so that $w \to -w$ when we go around at large $|w|$
   $n/2$ times.

 We can check that in the case that $n=3$ this determines all the Stokes matrices.
 For $n=3$ we get $\stokesparameter_1 = - \stokesparameter_2 = \stokesparameter_3 =1$.
  In the case $n=4$ there is
 a one parameter family of solutions (for each $\wshifted$)
 \eqn\soluts{
 \stokesparameter_2 = -(1 + e^{- (\wshifted + \bar \wshifted) } )/\stokesparameter_1 ~,~~~~~\stokesparameter_3 = e^{ \wshifted + \bar \wshifted } \stokesparameter_1
 ~,~~~~\stokesparameter_4 = - e^{ -(\wshifted + \bar \wshifted) }   (1 + e^{- (\wshifted + \bar \wshifted) } )/\stokesparameter_1
 }
 The left over parameter  simply corresponds to shifts in the origin of $w$ which
 leads to a rescaling of the parameters
  and can be viewed as a spacetime conformal transformation.
  On the
 other hand we will see that $\wshifted$ parametrizes the  single non-trivial cross ratio for this problem
 $(n=4)$.
  For general even $n$
 there is one cross ratio that we can compute in this fashion.
  This is computed as follows. First note that we can use the approximate solutions \simples\ to
  compute the inner products $s_i \wedge s_{i+1}$ namely the product of two small solutions on two
  consecutive cusps. We can take the normalization of these solutions to be the one in \simples\ in
  sector $[i,i+1]$. Then we would get $s_i \wedge s_{i+1} = \pm 1$ depending on whether the sector is
  in the region $Im(w) >0$ or $Im(w) < 0$ (if we consider the left problem). The only subtlety
  is in the computation of $s_n \wedge s_1 $. In this case we need to use \condit\ so that
  $s_n \wedge s_1 = - (-1)^{n/2} e^{ w_s + \bar w_s } $. We also know from \formdiff\ that
  $s_i \wedge s_{i+1} \sim x_{ij} $ up to factors that depend only on $i$ and a factor depending
   only on $j$. These factors cancel in the expression
   \eqn\crossra{
   { ~~~~~~~~(s_2 \wedge s_3 ) (s_4 \wedge s_5 )  \cdots  (s_n \wedge s_1 )
   \over (s_1 \wedge s_2 )  (s_3 \wedge s_4 ) \cdots (s_{n-1} \wedge s_n ) } = - e^{ w_s + \bar w_s } =
   { ~x_{23}^+ x^+_{45 } \cdots x^+_{n1}
   \over x^+_{12} x^+_{34} \cdots x^+_{n-1,n} }
   }
   where the $(-1)^{n/2}$ factor got canceled by
   the fact that for half of the sectors $s_i \wedge s_{i+1}$
   has one sign and for the other half the other sign.
   This is the derivation of \crosssimp .

 Finally, let us note that we can compute the $\stokesparameter_i$
 for the regular  polygon from \stokesident . For a regular
 polygon we expect that all the $\stokesparameter_i$ are equal. In fact,
 being more careful about the minus signs we get $\gamma_i =
 (-1)^i \stokesparameter$. One can then determine  where $\stokesparameter
  = 2 \cos { \pi \over n } $.

   Due to the  gauge transformation \fullg\
   the final expression for the spacetime coordinates, which
   is not gauge invariant, is
   \eqn\finalexpr{ \eqalign{
   Y_{a \dot a } = & \hat \psi^L_{\alpha a } \hat M_1^{\alpha \dot \beta}  \hat
    \psi^R_{\dot \beta \dot a }
   ~,~~~~~~
   \cr \hat M_1 = &
   ( e^{ - { 1 \over 8 } \log { p \over \bar p} \sigma^3 }
    e^{-i { \pi \over 4 } \sigma^2 } e^{-i { \pi \over 4 } \sigma^3 } )^t U
     e^{ - { 1 \over 8 } \log { p \over \bar p} \sigma^3 } e^{-i { \pi \over 4 } \sigma^2}
      e^{-i { \pi \over 4 } \sigma^3 }   =
     { 1 \over \sqrt{2}} \pmatrix{ 1 &  1 \cr -1 & 1 }
 }}
   where $U$ is given in
    \leftrightrel
   and $\hat \psi^{L,R}$ are the solutions to the left and right problems where the
   connection has been diagonalized for large $w$ as in \linpro .
   In \finalexpr\ we have  expressed
    \inversemap\
    in the gauge that simplifies the asymptotic form of the solutions.
    It is convenient to introduce the
    new coordinates
   \eqn\newcoord{
   u = { w + \bar w  }  + { w - \bar w \over i  } ~,~~~~
   v = - ( w + \bar w   )  + { w - \bar w \over i  }
   }
   Then,    in a region where the
   pair of   left and right solutions have the expressions
   \eqn\leftrigh{
   \psi^L_a = c^{L,+}_a \eta^L_+ + c^{L,-}_a \eta^L_- ~,~~~~~~~
   \psi^R_{\dot a}  = c^{R,+}_{\dot a} \eta^R_+ + c^{R,-}_{\dot a} \eta^R_-
   }
   we find that the spacetime coordinates,  \finalexpr , is
   \eqn\spacetimec{ \eqalign{
   Y_{a \dot a } = & { 1 \over \sqrt{2} } \left[ c^{L,+}_a  c^{R,+}_{\dot a}
   e^{ u }
   + c^{L,-}_a  c^{R,-}_{\dot a}
   e^{ -u  } -
   c^{L,-}_a  c^{R,+}_{\dot a}
   e^{ v} +
    c^{L,+}_a  c^{R,-}_{\dot a}
   e^{ - v }
   \right]
   }}
   Note that in each of the quadrants of the $w$ plane only one of these terms  dominates and
    determines the spacetime coordinates.  In particular,
   for the solution with $n=2$, which corresponds to the four sided Wilson loop
   discussed in  \refs{\AldayHR,\KruczenskiGT},
    the Stokes matrices are the identity   this is the exact solution. A simple choice
   for the $c$'s leads to
   \eqn\solnfour{
   Y_{a \dot a } = { 1 \over \sqrt{2} } \pmatrix{ e^u & e^{-v} \cr -e^v & e^{-u} }
   }
   But in general, for $n>2$,  \spacetimec\ it is only giving the asymptotic
   form of the solution. The matrices $c^{L,\pm}_a$ and $c^{R \pm}_{\dot a}$ change as in
   \changec\ when we
   cross Stokes lines. Also the various exponentials in \spacetimec\ change
   dominance as we cross anti-Stokes lines.

    Finally, let us mention that the area element in the new coordinates \newcoord\
  is  $4 \int d^2 w = { 1 \over 2 }  \int du dv $.

  \subsec{ Tracking a solution across Stokes lines }

  In this subsection we will track the solutions across Stokes lines.
  Our purpose in doing this is to see more clearly how the Stokes data determines
  the spacetime solution, the location of the cusps, and the asymptotic form  for the
  radial coordinate.

  \ifig\stokeswplane{ For the left problem (a), $Re(w)=0$ represents the anti-Stokes lines while the Stokes lines are at $Im(w)=0$. For instance, for $Re(w)<0$ a given solution dominates and  the whole region corresponds to a single value of $x^+$. For the right problem (b) the Stokes and anti-Stokes lines are interchanged.
 } {\epsfxsize2.7in\epsfbox{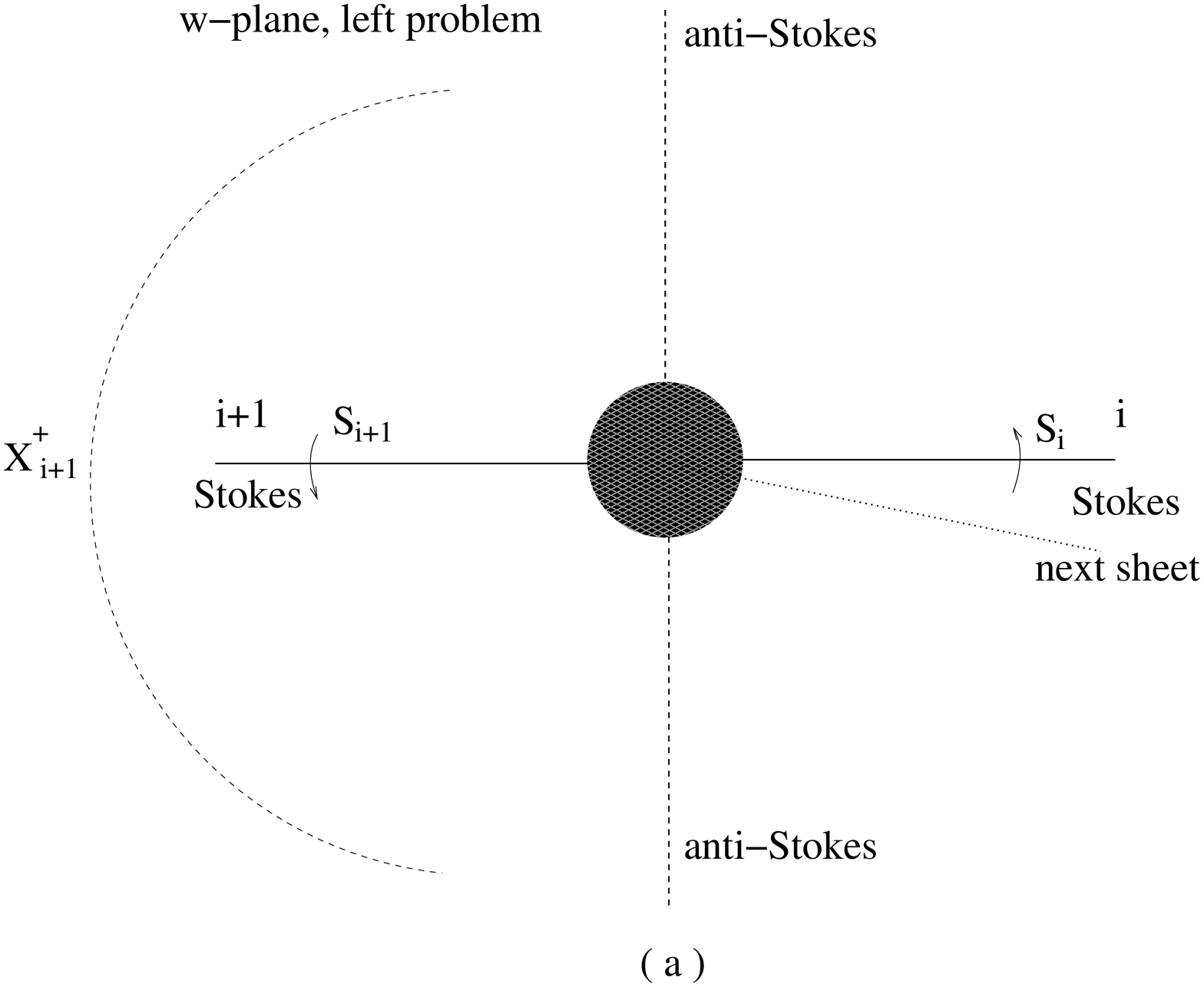} \epsfxsize2.7in\epsfbox{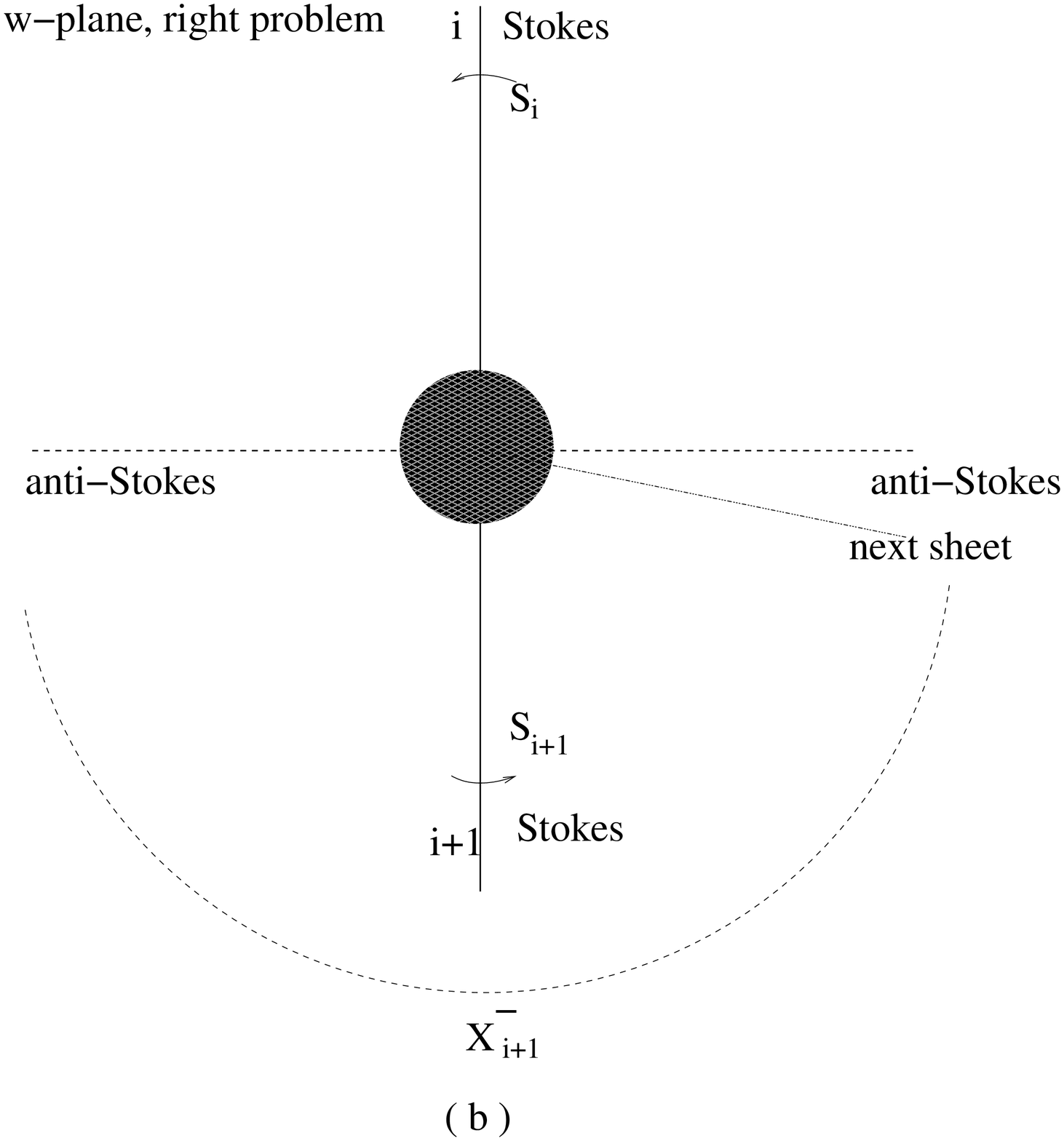}}

 Let us now track the behavior of a pair of solutions around various Stokes sectors.
  Let us
  start in the first Stokes sector with the matrix
   given by $c^{L ~ r}_{[12],a} = b^r_{  a } $, obeying
   \eqn\normcond{
   b^+_a b^-_b - b^-_a b^+_b= \epsilon_{ab}
   }
   Then the coefficients in the next four
   Stokes sectors can be found using \changec
   \eqn\coeffisec{ \eqalign{
  [12]: & ~~~~ c^{~+}_{[12],a} =  b^+_a ~,~~~~~~~~~~~~~~~~~~c^-_{[12],a} = b^-_a
  \cr
  [23]:& ~~~~ c^{~+}_{[23],a} = b^+_a + \stokesparameter_2 b^-_a  ~,~~~~~~~~~c^-_{[23],a} = b^-_a
 \cr
 [34]: & ~~~~ c^{~+}_{[34],a} =   b^+_a + \stokesparameter_2 b^-_a ~,~~~~~~~~~~~~c^-_{[34],a} = b^-_a + \stokesparameter_3 (  b^+_a + \stokesparameter_2 b^-_a)
    ~,~~~~~~~
    \cr
  [45]: & ~~~~ c^{~+}_{[45],a} =   b^+_a + \stokesparameter_2 b^-_a  + \stokesparameter_4[ b^-_a + \stokesparameter_3 (  b^+_a + \stokesparameter_2 b^-_a) ]
    ~,~~~~c^-_{[45],a} = b^-_a + \stokesparameter_3 (  b^+_a + \stokesparameter_2 b^-_a)
   }}
   Notice that the coefficients of the large terms never change when we cross the line.
    Only the small coefficients change.
   However, by the time we go once around the $w$ plane, we see that the Stokes data starts
   to show up in the large solutions. Namely, this is the $\stokesparameter_2$ term in $c_{[23]a}^+$
   when we are near the 3rd Stokes line.
    We can now do the same for the right problem. We label the corresponding quantities with
    tildes.
   We can now write down the asymptotic form of the solution
   in four consecutive cusps. We label the cusps by two indices, $(i,i)$ or $(i+1,i)$.
   The first index labels the left problem and the second index labels the right problem.
   Thus, these cusps sit at $(x^+_i, x^-_i)$ and $(x_{i+1}^+,x_i^-)$ respectively.
   The expressions for the $c$'s for the left and right problems will have the same form as in
   \coeffisec\ with the Stokes parameters for the left or the right problems.
    Each
   of the indices changes when we cross an anti-Stokes line. When we cross a Stokes line we
   do not change the leading asymptotic form of the solution so we have the same index $i$.
   We have that the solution in the following cusps behaves as
   \eqn\cuspsol{ \eqalign{
   (i, i):&~~~~ { 1 \over r} = b_1^+ \tilde b_1^+ e^u ~,~~~~~~~~ x^+_i = {b_2^+ \over b_1^+ } ~,~~~~~
x^-_i = { \tilde b^+_2 \over \tilde b^+_1  }
\cr
(i+1, i):&~~~~ { 1 \over r} =- b_1^- \tilde b_1^+ e^v ~,~~~~~~~~ x^+_{i+1}= { b^-_2 \over b^-_1 } ~,~~~~~
x^-_i = { \tilde b^+_2 \over \tilde b^+_1 }
\cr
(i+1,  i+1):&~~~~ { 1 \over r} =  b_1^- \tilde b_1^-
e^{-u} ~,~~~~~~~~ x^+_{i+1} = { b_2^- \over b_1^- } ~,~~~~~
x^-_{i+1} = { \tilde b^-_2 \over \tilde b^-_1 }
\cr
(i+2,  i+1):&~~~{1 \over r}=( b_1^+ + \stokesparameter^L_{i+1} b_1^-)\tilde b_1^-
e^{-v}~,~~~~
x^+_{i+2}={( b_2^+ + \stokesparameter^L_{i+1} b_2^-) \over ( b_1^+ + \stokesparameter^L_{i+1} b_1^-) }~,~~~~
x^-_{i+1} = { \tilde b^-_2 \over \tilde b^-_1 }
\cr
(i+2,i+2):& ~~~{1 \over r}=( b_1^+ + \stokesparameter^L_{i+1} b_1^-)( \tilde b_1^+ + \stokesparameter^R_{i+1} \tilde b_1^-)
e^{u}~,~~~~
\cr &~~~~~~~~~~~~~~~~~~~
x^+_{i+2}={( b_2^+ + \stokesparameter^L_{i+1} b_2^-) \over ( b_1^+ + \stokesparameter^L_{i+1} b_1^-) }~,~~~~
 x^-_{i+1} = {( \tilde b_2^+ + \stokesparameter^R_{i+1} \tilde b_2^-) \over ( \tilde b_1^+ + \stokesparameter^R_{i+1} \tilde b_1^-) }
 }}
Notice that the spacetime indices,  $a$,   of   $b^\pm_a$   label
the different components of the target space coordinates. The upper indices, $\pm$,  are related
to the sign of the exponential in \simples . The $b$'s that appear in \cuspsol\ are the ones
appearing in the dominant exponentials at each cusp. Note that the Stokes data appears only after a
few steps. Notice that the values for the $x^\pm_j$ are independent of the worldsheet coordinates
$u,v$ which is simply
saying that we are sitting at one of the cusps. Of course, the value of $1/r$ does depend on
$u$ and $v$ and this is what will enable us to introduce a cutoff.
Let us record the values of some of the kinematic variables
\eqn\kinemva{\eqalign{
x_{i+1}^+ - x_i^+ = & { b_2^- b_1^+ - b_2^+ b_1^- \over b_1^+ b_1^- } = { 1 \over b_1^+ b_1^-}  ~,~~~~~~~
x_{i+2}^+ - x_i^+ =  {  \stokesparameter^L_{i+1}     \over
( b_1^+ + \stokesparameter^L_{i+1} b_1^-) b_1^+ }
\cr
x_{i+1}^- - x_i^- = &   { 1 \over \tilde  b_1^+  \tilde b_1^-}  ~,~~~~~~~~~~~~~~~~~~~~~~~~~~
x_{i+2}^- - x_i^- =  {  \stokesparameter^R_{i+1}     \over
( \tilde b_1^+ + \stokesparameter^R_{i+1} \tilde b_1^-) \tilde b_1^+ }
}}
We will use these expressions when we regularize the area.

\subsec{Computing the regularized area }

Once we introduce the cutoff $r \geq \mu$ we find that, for example, $u$ in the first cusp,
labelled by $(i,i)$ in \cuspsol\ cannot become too big.
By looking at the expression for $r$ at the cusp $(i,i)$ we find that $r\geq \mu$ translates
into
\eqn\restr{
  u \leq  - \log \mu + \delta u_i ~,~~~~~~~\delta u_i \equiv - \log( b_1^+ \tilde b_1^+)
  }
  Similarly, we could consider the cusp $(i+1,i+1)$ and then we find
  \eqn\finrelx{
   u \geq - \left( - \log \mu + \delta u_{i+1} \right) ~,~~~~~~
   \delta u_{i+1} \equiv - \log( b_1^- \tilde b_1^-)
  }
  We think of $\delta u_i$ as the shifts in the position of the cutoff for the integral in the
  $w$ plane which depends on the kinematics. They are defined in such a way that a positive
  $\delta u_i$ would increase the area, both in the positive $u$ regions and the negative $u$
  regions.
  Notice that using \kinemva\ we obtain the useful relation
  \eqn\deltaxrel{
   \delta u_i + \delta u_{i+1} = \ell^+_i + \ell^-_i ~,~~~~~~\ell_i^\pm \equiv
    \log(x_{i+1}^\pm - x_i^\pm )
  }
 This relation will enable us to express most of the dependence of the area in terms
 of the physical spacetime quantities.
We can similarly introduce $\delta v_i$ at the cusp $(i+1,i)$ and $\delta v_{i+1} $ at cusp
$(i+2,i)$. We again find a formula similar to \deltaxrel\
\eqn\deltayrel{
  \delta v_{i} + \delta v_{i+1} = \ell^+_{i+1} + \ell^-_i
  }
  Here we have ignored an $i \pi$ that would arise from the minus sign in the expression for
  $1/r$ at cusp $(i+1,i)$ in \cuspsol . Such terms can be fixed at the end using the spacetime conformal
  Ward identity \anomward\
  and we will not keep track of them.
   Both of \deltaxrel\ and \deltayrel\ allow us to fix all the $\delta u_i, ~\delta v_i$ in
   terms of just one of them.
   Going all around the surface we get that
   \eqn\getequ{ \eqalign{
   \delta u_{n+1} = &  \delta u_{n+1} + \delta u_n - ( \delta u_{n-1} + \delta u_{n-2} ) +
   \cdots  + ( \delta u_2 + \delta u_1 )-  \delta u_1 ~,~~~~~~~~~~n ~~~{\rm odd}
   \cr
   \delta u_{n+1} = &  \delta u_{n+1} + \delta u_n - ( \delta u_{n-1} + \delta u_{n-2} ) +
   \cdots  - ( \delta u_2 + \delta u_1) +  \delta u_1 ~,~~~~~~~~~~n ~~~{\rm even}
    }}
    In the case that $n$ is odd,  we choose the origin in the $w$ plane so that $w\to -w$ when we
   go around the  $w$ plane $n/2$ times. This implies  that
 $\delta u_{n+1} = \delta u_1$. Then the first equation \getequ , together with \deltaxrel , \deltayrel,
  can be used to determine $\delta u_1$ in terms of spacetime quantities.
  In the case that $n$ is even we will have that $\delta u_{n+1} = \delta u_1 +  \shiftedu $.
  Then the second equation in \getequ\  determines $ \shiftedu$ in terms of spacetime quantities.
   This relation, together with a similar relation for $\delta v_{n+1}$,  implies \crosssimp .
   Here we are denoting $\shiftedu , ~\shiftedv$ as the values of $w_s$ in the coordinates in
   \newcoord .

   In the $n$ even case we have not yet determined $\delta u_1$.
    Note that the
   combinations in \deltaxrel\ and \deltayrel\ are invariant under a translation in the $w$
   plane (a translation changes $\delta u_i \to \delta u_i + (-1)^i \epsilon $).
   In order to fix the overall
    magnitude of the $\delta u_i$, and $\delta v_i$ we use the relation
   \eqn\newrels{\eqalign{
   \log (x^+_{i+2 } - x^+_i) (x^-_{i+1} - x^-_i)  = & \log \stokesparameter^L_{i+1} + \delta v_{i+1} +
   \delta u_i
   \cr
   \log (x^-_{i+2 } - x^-_i) (x^+_{i+2} - x^+_{i+1})  = &
    \log \stokesparameter^R_{i+1} + \delta u_{i+2} +
   \delta v_i
 }}
 where we used \kinemva .
 This combination of $\delta u_j$, $\delta v_j$ is not translation invariant in the $w$ plane
 and it can be used to remove completely all $\delta u_j$ at the expense of introducing
 (at least) one Stokes parameter $\stokesparameter^L$ and one $\stokesparameter^R$.
  This will be necessary in the case of $n$ even.

\subsec{Computing the regularized area for $n$ odd }

 Let us start with the case of $n$ odd. We will now compute the area in the following way.
 We first assume that the $w$ surface  is a simple branched cover over the $w$ plane,
 as we had for the regular polygon. We set this branch point at the origin of $w$.
 When we go around the $z$ plane we are going around the $w$ plane $n/2$ times and we map
 $w \to - w $.

 \ifig\xyarea{ Here we compute the part of the area that depends on the physical regulator. We
 replace the $w$ Riemman surface by the one we had for the regular polygon. Each line corresponds to
 a cusp. Note that since we are using the $u,v$ coordinates in \newcoord\ the figure is rotated by
 45 degrees relative
 to the figure in the $w$ plane. We compute the area by summing the area of each triangle, being careful
 to change the sheet as we go around the $w$ plane. } {\epsfxsize3.0in\epsfbox{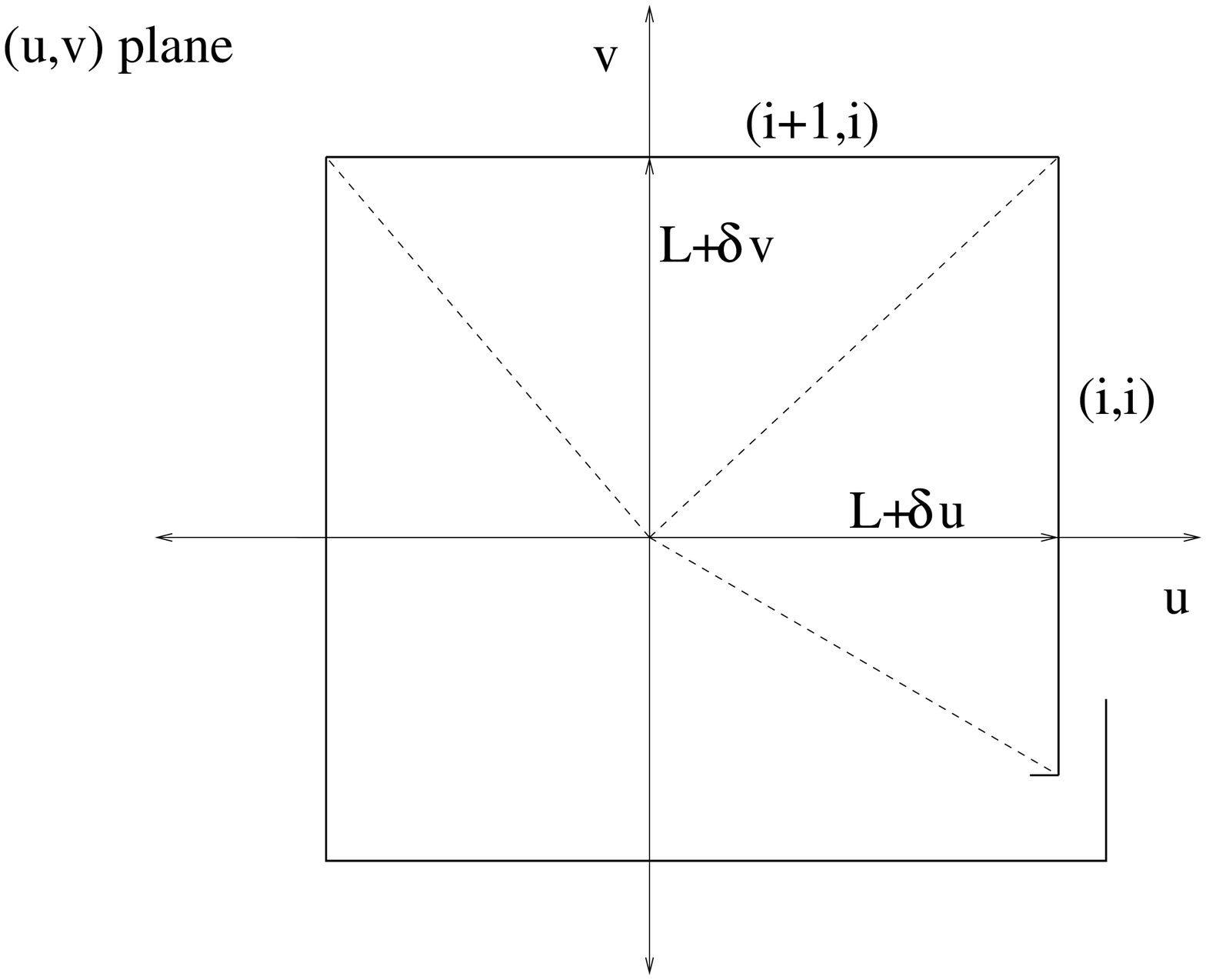}}

We can now compute the area. We divide the plane into triangles, as in \xyarea  .
 The height of the triangle involving cusp $(i,i)$  is
  given by $L + \delta u_i$, where $L = - \log \mu $.
 The base of the triangle is given by
 $  2 L + \delta v_i + \delta v_{i-1} $. We also have similar expressions for the other triangles.
 Adding the areas of all $2n$ triangles we get
 \eqn\areatri{
  A_{cutoff} = { 1 \over 4 } \sum_{i=1}^n ( L + \delta u_i ) ( 2 L + \delta v_{i+1} + \delta v_i ) + ( L + \delta v_{i+1} ) ( 2 L +
   \delta u_{i+1} + \delta u_{i} )
   }
 Recall that  $\delta u_{n+1} = \delta u_1$, $\delta v_{n+1} = \delta v_1$ and $L =- \log \mu$.
 We can rewrite this as
  \eqn\areawtot{ \eqalign{
    A_{cutoff}  = &  A_{div} + A_{BDS-like}
  \cr
  A_{div} = & \sum_i   { 1 \over 8 } [ 2 L + \delta u_{i+1} + \delta u_i ]^2 +
   { 1 \over 8 } [ 2 L + \delta v_{i+1} + \delta v_i ]^2
   \cr
   A_{BDS-like} = & { 1 \over 4 }\left[  \sum_{i=1}^n  - (  \delta u_i)^2 -
     ( \delta v_i )^2 + \right.
   \cr  & ~~~~~~~ \left. +
    \delta u_i  (  \delta v_{i+1} + \delta v_i  - \delta u_{i+1}) +   \delta v_{i+1}  (
   \delta u_{i+1} + \delta u_{i} -\delta v_{i}  ) \right]
  }}
  We see that the divergent term $A_{div}$ can be written as \areaf\ using \deltaxrel \deltayrel .
  As we explained above, for $n$ odd, we can use \deltaxrel , \deltayrel\ and \getequ \
  to  determine all
  the $\delta u_i$ and $\delta v_i$ in terms of spacetime quantities. Inserting those expressions
  in \areawtot\ we can see that $A_{BDS-like}$
   can be rewritten as \bdslike . Notice that $A_{BDS-like}$  \bdslike\ involves only distances
  between one cusp and the next one, only differences between neighboring values of $x^+$ or
  $x^-$ appear in \areawtot . This should be contrasted to the BDS expression
  \resabd\ which involves differences
  that are not nearest neighbors.

  Let us mention a couple of subtleties. First note that there are some $i \pi $'s which we have
  neglected in the expression for some of the $\delta v_i$. Such terms would lead to ambiguities
  involving only the $\ell_i^+$. The ward identity \anomward\ would fix such terms, since we cannot
  write cross ratios purely with the $\ell_i^+$ (for $n$ odd). The second issue is the following.
  Suppose that we order the $x_i^+$ in such a way that $x_i^+ < x^+_{i+1}$, then most of
  the $\ell_i^+ = \log(x^+_{i+1} - x^+_i ) $ are real. However, $\ell_n^+ $ will be the log of
  a negative number and it will contain a $ \pm i \pi $ which would make \bdslike\ complex . This
  is related to the fact that we cannot embed a Wilson null polygon in $R^{1,1}$ with all cusps forward
  or backward directed. One option is to send $x^\pm_{n} \to \infty$, throwing away all divergent terms
  together with the $i\pi$'s. A similar issue occurs with the BDS expression \resabd . However,
  we are interested in the remainder function which involves $A_{BDS-like} - A_{BDS}$ which
  is a conformal invariant function of the cross ratios. This expression is well defined and
  does not have any of these problems and it is a function of the cross ratios. In fact,
  this difference is computed in appendix D.

 So far, we have computed the area in the $w$ space neglecting the structure that occur in the
 interior. We simply inserted a branch cut at the origin.
 Now we would like to take into account the structure of cuts near the interior.
 One way to think about this is the following. The finite piece can be written as
 \eqn\integr{
 A_{periods} = 4 \int_{|w| < \Lambda}  d^2 w  - { i  }  \int_{|w| = \Lambda} (  w d \bar w - \bar w d w )
 }
 where the first piece is the integral with all branch cuts included and the second piece
 is the integral we would have obtained if we had neglected them, except that we wrote it purely in
 terms of the behavior of the Riemann surface for large $w$.
 The second piece is canceling the divergence that we have in the first term, and it is precisely
 of the form already included in \areawtot .
 One can show the finite expression \integr\  can  be expressed in terms of periods. For this
 purpose we can choose a basis of electric and magnetic cycles whose intersection numbers
 are canonical $\gamma^e_{r} \wedge \gamma^{m s} = \delta^s_r$. For $n$ odd we have
 ${ n -3 \over 3}$ cycles of each kind and they are all compact cycles, see \cycleseven .
 We can denote by
 $w^e_r$ and $w^{m r}$ the corresponding integrals. We omit the details of the derivation,
 but  we obtain the formula cited in the
 main text \aprep .
 The polynomial $p$ has $n-3$ non-trivial complex parameters which translate into the
 complex values of the periods we had above\foot{This Riemann surface, $y^2 = p(z)$, also appears
 in the description of some ${\cal N}=2$ systems in four dimensions. In that case one often
 separates the $(n-3)/2$ ``non-normalizable'' parameters in $p$
from the ``normalizable'' ones. In our problem   they all  appear on a equal footing and
the cutoff is introduced in a different way.   }.

 \subsec{Computing the regularized area for  $n$ even }

 The computation of the area in the case that $n$ is even is a bit more complicated.
 In this case we cannot simply approximate the structure of the $w$ plane at infinity by
 that of the regular polygon because we have the shift by $w\to w + \wshifted$ when we go around.
 \ifig\wplaneeven{ Possible example of $w-$plane. This corresponds to the case $n=10$ and a particular location of the zeroes. The total shift when we come back to the first sheet is $w_s$.
 } {\epsfxsize3.0in\epsfbox{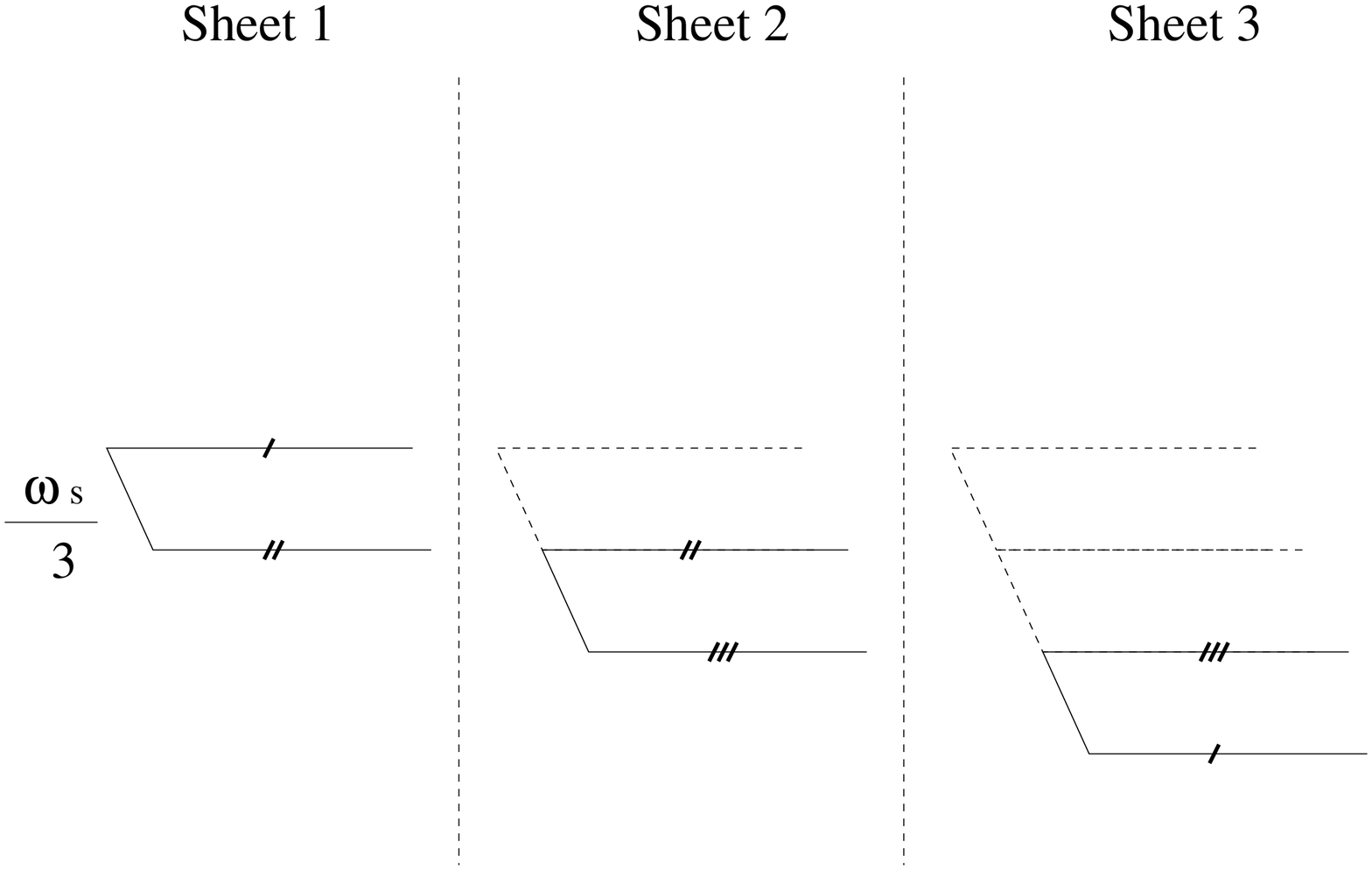}}
 The structure of the $w$ plane is that of a simple branched cover with a sliver missing.
 See for instance figure \wplaneeven  .  We can put this sliver along the first cusp.

 In order to compute the area it is convenient to separate the problem in two parts.
  First we consider the computation of the area in the large $w$ region which is sensitive
  to the physical regulator. We call this $A_{cutoff}$ as before.
   In order to compute this part we choose a simple reference
  Riemann surface which has the same structure at infinity. The Riemann surface under
  consideration has $(n-4)/2$ pairs of compact electric and magnetic cycles. In addition, it has
  an electric cycle going around all zeros at large $z$ whose period, $\wshifted$, can still be felt at
  infinity. Its dual magnetic cycle is non-compact. We take a
   reference Riemann surface where we shrink all the compact cycles but we leave the $\wshifted$ period.
   This implies that the Riemann surface has a structure such as the one
    summarized in \wplaneevenref .

   \ifig\cyclesevenref{In (a) we have a generic configuration for the zeros of the polynomial
   for the $n$ even case. In (b) we consider  a reference configuration where we move the zeros
   to a location such that we shrink all the periods except for the one corresponding to
   $\gamma_1^e$, the cycle that goes around all zeros.
 } {\epsfxsize4.5in\epsfbox{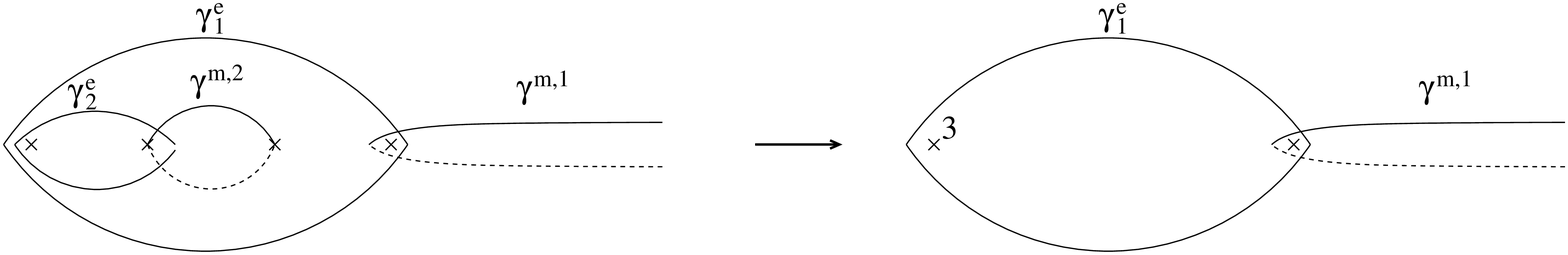}}

    We
   can equivalently say that we put all the zeros of the polynomial at the same location except
   for one of the zeros whose position is chosen so as to give the same $w_s$ as the original
   polynomial.
   We choose the origin of $w$ at one of the zeros.
  We can choose the location of the branch cuts so
  that the sliver points in the direction corresponding to the first cusp, the one characterized
  by  $\delta u_1$.
  \ifig\shiftedarea{Computation of the area for the case of $n$ even with a monodromy $w_s$.  In the figure for the particular case $n=4$. Each sheet is missing an equal sliver and we choose the origin of the $w$ plane to coincide with one of the zeroes of $p$. When we return to the first sheet, the coordinates $u$ and $v$ shift by $u_s$ and $v_s$.
 } {\epsfxsize4in\epsfbox{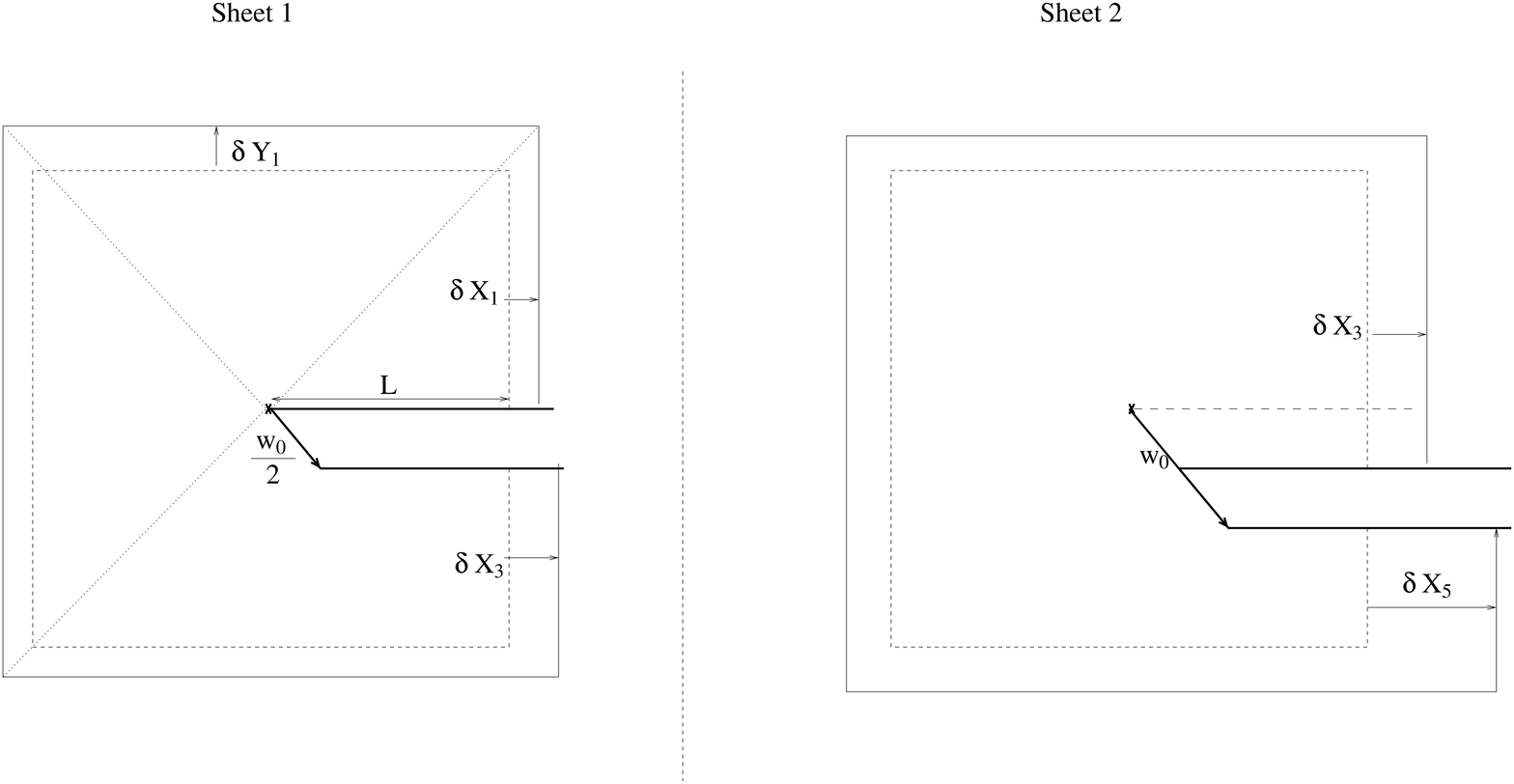}}
  In this case we obtain the following expression for the area
  \eqn\areacuto{ \eqalign{
  A_{cutoff} = &  { 1 \over 4 } \left[
  \sum_{i=1}^n ( L + \delta v_i) ( 2 L + \delta u_{i+1} + \delta u_i )
+   \sum_{i=2}^n ( L + \delta u_i) ( 2 L + \delta v_i + \delta v_{i-1} ) +
 \right.
  \cr
&\left. +  ( L + \delta u_1) ( L + \delta v_1) + (L + \delta u_{n+1}) (L + \delta v_n)
 + 2 (L + \delta u_{n+1} ) v_s - { u_s v_s   } \right]
}}
In this expression we should set $\delta u_{n+1} = \delta u_1 + \shiftedu$, $\delta v_{n+1} = \delta v_1 + \shiftedv$,
wherever they appear. The first line in \areacuto\ represents the area of all the
triangles associated to the cusps except for the first one. The last term is the area of the
first triangle.  We denoted some of these triangles in \shiftedarea .

It is interesting to note that this expression is sensitive to the origin that we choose
for $w$. In other words, if we were to change $\delta u_i \to \delta u_i + (-1)^{i} \epsilon_u$ and
$\delta v_i \to \delta v_i + (-1)^i \epsilon_v $, then we see that the area changes as
\eqn\change{
A_{cutoff} \to A_{cutoff} + { 1 \over 2 } ( \epsilon_v \shiftedu - \epsilon_u \shiftedv  )
}
This means that the area depends on the origin we choose for the wedge that we have in the
reference Riemann surface. This ambiguity should cancel once we add the result of the
integral on the Riemann surface with the proper cut structure, which we call $A_{periods}$.
We could also say that when we choose the reference Riemann surface, we want it to
have the same value for the non-compact magnetic period as the original. Of course, both are
infinite, but we want the finite pieces to be the same.

We can now use \deltaxrel , \deltayrel\ to determine all the $\delta u_i$ and $\delta v_i$ in
terms of physical quantities and in terms of $\delta u_1$ and $\delta v_1$. We can
now use \newrels\ to solve for $\delta u_1, ~\delta v_1$. The result for $\delta u_1, ~\delta v_1$
involves one of the Stokes parameters and one non-consecutive
 difference of the $x^+_i$.

 We can then express everything in terms of spacetime quantities and the two Stokes factors,
 one left and one right.
 This leads to
 $A_{cutoff} = A_{div} + A_{BDS-like-even} + A_{extra} $, where
 $A_{BDS-like-even}$ is written in \abdslikeeven ,   $A_{extra}$  in \aextra\ and $A_{div}$ in  \areaf .

\eqn\resar{ \eqalign{
 A_{cutoff} = & A_{div} + \sum_{i,j} \ell^+_i M_{ij}^{(1)} \ell^-_j -(\sum_{i=1}^n (-1)^i \ell_i^+ )^2-(\sum_{i=1}^n (-1)^i \ell_i^- )^2+
  \cr +&   { 1 \over 2 }
 ( \log \gamma_1^L - \hat \ell^+_1 )
 \sum (-1)^i \ell^-_i -
   { 1 \over 2 } (\log \gamma_1^R - \hat \ell^-_1 )
 \sum (-1)^i \ell^+_i
 \cr
 \sum_{i,j}  \ell^+_i M_{ij}^{(1)} \ell^-_j = &
  - { 1 \over 2 }  \sum_{i=1}^n \sum_{j =1}^{i-1}   (-1)^{i+j} \ell^+_i \ell^-_{j}
  + { 1 \over 2 }  \sum_{i=1}^n \sum_{j =i}^{n}   (-1)^{i+j} \ell^+_i \ell^-_{j}
  -
  \cr
  &~~~~ { 1 \over 4 } \sum_{i=1}^n \ell^+_i ( \ell^-_i + \ell^-_{i-1} )
 \cr
  \hat \ell_i^\pm  \equiv & \log ( x^\pm_{i+1} - x^\pm_{i-1} ) - \ell_i^\pm - \ell^\pm_{i-1}
   }}
  Note that only $\hat \ell_1^\pm$ appear in the expression for the area.
  This form of writing it emphasizes that the first cusp is treated differently, since it is
  there that the sign changes in the sum. We could define a similar sum but with
  $M^{(2)}$ which is defined as $M^{(1)}  $ except that the lower and upper
   limits of both sums that define $M^{(1)}$  are changed as
  $1 \to 2$ and $n\to n+1$, with the rest of the sum remaining the same. We find
  \eqn\twoexp{
  \sum_{i,j} \ell^+_i M_{ij}^{(2)} \ell^-_j - \sum_{i,j} \ell^+_i M_{ij}^{(1)} \ell^-_j =
 \ell^+_1 \sum_{i=1}^n (-1)^i \ell^-_i -  \ell^-_1 \sum_{i=1}^n (-1)^i \ell^+_i
  }
  We find that this cancels the change in the combination
  \eqn\combinch{ \eqalign{
  \log \gamma_2^L - \hat \ell^+_2 = &  - ( \log \gamma_1^L - \hat \ell_1^+ ) + 2 \ell_1^+
  \cr
  \log \gamma_2^R - \hat \ell^-_2 = &  - ( \log \gamma_1^R - \hat \ell_1^- ) + 2 \ell_1^-
  }}
  This is the change we would have in the other term of the area if we wrote it privileging
  the first cusp. Of course one could also take the average of all these expressions and
  write something which does not privilege any cusp.

   Finally we can compute the piece that involves the structure of cuts in the $w$ plane.
   This is very similar to the $n$ odd case, except that now we have $(n-4)/2$ pairs of
 compact electric and magnetic cycles, while there is one electric cycle going around all the
 zeros, whose period is $\wshifted$, and whose dual magnetic cycle goes to infinity. We can choose
 this dual magnetic cycle to start at one particular zero of the polynomial, defining the origin
 of $w$ at this point. The difference between the area computed so far and the true area
  of the $w$ space is given by \aprepeven .

\appendix{C}{Relation between the area and the metric in the moduli space}

In this appendix we derive the formula \metricarea\ relating the metric in moduli
space for the Hitchin equations and the area of the surface in $AdS$.

First let us review how we compute the metric in moduli space.
We
consider the Hitchin equations \hitchineq\
and we consider a small deformation of the solution $\delta A_\mu
$ and $\delta \Phi_\mu$ which continues to obey the equations. The
metric is computed using
 \eqn\metricc{
 \delta s^2 = \int d^2 z Tr[
- \delta A_z \delta A_{\bar z} + \delta \Phi_z \delta \Phi_{\bar z} ]
  }
 the
minus sign arises because we are considering anti-hermitian $A$'s.
However, before computing the metric, we must ensure that the
variation we are considering is orthogonal to gauge
transformations. This implies that the small variation should obey
the equations
  \eqn\smallv{
 -D_z \delta A_{\bar z} - D_{\bar z} \delta A_z + [\Phi_z ,\delta\Phi_{\bar z}]  + [ \Phi_{\bar z} ,
 \delta \Phi_{z} ] =0
 }

We know that the moduli space is parametrized by the polynomial $p$. For each choice of the
polynomial there is a unique solution for $\alpha$ obeying the boundary conditions.
We  parametrize the small variation of the polynomial as $\delta p$.
This leads to an equation for  $\delta \alpha $ which comes from linearizing
  \gensinh
\eqn\varequ{
\partial \bar \partial \delta \alpha - 2  \delta \alpha ( e^{2 \alpha }
+ e^{ - 2 \alpha } |p|^2 )  + e^{-2 \alpha} ( \bar p \delta p + \delta
\bar p  p ) = 0 }
A first approximation for  $\delta A$ and $\delta \Phi$ can be obtained by performing a small variation
$\delta \alpha, ~\delta p,~ \delta \bar p $ on the original expressions for $A$ and $\Phi $
\condec , \Bgeneric .
 However,  in order to obey \smallv\ we should also add the result of
performing
a gauge transformation with parameter $\gamma$,
 $\Phi \to \gamma [ \sigma^3, \Phi]$ and $A \to A - d \gamma \sigma^3 $.
 Imposing \smallv\ we get an equation determining $\gamma$.
\eqn\orthog{
   0 =   2 \partial \bar \partial \gamma  +   e^{ - 2 \alpha } ( - \delta \bar p p + \bar p \delta p ) -
   { 4 \gamma} ( e^{2 \alpha} + e^{- 2 \alpha } |p|^2 )
   }
  In order to solve the equation for $\gamma$ it is convenient to consider first
  a holomorphic variation $\delta p$ of the polynomial with $\delta \bar p =0$. Since the equations
  are linear, there is no problem in assuming that these two variations are independent.
  In this case one can see that we can choose
  $\gamma = \delta \alpha /2$.
  We can now do the same for an antiholomorphic variation $\bar \delta p =0$, $\bar \delta \bar p \not =0$.
  This leads to a solution $\bar \delta \alpha $. From now on the expression
  $\delta \alpha$ will denote the solution of \varequ\ with $\delta \bar p =0$ and $\bar \delta \alpha$ is
  the same but with $\delta p =0$ \foot{
  More explicitly, the equation for $\delta \alpha$ is $\partial \bar \partial \delta \alpha - 2  \delta \alpha ( e^{2 \alpha }
+ e^{ - 2 \alpha } |p|^2 )  + e^{-2 \alpha} \bar p \delta p  = 0 $ and the
equation for $\bar \delta \alpha$ is $ \partial \bar \partial \bar \delta \alpha - 2  \bar \delta \alpha ( e^{2 \alpha }
+ e^{ - 2 \alpha } |p|^2 )  + e^{-2 \alpha}   \bar \delta
\bar p  p  = 0 $.}. For the anti-holomorphic variation we see that
  $\gamma = - \bar \delta \alpha/2$. Combining the two we can say that
  $\gamma$ is simply the sum $\gamma = \delta \alpha/2 - \bar \delta \alpha/2$.
  We can now insert this in the expression for the metric \smallv\ and we
  obtain
  \eqn\metricog{
   ds^2 = \int dz^2  \left[
   e^{ - 2 \alpha} \delta p \bar \delta \bar p  - e^{ - 2 \alpha } ( p \delta \alpha
   \bar \delta \bar p  + \bar p \bar \delta \alpha \delta p ) \right]
   }
We would now like to find an equation for the Kahler potential. We propose that the
following expression gives us a Kahler potential for the metric \metricog\
\eqn\kahme{
K =  \int (e^{2 \alpha } + e^{-2\alpha } |p|^2 ) +   \int \partial \alpha \bar \partial \alpha
}
In order to check that the second derivative of $K$ gives the metric \metricog\ we first compute
 the holomorphic derivative by considering the holomorphic variation
\eqn\holder{\eqalign{
  \delta K = & \int  \left[  2  \delta \alpha ( e^{2 \alpha} - e^{-2 \alpha } p \bar p) +
  e^{-2 \alpha } \delta p \bar p -  2  \delta \alpha  \bar \partial \partial \alpha \right]
\cr
  \delta K = &   \int  \left[
  e^{-2 \alpha } \delta p \bar p \right]
}}
where   we have used the equation of motion for $\alpha$.
We can now consider the anti-holomorphic variation
\eqn\variat{\eqalign{
\bar \delta \delta K = &  \int
  e^{-2 \alpha } \delta p \delta \bar p  -2  e^{ - 2\alpha} \bar \delta \alpha  \delta p  \bar p
\cr
\bar \delta \delta K = &   \int
  e^{-2\alpha } \delta p \delta \bar p  -   e^{ - 2 \alpha} ( \bar \delta \alpha  \delta p  \bar p
+ \delta \alpha p \delta \bar p ) }}
We have used
that $\int e^{-\alpha } \bar \delta \alpha \delta v \bar v =
\int e^{-\alpha}  \delta \alpha  v \delta \bar v $
which can be proved by using the equation of motion for $\delta \alpha$ first and then the one for
$\bar \delta \alpha$.

Now imagine writing the polynomial $p_{n-2}(z) = \lambda^{n-2}  \prod_{i=1}^{n-2} ( z - z_i)$,
with $\sum z_i =0$.
 By performing a rescaling of $z_i$ we could remove $\lambda$ and absorb it into the $z_i$.
 So the physics only depends on $\lambda z_i $.
 However, it is convenient, for a moment, to think in terms of this   modulus, $\lambda$,
  since
 the derivative of the Kahler potential with respect to it is closely related to the area
 we would like to compute
 \eqn\areacomp{
  \partial_\lambda K |_{\lambda =1} = (n-2)  \int d^2 z e^{- 2 \alpha} p \bar p
  }
  Using the equation for $\alpha$ we see that
  the right hand side  differs from the area only by a simple constant
  \eqn\diffa{
  \int d^2 z ( e^{2 \alpha} -  e^{- 2 \alpha} p \bar p   ) = \int d^2 z \partial \bar \partial \alpha =
  \int_{|z|\gg 1 } { 1 \over 2 i } ( dz \partial \alpha - d \bar z \bar \partial \alpha )= { \pi \over 2 } (
   n -2 )
  }
  We used that
  $\hat \alpha$ vanishes at infinity, which implies that $\alpha \sim { 1 \over 4 } \log p \bar p$ at
  infinity, which gives the constant. This is a
  simple moduli independent constant that we will drop. We will be able to fix
  this overall constant in the
  area by comparing to the result for the regular polygon.

  As we explained above, the Kahler potential is expected to be a function only
  of $\lambda z_i $. Thus we can replace the left hand side in \areacomp\ by
  $ \sum_i z_i \partial_{z_i} K $. Of course we could have done the same using the
  anti-holomorphic part.
  Thus we can write the final expression for the area as in
 \metricarea .
 The proportionality constant between this expression and the area depends on the precise
 way we normalized the metric in the moduli space. However, one can fix this constant by going
 to some limits where one can perform the computation. Here we
 have assumed that the freedom for choosing $K$ is fixed by
 demanding rotational invariance $z_i \to e^{i \varphi } z_i $.
 In fact,   $D =  \sum_i z_i \partial_{z_i} K =  \sum_i {\bar z}_i \partial_{\bar z_i} K$
 is the moment map for the rotational symmetry. In other words, this symmetry is generated by
  $ \delta z_l =
 \sum_j i g^{z_l \bar z_j} \partial_{\bar z_j} D  = i z_l $, $ \delta \bar z_l = -
 \sum_j i g^{\bar z_l  z_j} \partial_{ z_j} D  = - i \bar z_l $.
 In fact, if we imagine a supersymmetric theory where the target space is given by the $z_i$, viewed
 as chiral superfields, then $D$ would be the D-term associated to the weak gauging of the $U(1)$
 rotation symmetry.

\appendix{D}{Large number of cusps}

In this appendix we study the regular polygon solution in the limit of a large number of cusps. This makes contact with the zig-zag world-sheet considered in \AldayHE . We make the following choice for the holomorphic polynomial
\eqn\equts{
 p(z)= \kappa_n z^{n-2},~~~~~~~~{ 1 \over 4} (\alpha'' + \alpha'/\rho ) - e^{2\alpha} +  \kappa_n^2 \rho^{2(n-2)}e^{-2\alpha } =0
}
The constant $\kappa_n$ has been introduced in such a way that $\alpha=0$ at the origin. Using the results of \McCoyCD\ we find $\kappa_n=4n^2 \left(-{\Gamma(1/n)\over \Gamma(-1/n)}\right)^n$. We will study solutions to this equation in the limit $n \rightarrow \infty$. In the range $\rho<1$, the last term of equation \equts\ vanishes and we are left with the Liouville equation, which can be easily solved to give
\eqn\solune{
 e^{2 \alpha } = { 1 \over (1 - \rho^2 )^2 }
 }
The constants of integration have been fixed in such a way that $\alpha$ is regular for $\rho<1$ and blows as $\rho \rightarrow 1$. In order to analyze the $\rho >1$ it is convenient to make a change of coordinates
\eqn\scaling{\eqalign{
  z = e^{ \epsilon/n} ~,~~~~~~  \tilde \kappa = \lim_{n \rightarrow \infty} {\kappa \over n^2} =4 e^{-2\gamma_e} \cr
w = \sqrt{\kappa_n} { 2 z^{ n/2} \over n } \to w = 2 \sqrt{ \tilde \kappa} e^{ \epsilon/2},~~\bar w = 2 \sqrt{ \tilde \kappa} e^{ \bar{\epsilon}/2}
}}
where $\gamma_e$ is the Euler gamma constant. Notice that as we take the $n \rightarrow \infty$ limit, $w$ remains finite. On the other hand, $\hat \alpha(w,\bar w)$ is also independent of $n$ and satisfies the usual sinh-Gordon equation, supplemented with somewhat peculiar boundary conditions at the origin \foot{Notice that the whole $\rho<1$ region is mapped to the origin of the $w-$plane.}
 \eqn\boundcont{
  e^{2\hat \alpha }  \sim  { 1 \over 4 \tilde \kappa  (Re(\epsilon))^2 } e^{ - Re(\epsilon) } \sim { 1 \over 4 |w|^2
  \log( {|w| \over 2 \sqrt{\tilde \kappa} } ) }
 }
This kind of boundary conditions was also considered in \McCoyCD\ and \CecottiME\ . In order to find the space time solution corresponding to the zig-zag, one should perform the inverse map. We can write $\epsilon=x+i y$, with $-\infty < x,y < \infty$. Then $x$ is related to $|w|$ while $y$ depends on its phase. As $n$ tends to infinity, the coordinate $y$ becomes non compact, however the solution will be periodic in this coordinate $y$. The boundary of $AdS$ is reached when $x \rightarrow \infty$, where the radial $AdS$ coordinate should vanish and the solution should look like a zig-zag. On the other hand, for $x \rightarrow -\infty$, the world-sheet should approach a straight line, sitting at $t=0$ and extending along the other two coordinates of $AdS_3$.

In order to solve for the inverse map, we can solve the linear problem in $w-$coordinates, {\it e.g.} using \linpro . The solution obtained this way, will be related to the zig-zag solution by conformal transformations.

As already mentioned, for the zig-zag solution, the World-sheet approaches the straight line as $x \rightarrow - \infty$. The solution corresponding to the straight line can be reduced to the generalized sinh-Gordon model and it corresponds to
 \eqn\straightline{p(z)=0,~~~~~e^{2\alpha}={1 \over (z+\bar{z})^2}
 }
 which is of course essentially the same as \solune .
It is then very simple to solve the linear problem and find the holonomies $\Omega^{L,R}$ up to a set of undetermined constants. Requiring the solution to have the correct properties, fixes most of these constants. We find
\eqn\straightlineOm{\Omega^L=\left(\matrix{{i \over  c \sqrt{2 x}} &
 { c(y-i x) \over \sqrt{2x}} \cr -{i \over c \sqrt{2 x}} & {-c(y+i x)\over \sqrt{2x}} }\right),~~~~~~~\Omega^R=\left(\matrix{{1 \over c \sqrt{2 x}} &
 { c(x-I y) \over \sqrt{2x}} \cr -{1 \over c \sqrt{2 x}} & {c(x+i y)\over \sqrt{2x}} }\right)
 }
Where we have written $z=x+i y$. The space time element in simply given by
\eqn\straightlineY{Y=\left(\matrix{Y_{-1}+ Y_2 &
 Y_1- Y_0\cr Y_1+ Y_0 & Y_{-1}- Y_2 }\right)=(\Omega^L)^T \Omega^R=\left(\matrix{{-i \over c^2 x}&
 -{y \over x}\cr -{y \over x} & {i c^2(x^2+y^2)\over x}}\right)
 }
Which has the correct features of the straight line solution.

\appendix{E}{Relation between one loop and strong coupling results}

As seen in the body of the paper, the area at strong coupling contains a piece
``$A_{BDS-like}$".
 For   $n$ odd   this is given by
\eqn\bdslikeap{\eqalign{
A_{BDS-like} = &   \ell^+_i M_{ij} \ell^-_j={ 1 \over 2 }
\sum_{i=1}^n \sum_{j=i}^{n+i-2}(-1)^{i+j}\ell_i^+ \ell_j^-+ { 1 \over 4 }
\sum_{i=1}^n \ell_i^+(\ell^-_{i-1}-\ell_i^-)
  \cr
  \ell_i^+ \equiv  & \log(x^+_{i+1} - x^+_i ) ~,~~~~~~~~~~\ell_i^- \equiv  \log(x^-_{i+1} - x^-_i )
  ~,~~~~~~\ell^\pm_{n+i} \equiv \ell^\pm_i
  }}
On the other hand,
 the BDS expression
   restricted to our particular kinematical configuration takes the very simple form
\eqn\resabdap{
A_{BDS} = - { 1 \over 4 } \sum_{i=1}^n \sum_{ j=1,   j \not = i,i-1 }^n  \log { x^+_j - x_i^+ \over x^+_{j+1} - x^+_i }
 \log { x^-_j - x^-_{i-1} \over x^-_j - x^-_i }
 }
and also satisfies the conformal Ward identities. As a result, the difference between the two answers should be a function of the invariant cross-ratios only and it will contribute to what is usually called in the literature, the remaining function.

\ifig\crossratio{We arrange the $x^+_i$ sequentially. Associated to two points, $1$ and $6$ in this case,
we can construct a unique cross section which involves $x^+_{16}$ and consecutive
 differences $x^+_{i,i+1}$.
 In this figure, each highlighted segment represents
  a difference between the two end points,
  the $+$ sign are factors appearing in the numerator of the cross ratio and minus signs
 are factors appearing in the denominator. This diagram is
 presenting only the $x^+_i$. There is a similar one for $x^-_i$.
 } {\epsfxsize1.7in\epsfbox{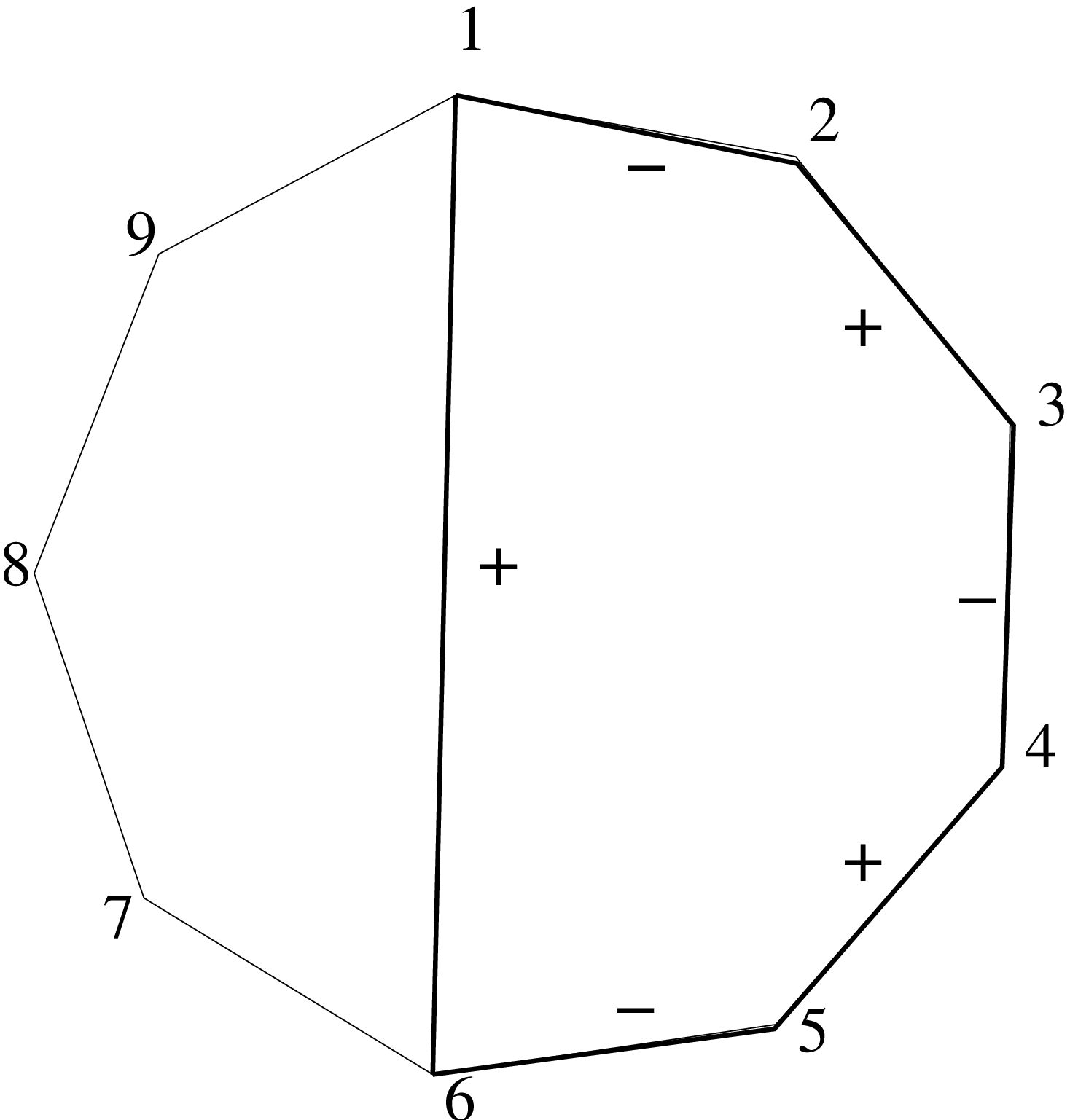}}

 Notice that the strong coupling answer \bdslikeap\ depends only on nearest neighbors distances while this is not the case of the one loop result. Given a distance $x_j-x_i$, with $|j-i|>1$, we can form a unique cross-ratio involving $x_{ji}$ and only nearest neighbors distances. Since $n$ is odd, there is a unique way to close the line going from $x_i$ to $x_j$ along the polygon in such a way that we have an even number of sides (see \crossratio ). We call the cross ratios formed in this way $c_{ij}$ \foot{Of course not all these cross-ratios are independent.}. In general we can think of a cross
 ratio as given by a closed path visiting an even number of points. We label the segments in these
 paths  with
 plus and minus signs sequentially and we interpret these signs as saying whether the differences
 appear in the numerator or the denominator. We can see that in the $n$ even case we can form
 a cross ratio involving only consecutive differences, while this is not possible for $n$ odd.

One then can express every non-nearest neighbor distance in the
BDS ansatz in terms of its corresponding cross-ratio and
neighboring distances. Since, as already mentioned, \bdslikeap\ is
the only combination of nearest neighboring distances which
satisfies the conformal Ward identities we must have
 \eqn\dif{
A_{BDS}-A_{BDS-like} = -{ 1 \over 4 }
\sum_{i=1}^n \sum_{ j=1,   j \not = i,i-1 }^n  \log { c^+_{i,j} \over c^+_{j+1,i}}
 \log {c^-_{i-1,j}  \over c^-_{i,j}}
 }
For the case $n=even$, non-nearest neighbors distances will enter in both answers and the
difference seems to be more complicated.
However, the particular case $n=4$,
 which is the most relevant for the present paper,
  can be worked out in detail and we obtain
 \eqn\diffbds{
 A_{BDS-like-even}-A_{BDS} =- { 1 \over 2} \log(1 + \chi^-) \log ( 1 + { 1 \over \chi^+ } )
 }
 where $\chi^\pm$ are the two cross ratios defined in \singlc .

\appendix{F}{Discrete symmetries }

In this appendix we
  comment on a couple of discrete symmetries of the problem.

The first is worldsheet parity.  It translates into
the fact that if $W$ is a solution of \orthbasis , then
$\sigma^1 \overline{ W} \sigma^1$ is also a solution to the problem, after we exchange $z \to \bar z $
and $p \to \bar p$. This symmetry is ensured by the fact that
$B_{\bar z}^L = \sigma^1 \overline{ B_z^L} \sigma^1 $,  and the same
 expression for $B^R$.

The second discrete symmetry is spacetime parity. We can view it as the symmetry that changes
the target space coordinate $Y_0 \to - Y_0$ leaving everything else the same. In Poincare coordinates
this corresponds to $r,x^+,x^- \to r,x^-,x^+$.
This changes $Y_{a\dot b} \to Y_{\dot b a }$. Thus we expect that this symmetry
exchanges the left and right problems.
Note that this symmetry reverses the sign of $N$, due to the epsilon symbol in \sinhg .
 In other words
$N_{a \dot b} \to - N_{\dot b a}$.
 In fact, it also maps $W \to \sigma^1 W^t \sigma^1$, where
$W^t$ is the transpose. Since there is an extra sign change for
the normal vector $N$, we see from \sinhg\ that we should also
change the signs of $p, ~ \bar p$,  $p \to - p $, $\bar p \to -
\bar p$. Then we see that this is a symmetry of the equations. It
amounts to changing $B^R \to {B'}^R =  \sigma^1 B^L  \sigma^1 |_{p
\to -p, ~\bar p \to - \bar p }$. The change in the sign of $p$ is
what implies that the left and right problems are related by a
change in the  spectral parameter   by a factor of $i$, as in
\leftrightrel . The reason is that the introduction of a spectral
parameter can also be viewed, up to a constant gauge
transformation, as the result of changing $p$ and $\bar p$ as $
p\to p/\zeta^2$ and $\bar p \to \bar p \zeta^2 $.

\listrefs

\bye